%% file: paper.tex
\documentclass[12pt]{article}

\usepackage[title]{appendix}
\usepackage{lscape}
\usepackage{amsfonts}
\usepackage{amssymb}
\usepackage{amsmath}
\usepackage{epstopdf}
\usepackage{float}
\usepackage{natbib}
\usepackage[margin=.75in]{geometry}
\usepackage[hang]{caption}
\usepackage{color}
\usepackage[pdftex]{graphicx}
\usepackage{psfrag}
\usepackage{rotating}

\newtheorem{theorem}{Theorem}[section]

\newtheorem{corollary}{Corollary}[section]

\newtheorem{lemma}{Lemma}[section]

\newtheorem{proposition}{Proposition}[section]
\newtheorem{remark}{Remark}

\begin{document}

\title{Is the diurnal pattern sufficient to explain intraday \\ variation in volatility? A nonparametric assessment\thanks{This paper was previously entitled ``Testing for heteroscedasticity in jumpy and noisy high-frequency data: A resampling approach.'' We appreciate the thoughtful feedback received from two anonymous referees, the associate editor, and Yacine A\"{i}t-Sahalia as part of the revision process, which helped to significantly improve the paper. We also thank for comments on an earlier version of this paper made by conference participants at the 10th Computational and Financial Econometrics (CFE) conference in Seville, Spain; the Vienna-Copenhagen (VieCo) 2017 conference in Vienna, Austria; the 10th Annual SoFiE meeting in New York, USA; the Canadian Econometric Study Group (CESG) 2017 meeting in Toronto, Canada; and at seminars in Aarhus, Aix-Marseille, Albany, California Polytechnic, Connecticut, Cornell, Erasmus (Rotterdam), Glasgow, Manchester, Nottingham and Queen's University. The authors acknowledge research funds from the Danish Council for Independent Research (DFF -- 4182-00050). In addition, Podolskij thanks the Villum Foundation for the grant ``Ambit fields: Probabilistic properties and statistical inference.'' This work was also supported by CREATES, which is funded by the Danish National Research Foundation (DNRF78). Please address correspondence to: khounyo@albany.edu.}}
\author{Kim Christensen\thanks{Aarhus University, Department of Economics and Business Economics, CREATES, Fuglesangs All\'{e} 4, 8210 Aarhus V, Denmark.}
\and Ulrich Hounyo\thanks{University at Albany -- State University of New York, Department of Economics, 1400 Washington Avenue, Albany, New York 12222, United States.} $^{\text{,}}$ \kern-0.15cm \footnotemark[2]
\and Mark Podolskij\thanks{Aarhus University, Department of Mathematics, Ny
Munkegade 118, 8000 Aarhus C, Denmark.} $^{\text{,}}$ \kern-0.15cm
\footnotemark[2]}
\date{March, 2018}

\maketitle

\vspace*{-0.5cm}

\begin{abstract}
In this paper, we propose a nonparametric way to test the hypothesis that time-variation in intraday volatility is caused solely by a deterministic and recurrent diurnal pattern. We assume that noisy high-frequency data from a discretely sampled jump-diffusion process are available. The test is then based on asset returns, which are deflated by the seasonal component and therefore homoskedastic under the null. To construct our test statistic, we extend the concept of pre-averaged bipower variation to a general It\^{o} semimartingale setting via a truncation device. We prove a central limit theorem for this statistic and construct a positive semi-definite estimator of the asymptotic covariance matrix. The $t$-statistic (after pre-averaging and jump-truncation) diverges in the presence of stochastic volatility and has a standard normal distribution otherwise. We show that replacing the true diurnal factor with a model-free jump- and noise-robust estimator does not affect the asymptotic theory. A Monte Carlo simulation also shows this substitution has no discernable impact in finite samples. The test is, however, distorted by small infinite-activity price jumps. To improve inference, we propose a new bootstrap approach, which leads to almost correctly sized tests of the null hypothesis. We apply the developed framework to a large cross-section of equity high-frequency data and find that the diurnal pattern accounts for a rather significant fraction of intraday variation in volatility, but important sources of heteroskedasticity remain present in the data.


\end{abstract}

\setlength{\baselineskip}{18pt}\setlength{\abovedisplayskip}{10pt} %
\belowdisplayskip \abovedisplayskip \setlength{\abovedisplayshortskip }{5pt} %
\abovedisplayshortskip \belowdisplayshortskip \setlength{%
\abovedisplayskip}{8pt} \belowdisplayskip \abovedisplayskip%
\setlength{\abovedisplayshortskip }{4pt}

\vfill

\thispagestyle{empty}

\pagebreak

\section{Introduction} \setcounter{page}{1}

There is a widespread agreement in the literature that any dynamic model of volatility should---at a minimum---account for two distinct features in order to explain the formation of diffusive risk in financial markets. On the one hand, a mean-reverting but highly persistent stochastic component is needed at the \textit{interday} horizon to capture volatility clustering \citep*[e.g.,][]{fama:65a, mandelbrot:63a}. On the other, a pervasive diurnal effect is required as one of the most critical determinants to describe the recurrent behavior of \textit{intraday} volatility. In stock markets, for example, there is a tendency for absolute (or squared) price changes during the course of a trading day to form a so-called ``U''- or reverse ``J''-shape with notably larger fluctuations near the opening and closing of the exchange than around lunch time \citep*[see, e.g.,][for early-stage documentation of this attribute]{harris:86a, wood-mcinish-ord:85a}.\footnote{Moreover, diurnal effects are present in other financial variables, such as asset covariances and correlations, bid-ask spreads, trade durations, trading volume, and quote updates.} In addition to these effects, volatility may exhibit large, sudden shifts around the release of important economic news, such as macroeconomic information \citep*[e.g.,][]{andersen-bollerslev:98b}.

A recent strand of work, fueled by access to high-frequency data and complimentary theory for model-free measurement of volatility, has taken a more detailed close-up of these components and largely confirmed their presence.\footnote{A comprehensive list of papers in this field, including several reviews of the literature, is available at the webpage of the Oxford-Man Institute of Quantitative Finance's Realized Library: http://realized.oxford-man.ox.ac.uk/research/literature.} The diurnal U-shape, in particular, has emerged as a potent---if not predominant---source of within-day variation in volatility. It is therefore common to formulate \textit{parametric} models of time-varying volatility targeted for high-frequency analysis (be it in continuous- or discrete-time) as a composition of a stochastic and deterministic process (with suitable restrictions imposed to ensure the parameters are separately identified). A standard approach is to assume that the stochastic process is constant within a day but is evolving randomly between them (thus enabling volatility clustering), while the deterministic part is a smooth periodic function that is allowed to change within the day but is otherwise time-invariant (thus capturing the diurnal effect), see, e.g., \citet*{andersen-bollerslev:97b, andersen-bollerslev:98b, boudt-croux-laurent:11b, engle-sokalska:12a} and references therein.

Indeed, a major motivation behind the preferred use of realized measures of return variation that are temporally aggregated to the daily frequency is to avoid dealing with the diurnal effect, since it is widely believed to make them intrinsically robust against its presence. However, as stressed by \citet*{andersen-dobrev-schaumburg:12a,dette-golosnoy-kellermann:16a} diurnal effects inject a strong Jensen's inequality-type bias in some of these estimators; an effect that is reinforced and magnified with a high ``volatility-of-stochastic volatility'' \citep*[e.g.,][]{,christensen-oomen-podolskij:14a}. This can, for instance, alter the finite sample properties of jump tests designed to operate at \textit{either} the intraday \textit{or} interday horizon \citep*[e.g.,][]{andersen-bollerslev-dobrev:07a, barndorff-nielsen-shephard:06a, lee-mykland:08a} and make them significantly leaned toward the alternative and cause spurious jump detection as result.\footnote{The effects are deeply intertwined, however, because jumps can also induce substantial biases in and distort estimates of both integrated variance \citep*[e.g.,][]{barndorff-nielsen-shephard:04b, christensen-oomen-podolskij:14a} and the diurnal pattern \citep*[e.g.,][]{andersen-bollerslev-das:01c, boudt-croux-laurent:11b}.} As such, further investigation of diurnal effects appears warranted.

In this paper, we develop a \textit{nonparametric} framework to assess if diurnal effects can, in fact, explain all of the intraday variation in volatility, as stipulated by such a setup. A casual inspection of high-frequency data does not offer conclusive evidence about the validity of this conjecture. In concrete applications, inference is obscured by microstructure noise at the tick-by-tick frequency \citep*[e.g.,][]{hansen-lunde:06b} and the existence of price jumps that are potentially very small and highly active \citep*[e.g.,][]{ait-sahalia-jacod:12a}.\footnote{In view of this, a related topic is whether a diffusive component is needed in the first place to represent risk formation in financial asset prices. While the prevailing evidence is slightly mixed, it appears largely affirmative \citep*[see, e.g.,][]{ait-sahalia-jacod:12b,kolokolov-reno:24a,todorov-tauchen:10a}.} Moreover, even if stochastic volatility is truly present, in practice its components may be so persistent that it is acceptable (and convenient) to regard it as absent on small time scales.\footnote{Of course, the locally constant approximation of stochastic volatility is one of the most heavily exploited in the analysis of high-frequency data, see, e.g., \citet*{mykland-zhang:09a}.}

Consistent with the above, we model the asset log-price as a general arbitrage-free It\^{o} semimartingale, which is contaminated by microstructure noise. In this framework, the asymptotic theory is infill, i.e. the process is assumed to be observed on a fixed time interval with mesh tending to zero.

There are several existing tests of constant volatility available in the high-frequency volatility area \citep*[see, e.g.,][]{dette-podolskij-vetter:06a, dette-podolskij:08a, vetter-dette:12a}. To our knowledge, none allow for the joint disturbance of jumps and microstructure noise, nor do they directly study the extension to diurnal variation advocated here. We formulate a test on the back of log-returns that are only homoskedastic under the null, after they are filtered for diurnal effects. We then study a jump-robust version of the pre-averaged bipower variation, where we extend the bivariate central limit theorem of \citet*{podolskij-vetter:09a} to the jumpy setting \citep*[see][for further work on noise-robust volatility estimation]{barndorff-nielsen-hansen-lunde-shephard:08a, jacod-li-mykland-podolskij-vetter:09a, zhang-mykland-ait-sahalia:05a, zhang:06a}. The test is constructed via the asymptotic distribution implied by a transformation of such statistics and an application of Cauchy-Schwarz for a particular---but standard---choice of the parameters. As an aside, we add that a slightly different configuration of our $t$-statistic (based on the comparison of suitably non-truncated and truncated statistics) can serve as a basis for a jump test, which is robust to diurnal effects, but we do not pursue this idea in the present paper.

As the diurnal pattern is unknown in practice, we follow the ``two-stage'' approach of \citet*{andersen-bollerslev:97b}. In the first stage, the diurnal factors are pre-estimated. We propose a nonparametric estimator, which is both inherently jump- and noise-robust and therefore applicable in practice. It extends previous work of, e.g., \citet*{andersen-bollerslev:97b, andersen-dobrev-schaumburg:12a, boudt-croux-laurent:11b, taylor-xu:97a, todorov-tauchen:12a} to the noisy setting \citep*[see also][]{hecq-laurent-palm:12a}. In the second stage, the estimator is inserted and therefore replaces the true value in the deflation step. The estimate contains a sampling error, however, which afflicts the calculation and may propagate through the system and invalidate the analysis. This problem has largely gone unnoticed (or at least been ignored) in previous work.\footnote{A notable exception is \citet*{todorov-tauchen:12a}. In the supplemental material to that paper, available at Econometrica's website, the authors treat the impact of diurnal filtering on their realized Laplace transform estimator of volatility, but microstructure noise is supposed to be absent.} In a double-asymptotic setup with mesh going to zero, as before, and time span now increasing to infinity, we show our feasible statistic has a sampling error of sufficiently small order to not affect the previous infill asymptotic theory. As such, the inference procedure only applies if the researcher has high-frequency data over a large number of days, but is interested in the behavior of volatility on smaller time scales, which is an inherent but unavoidable limitation of our approach.

A simulation study reveals that this substitution also has no discernable impact in finite samples. The test is, however, severely distorted by the presence of small infinite-activity price jumps, which it understandably appears to confuse with stochastic volatility. To improve inference, we suggest a new bootstrap approach. It is of independent interest and can be viewed as an overlapping version of the wild blocks of blocks bootstrap by \citet*{hounyo-goncalves-meddahi:17a}. We prove the first-order validity of the bootstrap, while in simulations it helps to restore an almost correctly sized test.

The paper is structured as follows. In Section \ref{section:theory}, we introduce our theoretical framework and the assumptions. We extend the asymptotic theory of the pre-averaged bipower variation and construct a jump- and noise-robust test of the hypothesis that all intraday variation in volatility is captured by the diurnal pattern. In Section \ref{section:diurnal-variance}, we propose an estimator of the latent diurnal pattern and show that the sampling error of the feasible statistic is sufficiently small to not affect the previous results. In Section \ref{section:bootstrap}, we introduce the bootstrap and show its consistency for testing the null hypothesis in a noisy jump-diffusion setting. In Section \ref{section:montecarlo}, we present the Monte Carlo results, while an empirical illustration is conducted in Section \ref{section:empirical}. We conclude in Section \ref{section:conclusion}. The proofs and some auxiliary results are relegated to an online-only appendix available at the journal website.

\section{Theoretical setup} \label{section:theory}

We let $X$ denote a latent efficient log-price defined on a filtered probability space $( \Omega, \mathcal{F}, ( \mathcal{F}_{t})_{t\geq 0},P)$ and recorded in the window $[0,T]$, where $T$ is the number of days in the sample and the subinterval $[t-1,t]$ is the $t$th day, for $t = 1, \ldots T$. Throughout, $T$ is mainly fixed and the asymptotic theory is infill, so we often impose $T = 1$ as a normalization, but please note the important digression in Section \ref{section:diurnal-variance}, where a long-span analysis ($T \rightarrow \infty$) is required, and there the additional notation and interpretation of $[0,T]$ is helpful.

As consistent with the no-arbitrage restriction \citep*[e.g.,][]{delbaen-schachermayer:94a}, we model $X$ as an It\^{o} semimartingale:
\begin{equation} \label{equation:one}
X_{t}=X_{0}+\int_{0}^{t}a_{s}\text{d}s+\int_{0}^{t}\sigma _{s}\text{d} W_{s}+J_{t}, \quad t \in [0,T],
\end{equation}
where $(a_{t})_{t\geq 0}$ is a predictable, locally bounded drift process, $(\sigma _{t})_{t\geq 0}$ is an adapted, c\`{a}dl\`{a}g volatility process, while $(W_{t})_{t\geq 0}$ is a Brownian motion.

$J_{t}$ is a jump process defined by the equation:
\begin{equation} \label{equation:one-Jump-part}
J_{t}=\big(\delta 1_{\{|\delta |\leq 1\}}\big)\star \Big(\underline{\mu }_{t}-\underline{\nu }_{t}\Big)+\big(\delta 1_{\{|\delta |>1\}}\big)\star \underline{\mu }_{t},
\end{equation}
where $\underline{\mu }$ is a Poisson random measure on $\mathbb{R}_{+} \times \mathbb{R}$ and $\underline{ \nu}$ is a predictable compensator of $\underline{ \mu}$, such that $\underline{\nu }(\text{d}s, \text{d}x) = \text{d}s \otimes \lambda (\text{d}x)$ and $\lambda$ is a $\sigma$-finite measure.

As explained above, the main idea of the paper is to construct a test, which tells whether a diurnal component is adequate to describe the evolution of within-day volatility.\footnote{We do not speak about the dynamics of volatility during market close. Thus, our framework is consistent with random changes in between-day volatility.} To this end, we need to put some structure on the problem, starting with: \\[-0.25cm]

\noindent \textbf{Assumption (D1)}: $\quad$ $\sigma_{t} = \sigma_{sv,t} \sigma _{u,t}$. \\[-0.25cm]

\noindent $\sigma_{sv,t}$ and $\sigma_{u,t}$ represent two distinct sources of time-varying volatility in many financial return series. The first term, $\sigma_{sv,t}$, denotes a stochastic process, which allows for randomness in the evolution of $\sigma_{t}$ over time. The second term, $\sigma_{u,t}$, is a deterministic seasonal component that represents the diurnal pattern.

The multiplicative structure means our test can be formed by deflating the log-return series with $\sigma_{u,t}$ and checking if the outcome is homoskedastic, as we do in Section \ref{section:test}. \\[-0.25cm]

\noindent \textbf{Assumption (D2)}: $\quad$ $\big( \sigma_{sv,t}^{2} \big)_{t \geq 0}$ is stationary with $E( \sigma_{sv,t}^{2} ) = \sigma^{2}$ and $\sum_{k=0}^{ \infty} \text{cov} \big( \sigma_{sv,t}^{2}, \sigma_{sv,t+k}^{2} \big) < \infty$. \\[-0.25cm]

\noindent \textbf{Assumption (D3)}: $\quad$ $( \sigma _{u,t}^{2})_{t \geq 0}$ is a continuously differentiable 1-periodic function with bounded derivative for $t \rightarrow 0$ and $t \rightarrow 1$ and normalized such that $\int_{0}^{1} \sigma _{u,s}^{2} \text{d}s = 1$. \\[-0.25cm]

\noindent Assumption (D2) -- (D3) are sufficient to ensure identification of both volatility components from the data.\footnote{The normalization in Assumption (D3) is known as a ``standardization condition,'' which ensures that the decomposition in Assumption (D1) is unique under (locally) constant volatility, see, e.g., \citet*{andersen-bollerslev:97b, boudt-croux-laurent:11b, taylor-xu:97a}.} Apart from the stationarity of the stochastic volatility, the former also restricts its memory, which implies the process is ergodic. The latter says the diurnal component has to be recurrent, so that we can gradually infer it from gathering a larger sample.\footnote{It can be extended to accommodate a day-of-the-week effect \citep*[e.g.,][]{andersen-bollerslev:98b}.} Taken together, the conditions imply that an average of (an estimate of) volatility sampled at a fixed time of the day $s \in (0,1)$, i.e. $T^{-1} \sum_{t = 1}^{T} \sigma_{t-1 + s}^{2} =  \sigma_{u,s}^{2} T^{-1} \sum_{t = 1}^{T} \sigma_{sv,t-1+s}^{2} \overset{p}{ \rightarrow} \sigma_{u,s}^{2} \sigma^{2}$, as $T \rightarrow \infty$, thereby delivering $\sigma_{u,s}^{2}$ after suitable normalization.

In Section \ref{section:diurnal-variance}, we use this idea to recover $\sigma_{u}$ from noisy high-frequency data, so that we can compute the test in practice, but---for the moment---we treat it as observed. \\[-0.25cm]

\noindent We further rule out jumps in $\sigma_{sv,t}$: \\[-0.25cm]

\noindent \textbf{Assumption (V)}: $\quad \sigma_{sv,t}$ is of the form:
\begin{equation}
\sigma_{sv,t} = \sigma_{0} + \int_{0}^{t} \tilde{a}_{s} \text{d}s + \int_{0}^{t} \tilde{ \sigma}_{s} \text{d}W_{s} + \int_{0}^{t} \tilde{v}_{s} \text{d}B_{s},
\end{equation}
where $( \tilde{a}_{t})_{t \geq 0}$, $(\tilde{ \sigma}_{t})_{t \geq 0}$ and $( \tilde{v}_{t})_{t\geq 0}$ are adapted, c\`{a}dl\`{a}g stochastic processes, while $(B_{t})_{t \geq 0}$ is a standard Brownian motion that is independent of $W$. \\[-0.25cm]

\noindent Assumption (V) is common in the realized volatility literature \citep*[see, e.g.,][]{barndorff-nielsen-hansen-lunde-shephard:08a, goncalves-meddahi:09a, mykland-zhang:09a, christensen-podolskij-vetter:13a, hounyo:17a}. It facilitates the control of some approximation errors in the proofs, but it can potentially be relaxed. In recent work, \citet*{christensen-podolskij-thamrongrat-veliyev:17a} operate with a power variation-based statistic and impose a weaker set of assumptions, which allows for rather unrestricted jump dynamics in volatility (see their equation (2.3) and Theorem (3.2), which is based on Assumption (H1) from \citet*{barndorff-nielsen-graversen-jacod-podolskij-shephard:06a}). It may be possible to extend our setting in that direction, but we leave a full exploration of it for future research.

In some of our results, we also assume that the volatility is bounded away from zero. In particular, we sometimes adopt the following condition: \\[-0.25cm]

\noindent \textbf{Assumption (V')}: $\quad$ $\sigma_{sv,t} > 0$ and $\sigma_{u,t} > 0$, for all $t\geq 0$. \\[-0.25cm]

\noindent At last, we impose that: \\[-0.25cm]

\noindent \textbf{Assumption (J)}: $\quad$ There exists a sequence of stopping times $( \widetilde{ \tau}_{n})_{n = 1}^{ \infty}$ increasing to $\infty$ and a deterministic nonnegative function $\tilde{ \gamma}_{n}$ such that $\int_{ \mathbb{R}} \tilde{ \gamma}_{n}(x)^{ \beta} \lambda ( \text{d}x) < \infty$ and $|| \delta ( \omega, t , x) || \wedge 1 \leq \tilde{ \gamma}_{n}(x)$, for all $( \omega, t, x)$ with $t \leq \tilde{ \tau }_{n}( \omega)$, where $\beta \in [0,2]$. \\[-0.25cm]

\noindent $\beta$ captures the activity of the jump process. As $\beta$ approaches two, the jumps are smaller but more vibrant. As explained by \citet*{todorov-bollerslev:10a}, the harder they are to distinguish from the diffusive part of $X$. Below, we impose Assumption (J) to hold for any $\beta \in [0,1)$, thus restricting attention to jump processes with sample paths of finite length.

\subsection{Microstructure noise}

The presence of market frictions (such as price discreteness, rounding errors, bid-ask spreads, gradual response of prices to block trades and so forth) prevent us from observing the true, efficient log-price process $X_{t}$. Instead, we observe a noisy version $Y_{t}$, which we assume is given by
\begin{equation}
Y_{t} = X_{t} + \epsilon_{t}, \label{equation:Y}
\end{equation}
where $\epsilon_{t}$ is a noise term that collects the market microstructure effects. We assume that $\epsilon_{t}$ is independently distributed and independent of $X_{t}$, such that
\begin{equation} \label{equation:noise-variance}
E( \epsilon_{t})=0 \quad \text{and} \quad E \big( \epsilon_{t}^{2} \big) = \sigma_{u,t}^{2} \omega^{2},
\end{equation}
for any $t$, where $Y_{t}$ is observed.

As consistent with, e.g, \citet*{bandi-russell:06a, kalnina-linton:08a}, the second moment of the noise is allowed to be heteroskedastic and exhibit diurnal variation. We assume it is identical to the volatility diurnality, which conveniently makes the detrended noise asymptotically i.i.d (cf. \eqref{equation:rescaled-increment}). We return to this later in Remark \ref{remark:noise-variance}, where we highlight the impact of weakening it to a general form of heteroskedasticity.

About the noise distribution, we follow \citet*{podolskij-vetter:09a}: \\[-0.25cm]

\noindent \textbf{Assumption (A)}: $\quad $ (i) $\epsilon $ is distributed symmetrically around zero, and (ii) for any $0>a>-1$, it holds that $E(|\epsilon _{t}|^{a})<\infty $. \\[-0.25cm]

\noindent \textbf{Assumption (A')}: $\quad $ Cramer's condition is fulfilled, that is $\lim \sup_{t\rightarrow \infty }|\chi (t)|<1$, where $\chi $ denotes the characteristic function of $\epsilon $.

\subsection{Test of heteroskedasticity} \label{section:test}

To develop a test of the ``no heteroskedasticity after diurnal correction'' assumption, we partition the sample space $\Omega $ into the following two subsets:
\begin{equation}
\Omega_{ \mathcal{H}_{0}} = \{ \omega : \sigma_{sv,t} \text{ is constant for } t \geq 0 \},
\end{equation}
and $\Omega_{ \mathcal{H}_{a}} = \Omega_{ \mathcal{H}_{0}}^{ \complement}$. The null hypothesis can then formally be defined as $\mathcal{H}_{0} : \omega \in \Omega_{ \mathcal{H}_{0}}$, whereas the alternative is $\mathcal{H}_{a}: \omega \in \Omega_{ \mathcal{H}_{a}}$.

Our goal is to find a test with a prescribed asymptotic significance level and with power going to one to test the hypothesis that $\omega \in \Omega_{ \mathcal{H}_{0}}$. The key challenge we address is how to construct such a test, when $X$---apart from being driven by a Brownian component---is subject to diurnal variation, potentially discontinuous and observed with measurement error. The solution is based on computing a set of estimators, which reveal information about the presence of time-variation in the stochastic
volatility $\sigma _{sv,t}$ robustly to the above features.

The differential form of \eqref{equation:one} scaled by $\sigma_{u,t}$ yields:
\begin{equation} \label{equation:deflated-increment}
\frac{ \text{d}X_{t}}{ \sigma_{u,t}} = \frac{a_{t}}{ \sigma _{u,t}} \text{d}t + \frac{ \sigma _{t}}{ \sigma_{u,t}} \text{d}W_{t} + \frac{ \text{d}J_{t}}{ \sigma_{u,t}},
\end{equation}
or
\begin{equation}
\text{d}X_{t}^{d} = a_{t}^{d} \text{d}t + \sigma_{sv,t} \text{d}W_{t} + \text{d}J_{t}^{d},
\end{equation}
where a superscript $d$ is used to represent a process that has been adjusted by the seasonal component of volatility.\footnote{Note that many parts of this paper can be applied to both the raw and deflated log-returns series (e.g., the pre-averaging theory in the next subsection). We base the theoretical exposition on the seasonally adjusted version to minimize the notational load, while we present results for both series in the empirical application.}

Then, we study the quadratic variation of $X^{d}$:
\begin{equation} \label{equation:qv}
[X^{d}]_{t} = \int_{0}^{t} \sigma _{sv,s}^{2}ds + \sum_{s \leq t} | \Delta X_{s}^{d}|^{2},
\end{equation}
where $\int_{0}^{t} \sigma_{sv,s}^{2} \text{d}s$ is the integrated variance of $X^{d}$, while $\sum_{s \leq t} | \Delta X_{s}^{d} |^{2} $ is the sum of the squared deflated jumps, where $\Delta X_{s}^{d} = X_{s}^{d} - X_{s-}^{d}$.

We note that if the stochastic volatility process is constant, say $\sigma_{sv,t} = \sigma$, \eqref{equation:one} reduces to
\begin{equation} \label{one-Constant-Vol}
X_{t} = X_{0} + \int_{0}^{t}a_{s} \text{d}s + \sigma \sigma _{u,t} \big(W_{t} - W_{0} \big) + \big( \delta 1_{ \{| \delta | \leq 1 \}} \big) \star \Big( \underline{ \mu}_{t} - \underline{ \nu }_{t} \Big) + \big( \delta 1_{ \{| \delta |>1 \}} \big) \star \underline{ \mu}_{t},
\end{equation}
while
\begin{equation} \label{equation:qv_constant}
[X^{d}]_{t} = \sigma^{2}t + \sum_{s \leq t} | \Delta X_{s}^{d}|^{2}.
\end{equation}
The construction of the $t$-statistic now progresses in three steps. Firstly, we account for microstructure noise by doing local pre-averaging of $Y^{d}$. Secondly, we tease out the continuous part of the quadratic variation by suitably removing the jump component in \eqref{equation:qv_constant}. Thirdly, we develop a fully feasible theory by proposing a statistic that can replace $\sigma_{u}$ in the computations.

\subsection{The pre-averaging approach}

In this section, we confine the clock to $t \in \lbrack 0,1]$, i.e. we set $T = 1$. In our simulations and empirical work, we implement the test ``day-by-day,'' so that here the unit interval is naturally interpreted as a trading day's worth of data.

The noisy log-price $Y_{t}$ is observed at regular time points $t_{i} = i/n$, for $i=0, \ldots, n$. Then, the deflated intraday log-returns (at frequency $n$) can be computed as:
\begin{equation} \label{equation:rescaled-increment}
\Delta_{i}^{n}Y^{d} \equiv Y_{i/n}^{d} - Y_{(i-1)/n}^{d}, \quad i = 1, \ldots, n.
\end{equation}
As $Y_{t}^{d} = X_{t}^{d} + \epsilon_{t}^{d}$, we can split $\Delta_{i}^{n}Y^{d}$ into
\begin{equation} \label{equation:return-decomposition}
\Delta_{i}^{n} Y^{d} = \Delta_{i}^{n} X^{d} + \Delta_{i}^{n} \epsilon^{d},
\end{equation}
where $\Delta_{i}^{n} X^{d} = X_{i/n}^{d} - X_{(i-1)/n}^{d}$ denotes the $n$-frequency return of the efficient log-price, while $\Delta_{i}^{n} \epsilon^{d} = \epsilon_{i/n}^{d} - \epsilon_{(i-1)/n}^{d}$ is the change in the microstructure component.

To lessen the noise, we adopt the pre-averaging approach of \citet*{jacod-li-mykland-podolskij-vetter:09a, podolskij-vetter:09a, podolskij-vetter:09b}. To describe it, we let $k_{n}$ be a sequence of positive integers and $g$ a real-valued function. $k_{n}$ represents the length of a pre-averaging window, while $g$ assigns a weight to those noisy log-returns that are inside it. $g$ is defined on $[0,1]$, such that $g(0)=g(1)=0$ and $ \int_{0}^{1}g(s)^{2}\text{d}s>0$. We assume $g$ is continuous and piecewise
continuously differentiable with a piecewise Lipschitz derivative $g^{\prime}$. A canonical function that fulfills these restrictions is $g(x)=\min(x,1-x)$.

We introduce the notation:
\begin{equation}
\phi_{1} (s) = \int_{s}^{1} g^{ \prime}(u) g^{ \prime}(u-s) \text{d}u \quad
\text{and} \quad \phi_{2} (s) = \int_{s}^{1} g(u) g(u-s) \text{d}u,
\end{equation}
and for $i = 1,2$, we let $\psi_{i} = \phi_{i}(0)$. For instance, if $g(x) = \min(x,1-x)$, it follows that $\psi_{1} =1$ and $\psi_{2} = 1/12$.

Also, we write:
\begin{equation} \label{equation:psi-n}
\psi_{1}^{n} = k_{n} \sum_{j=1}^{k_{n}} \Bigg(g \bigg( \frac{j}{k_n} \bigg) - g \bigg( \frac{j-1}{k_n} \bigg) \Bigg)^{2} \quad \text{and} \quad \psi_{2}^{n} = \frac{1}{k_{n}} \sum_{j=1}^{k_{n}-1} g^{2} \bigg( \frac{j}{k_n} \bigg).
\end{equation}
In the appendix, after freezing the volatility locally, $\psi_{1}^{n}$ and $\psi_{2}^{n}$ appear in the conditional expectation of the squared pre-averaged return in \eqref{equation:preavgY}. As $n \rightarrow \infty$,
\begin{equation}
\psi_{1}^{n} \rightarrow \psi_{1} \quad \text{and} \quad \psi_{2}^{n} \rightarrow \psi_{2},
\end{equation}
while $\psi_{i}^{n} - \psi_{i} = O \big( n^{-1/2} \big)$, for $i = 1,2$, so we can work with $\psi_{i}$ and not worry about the effect of this substitution in the asymptotic theory. In contrast, $\psi_{i}^{n}$ can differ a lot from $\psi_{i}$, if $k_{n}$ is small, so as a practical guide it is better to work with \eqref{equation:psi-n}.

The pre-averaged return, say $\Delta _{i}^{n} \bar{Y}^{d}$, is then found by computing a weighted sum of consecutive $n$-frequency deflated log-returns over a block of size $k_{n}$:
\begin{equation} \label{equation:preavgY}
\Delta_{i}^{n} \bar{Y}^{d} = \sum_{j=1}^{k_{n}-1}g \bigg( \frac{j}{k_{n}} \bigg) \Delta_{i+j-1}^{n}Y^{d}, \quad i=1, \ldots, n-k_{n}+2.
\end{equation}
As readily seen, pre-averaging entails a slight ``loss'' of summands compared to $n$. Thus, while the original sample size is $n$, there are only $n-k_{n}+2$ elements in $( \Delta_{i}^{n} \bar{Y}^{d})_{i=1}^{n-k_{n}+2}$. It follows from the decomposition in \eqref{equation:return-decomposition} that $\Delta_{i}^{n} \bar{Y}^{d} = \Delta_{i}^{n} \bar{X}^{d} + \Delta_{i}^{n} \bar{ \epsilon}^{d}$ and, as shown by \citet*{vetter:08a},
\begin{equation}
\Delta_{i}^{n} \bar{X}^{d} = O_{p} \Bigg( \sqrt{ \frac{k_{n}}{n}} \Bigg) \quad \text{and} \quad \Delta_{i}^{n} \bar{\epsilon}^{d} = O_{p} \bigg( \frac{1}{ \sqrt{k_{n}}} \bigg).
\end{equation}
Thus, the noise is dampened, thereby reducing its influence on $\Delta_{i}^{n} \bar{Y}^{d}$. As an outcome, we retrieve a basically noise-free estimate, which can substitute the efficient log-return $\Delta_{i}^{n}X^{d}$ in subsequent computations, taking proper account of the dependence introduced in $( \Delta _{i}^{n} \bar{Y}^{d})_{i=1}^{n-k_{n}+2}$.\footnote{If $k_{n}$ is even, it follows with the above definition of $g(x) = \min(x,1-x)$ that the pre-averaged returns in \eqref{equation:preavgY} can be rewritten as $\Delta_{i}^{n} \bar{Y}^{d} = \frac{1}{k_{n}} \sum_{j=1}^{k_{n}/2} Y_{ \frac{i+k_{n}/2+j}{n}}^{d} - \frac{1}{k_{n}} \sum_{j=1}^{k_{n}/2} Y_{\frac{i+j}{n}}^{d}$. Thus, the sequence $(2\Delta_{i}^{n} \bar{Y}^{d})_{i=1}^{n-k_{n}+2}$ can be interpreted as constituting a new set of increments from a price process that is constructed by simple averaging of the rescaled noisy log-price series, $(Y_{i/n}^{d})_{i=0}^{n}$, in a neighbourhood of $i/n$, thus making the use of the term pre-averaging and the associated notation transparent.} The reduction increases with larger $k_{n}$, but too much pre-averaging also impedes the accuracy of estimators of the quadratic variation, yielding a trade-off in selecting $k_{n}$. To strike a balance and get an efficient $n^{-1/4}$ rate of convergence, \citet*{jacod-li-mykland-podolskij-vetter:09a} propose to set:
\begin{equation} \label{equation:theta}
k_{n}=\theta \sqrt{n} + o \big(n^{-1/4} \big),
\end{equation}
for some $\theta \in (0,\infty )$. With this choice, the orders of $\Delta_{i}^{n} \bar{X}^{d}$ and $\Delta_{i}^{n} \bar{ \epsilon}^{d}$ are balanced and equal to $O_{p} \big(n^{-1/4} \big)$. An example of \eqref{equation:theta} used throughout this paper is $k_{n} = \big[ \theta \sqrt{n} \big]$.

\subsubsection{The pre-averaged bipower variation}

With the pre-averaged return series, $(\Delta _{i}^{n}\bar{Y}^{d})_{i=1}^{n-k_{n}+2}$, available, \citet*{podolskij-vetter:09a} propose the bipower variation statistic:
\begin{equation} \label{equation:preavgBV}
BV(Y^{d},l,r)^{n} = n^{ \frac{l+r}{4}-1} \frac{1}{ \mu_{l} \mu_{r}} \sum_{i=1}^{N_{n}} y(Y^{d},l,r)_{i}^{n},
\end{equation}
where $l,r \geq 0$, $y(Y^{d},l,r)_{i}^{n}=| \Delta_{i}^{n} \bar{Y}^{d}|^{l}| \Delta_{i+k_{n}}^{n} \bar{Y}^{d}|^{r}$, $N_{n} = n-2k_{n}+2$ and $\mu_{p} = E(|N(0,1)|^{p})$.\footnote{In order to avoid a finite sample bias in the construction of $BV(l,r)^{n}$,
we only divide it by $N_{n}$ (the number of summands in the estimator) in our simulations and empirical work. We stick with $n$ in the theoretical parts of the paper, as it involves less notation.} In the following, if we write $BV(l,r)^{n}$ and $y(l,r)_{i}^{n}$, we assume that they are implicitly defined with respect to $Y^{d}$. \citet*{podolskij-vetter:09a} show that under suitable regularity conditions, in particular that $X$ is a continuous It\^{o} semimartingale (i.e., $X$ follows (A.7) in the web appendix), then as $n \rightarrow \infty$
\begin{equation} \label{Convergence-Proba-BV}
BV(l,r)^{n} \overset{p}{ \rightarrow}BV(l,r) = \int_{0}^{1} \bigg( \theta \psi_{2} \sigma _{sv,s}^{2} + \frac{1}{ \theta} \psi _{1} \omega ^{2} \bigg)^{ \frac{l+r}{2}} \text{d}s,
\end{equation}
and
\begin{equation}
{n}^{1/4} \Bigg(
\begin{array}{c}
BV(l_{1},r_{1})^{n}-BV(l_{1},r_{1}) \\[0.25cm]
BV(l_{2},r_{2})^{n}-BV(l_{2},r_{2})%
\end{array}
\Bigg) \overset{d_{s}}{ \rightarrow}MN \big(0, \Sigma \big), \label{CLT-Joint}
\end{equation}
with $l_{1},r_{1},l_{2},r_{2}\geq 0$, where ``$\overset{d_{s}}{ \rightarrow}$'' is stable convergence, $\Sigma = \big( \Sigma _{ij}^{l_{1},r_{1},l_{2},r_{2}} \big)_{1 \leq i,j \leq 2}$ the conditional covariance matrix of the limiting process $n^{1/4} \big(BV(l_{1},r_{1})^{n},BV(l_{2},r_{2})^{n} \big)^{\intercal}$, and $^{\intercal}$ the transpose.\footnote{The formal definition of $\Sigma $ is given in Appendix A.}

\subsubsection{A truncated pre-averaged bipower variation}

The estimator in \eqref{equation:preavgBV} can also be made jump-robust in both the stochastic limit and its asymptotic distribution, but---as explained by \citet*{podolskij-vetter:09a}---this puts strong restrictions on $l$ and $r$. Firstly, the central limit theory in \eqref{CLT-Joint} is not valid for the popular choice $l = r = 1$. Indeed, \citet*{vetter:10a} shows that this estimator is not even mixed Gaussian, which severely constrains our ability to draw inference. Secondly, the version with $l = r = 2$ as implemented below, does not converge to the limit in \eqref{Convergence-Proba-BV}, if $X$ jumps, and while that is true for the pre-averaged (1,1)-bipower variation, asymptotically, it is well-known that the latter typically has a pronounced
upward bias in finite samples \citep*[e.g.,][]{christensen-oomen-podolskij:14a}. Thus, to achieve a better jump-robustness
and enlarge the feasible set of powers for which we can do hypothesis testing, we follow \citet*{corsi-pirino-reno:10a} in the no-noise and finite-activity jump setting by combining the bipower idea with the truncation approach of \citet*{mancini:09a, jacod-protter:12a, jing-liu-kong:14a}.

To introduce our $t$-statistic for the homoskedasticity test, we therefore start by deriving a result as above for a truncated pre-averaged bipower variation, which verifies that the probability limit and asymptotic distribution of this new estimator are identical to those given by \eqref{Convergence-Proba-BV} and \eqref{CLT-Joint} in the general setting, where $X$ follows the It\^{o} semimartingale in \eqref{equation:one}. Thus, we propose to set:
\begin{equation}  \label{estimator-trunc}
\check{BV}(l,r)^{n} = n^{ \frac{l+r}{4} - 1} \frac{1}{ \mu_{l} \mu_{r}} \sum_{i=1}^{N_{n}} \check{y}(l,r)_{i}^{n},
\end{equation}
where $\check{y}(l,r)_{i}^{n} = | \Delta_{i}^{n} \bar{Y}^{d} |^{l} 1_{ \left\{| \Delta_{i}^{n} \bar{Y}^{d} | < \upsilon_{n} \right\}} | \Delta_{i+k_{n}}^{n} \bar{Y}^{d} |^{r} 1_{ \left\{ | \Delta_{i+k_{n}}^{n} \bar{Y}^{d}| < \upsilon_{n} \right\} }$ and $1_{\left \{ \cdot \right \}}$ is the indicator function, which discards pre-averaged log-returns that exceed a predetermined level
\begin{equation} \label{equation:truncation}
\upsilon_{n} = \alpha u_{n}^{ \varpi}, \text{ for } \alpha > 0 \text{ and } \varpi \in (0, 1/2),
\end{equation}
such that $u_{n} = k_{n} / n$.

\begin{theorem}
\label{CLT-BV-Trunc} Let $l_{1},r_{1},l_{2}$ and $r_{2}$ be four positive real numbers and $X$ be given by \eqref{equation:one}. Suppose that Assumption (J) holds for some $\beta \in \lbrack 0,\min \{1,l_{1},r_{1},l_{2},r_{2}\})$ and that $\Big(\frac{l_{1}+r_{1}-1}{2(l_{1}+r_{1}-\beta )}\vee \frac{l_{2}+r_{2}-1}{2(l_{2}+r_{2}-\beta )}\Big) \leq \varpi <1/2$. Furthermore, we assume (D1), (V), (A), and impose the moment condition $E(|\epsilon _{t}|^{s})<\infty $, for some $s>(3\vee
2(r_{1}+l_{1})\vee 2(r_{2}+l_{2}))$. If any $l_{i}$ or $r_{i}$ is in $(0,1]$, we postulate (V$^{\prime })$, otherwise either (V$^{\prime }$) or (A$^{\prime }$). In addition, suppose that $k_{n}\rightarrow \infty $ as $n \rightarrow \infty $ such that \eqref{equation:theta} holds. Then, as $n \rightarrow \infty $,
\begin{equation}
{n}^{1/4}\Bigg(%
\begin{array}{c}
\check{BV}\left( l_{1},r_{1}\right) ^{n}-BV\left( l_{1},r_{1}\right) \\%
[0.25cm]
\check{BV}\left( l_{2},r_{2}\right) ^{n}-BV\left( l_{2},r_{2}\right)%
\end{array}%
\Bigg)\overset{d_{s}}{ \rightarrow}MN \big(0, \Sigma). \label{CLT-Joint-trunc}
\end{equation}
\end{theorem}
Theorem \ref{CLT-BV-Trunc} shows that \eqref{estimator-trunc} is robust to the jump part in its limiting distribution. Note that $\Sigma$ is identical to the matrix in \eqref {CLT-Joint}. To our knowledge, the result is new with the main innovations being the statistic is \eqref{estimator-trunc} and the underlying process is a general It\^{o} semimartingale given by \eqref{equation:one}. It extends Theorem 3 of \citet*{podolskij-vetter:09a} to discontinuous $X$ by establishing a joint asymptotic distribution, as in \eqref{CLT-Joint}, for the class of truncated pre-averaged bipower variation. In previous work, \citet*{jing-liu-kong:14a} prove---under some regularity conditions---the consistency and CLT for the truncated pre-averaged realized variance, i.e. the statistic of the form $\check{BV}(2,0)^{n}$, when $X$ follows \eqref{equation:one}. Our paper generalizes the latter article to the bipower setting with---subject to the above constraint---arbitrary powers.

The lower bound on $\varpi$ is determined by an interplay between the bipower parameters and the activity of the jump process. The crude intuition is that small jumps tend to resemble Brownian motion, so if the threshold vanishes too slowly, it can impair the jump-robustness, and this effect is aggravated for larger bipowers. We therefore normally work with $\varpi$ close to a half in practice. In our Monte Carlo, $l_{1} = r_{2} = 2$, $l_{2} = r_{2} = 1$ and $\beta = 0.5$, so that any $\varpi \in [1/3, 1/2)$ is valid. Throughout, we always set $\varpi = 0.49$.

The above enables extraction of an essentially noise-free and jump-robust estimate of the continuous piece of the quadratic variation in \eqref{equation:qv} and thus facilitates the construction of a test for the presence of time-variation in $\sigma_{sv,t}$. An implication of \eqref{CLT-Joint-trunc} is that for any $l_{1},r_{1},l_{2},r_{2} \geq 0$, which adhere to the conditions of Theorem \ref{CLT-BV-Trunc} and such that $l_{1}+r_{1} > l_{2}+r_{2}$, as $n \rightarrow \infty $,
\begin{equation} \label{equation:heteroskedasticity}
\begin{split}
& \check{BV}(l_{1},r_{1})^{n}-(\check{BV}(l_{2},r_{2})^{n})^{\frac{l_{1}+r_{1}}{l_{2}+r_{2}}}\overset{p}{ \rightarrow} BV(l_{1},r_{1})-(BV(l_{2},r_{2}))^{\frac{l_{1}+r_{1}}{l_{2}+r_{2}}} \\[0.25cm]
& =\int_{0}^{1}\bigg(\theta \psi _{2}\sigma _{sv,s}^{2}+\frac{1}{\theta } \psi_{1} \omega ^{2}\bigg)^{\frac{l_{1}+r_{1}}{2}}\text{d}s-\Bigg[ \int_{0}^{1}\bigg(\theta \psi _{2}\sigma _{sv,s}^{2}+\frac{1}{\theta }\psi
_{1}\omega ^{2}\bigg)^{\frac{l_{2}+r_{2}}{2}}\text{d}s\Bigg]^{\frac{l_{1}+r_{1}}{l_{2}+r_{2}}}\geq 0,
\end{split}
\end{equation}
with equality if and only if $\sigma_{sv,t}$ is constant. We thus build a test of $\mathcal{H}_{0}$ via the infeasible $t$-statistic:
\begin{equation} \label{CLT-Homoskedasticity-Test}
T_{\text{inf.}}^{n} = \frac{{n}^{1/4} \Big( \check{BV}(l_{1},r_{1})^{n} - ( \check{BV}(l_{2},r_{2})^{n})^{ \frac{l_{1}+r_{1}}{l_{2}+r_{2}}} \Big)}{ \sqrt{V}} \overset{d}{ \rightarrow} N(0,1),
\end{equation}
where
\begin{align}
\begin{split} \label{equation:V}
V &= \Sigma_{11} - 2 \bigg( \frac{l_{1}+r_{1}}{l_{2}+r_{2}} \bigg) \big( \check{BV}(l_{2},r_{2})^{n} \big)^{ \frac{l_{1}+r_{1}}{l_{2}+r_{2}}-1} \Sigma_{12} + \bigg( \frac{l_{1}+r_{1}}{l_{2}+r_{2}} \bigg)^{2} ( \check{BV}(l_{2},r_{2})^{n})^{2 \left( \frac{l_{1}+r_{1}}{l_{2}+r_{2}}-1 \right)} \Sigma_{22}.
\end{split}
\end{align}
Note that the convergence in \eqref{CLT-Homoskedasticity-Test} holds only under $\mathcal{H}_{0}$, while under $\mathcal{H}_{a}$ it follows from \eqref{equation:heteroskedasticity} that ${n}^{1/4} \Big( \check{BV}(l_{1},r_{1})^{n} - ( \check{BV}(l_{2},r_{2})^{n})^{ \frac{l_{1}+r_{1}}{l_{2}+r_{2}}} \Big) \rightarrow \infty$. This way we can determine if $X^{d}$ has homoskedastic or heteroskedastic volatility with asymptotically correct size and power tending to one, as $n \rightarrow \infty$. To render the test feasible, we propose a consistent estimator of $\Sigma$ in Section \ref{section:bootstrap}, which can be plugged into \eqref{equation:V}. It is both inherently robust to heteroskedasticity and positive semi-definite.

\section{A local estimator of diurnal variance} \label{section:diurnal-variance}

In the previous section, we pretended the diurnal component of volatility was available to deflate the noisy log-return series (i.e., \eqref{equation:deflated-increment}). In practice, $\sigma_{u}$ is unobserved. We here propose a nonparametric jump- and noise-robust estimator of it and state appropriate conditions, under which the sampling error---induced by this estimation---is asymptotically negligible, so that it does not thwart the results in Section \ref{section:theory} (and \ref{section:bootstrap}).

It turns out to be impossible to recover the latent diurnal variance on a fixed time interval. We thus resort to a long-span asymptotic theory, which extracts information about it by pooling high-frequency data across days.

As above, we suppose that on day $t$ we record $Y$ at equidistant time points $t_{i} = t-1 + i/n$, for $i = 0, 1, \ldots, n$ and write the associated $n$-frequency log-returns as:
\begin{equation}
\Delta_{(t-1)n + i}^{n}Y \equiv Y_{t-1 + i/n} - Y_{t-1 + (i-1)/n}, \quad \text{for } t = 1, \ldots, T \text{ and } i =1, \ldots, n.
\end{equation}
As in \citet*{zhang-mykland-ait-sahalia:05a}, we operate within a two time scale framework, where the ``slow'' scale uses a coarser set of $m$-frequency returns, where $m < n$, i.e. $\Delta_{(t-1)m + j}^{m}Y = Y_{t-1 + j/m} - Y_{t-1 + (j-1)/m}$, for $t = 1, \ldots, T$ and $j = 1, \ldots, m$, which is reserved for diurnal variance estimation, while the ``fast'' scale is based on all observed $n$-frequency returns and is intended for a bias-correction. Throughout, we assume $m$ is a divisor of $n$, so that $\{ j/m \}_{j = 0}^{m} \subseteq \{ i/n \}_{i = 0}^{n}$.

In the following, we say a process $(b_{t})_{t \geq 0}$ is bounded in $L^{p}$, if
\begin{equation}
\underset{t \in \mathbb{R}_{+}}{\sup }E \big[ |b_{t}|^{p} \big] < \infty.
\end{equation}

\noindent \textbf{Assumption (D4)}: $\quad$ $(a_{t})_{t\geq 0}$, $( \tilde{a}_{t})_{t \geq 0}$, $(\tilde{ \sigma}_{t})_{t \geq 0}$ and $( \tilde{v}_{t})_{t\geq 0}$ are bounded in $L^{4}$. \\[-0.25cm]

Assumption (D4) adds some regularity to the driving processes in $X$, which is necessary here as $T \rightarrow \infty$ in the asymptotic theory, and so we cannot appeal to the standard ``localization'' procedure \citep*[e.g.,][]{jacod:08a} to bound various terms in the proofs.

Now, we set:
\begin{equation} \label{equation:diurnal-variance-estimator}
\hat{ \sigma}_{u,s}^{2} = \frac{1}{T} \sum_{t=1}^{T} ( \sqrt{m} \Delta _{(t-1)m + j}^{m}Y)^{2} - \frac{m}{T} \sum_{t=1}^{T}[ \hat{ \text{var}}( \epsilon_{t-1 + (j-1)/m}) + \hat{ \text{var}}( \epsilon_{t-1 + j/m})], \quad \text{ for }s \in \lbrack (j-1)/m,j/m),  \end{equation}
where $\hat{ \text{var}}( \epsilon_{t-1+(j-1)/m})$ is a consistent estimator of $\text{var}( \epsilon_{t-1 + (j-1)/m})$, which has to converge at a rate faster than $m^{-1}$, e.g.
\begin{equation} \label{equation:noise-variance-i.i.d.}
\hat{ \text{var}} ( \epsilon_{t-1 + (j-1)/m}) = - \frac{1}{T} \frac{1}{n/m - 1} \sum_{t=1}^{T} \sum_{i=1}^{n/m} \bigg[ \Big( \Delta _{i + \left[t-1 + \frac{j-1}{m} \right] n}^{n}Y \Big) \Big( \Delta _{i-1 + \left[t-1 + \frac{j-1}{m} \right]n}^{n}Y \Big) \bigg].
\end{equation}
As readily seen, $\hat{ \sigma}_{u,s}^{2}$ is based directly on the raw noisy high-frequency data. It does not require jump-truncation nor pre-averaging and is therefore trivial to compute.\footnote{It is naturally also possible to pre-average and follow up with jump-truncation. This may lead to a better rate of convergence for the diurnal variance estimator. Here, we do not follow this line of thought, as it requires extra tuning parameters, and because the current setup appears to work reasonably well in practice.}  Due to its reliance on the squared normalized noisy high-frequency increment, however, it accumulate a bias from the microstructure noise, which the second term in \eqref{equation:diurnal-variance-estimator} cancels out by computing a local block-wise estimator of the noise variance.

While it appears counterintuitive, $\hat{ \sigma}_{u,s}^{2}$ is also jump-robust, as we show below. The intuition is that in our model, there are no fixed points of discontinuity in $X$, so that the influence of any jumps is intrinsically averaged away, as $m \rightarrow \infty$, $T \rightarrow \infty$ and $n \rightarrow \infty$.

\begin{proposition} \label{proposition:diurnal-component} Assume that $X$ is given by \eqref{equation:one}. Moreover, we suppose Assumption (D1) -- (D4), (V), (V'), (A) and (A'). If $m \rightarrow \infty$ and $T \rightarrow \infty$, as $n \rightarrow \infty$, then it holds that
\begin{equation}
\hat{ \sigma}_{u,s}^{2} = \sigma_{u,s}^{2} + O_{P}( m T^{-1/2}) + O_{P}( m^{3/2} n^{-1/2} T^{-1/2}).
\end{equation}
\end{proposition}
Next, we note that:
\begin{equation}
\sqrt{\hat{ \sigma}_{u,s}^{2}} - \sqrt{ \sigma_{u,s}^{2}} = \hat{ \sigma}_{u,s} - \sigma_{u,s} \simeq \frac{1}{2 \sqrt{ \sigma_{u,s}^{2}}} \big( \hat{ \sigma}_{u,s}^{2} - \sigma_{u,s}^{2} \big),
\end{equation}
with $\sigma_{u,s}^{2}$ bounded away from $0$. Thus, we can write:
\begin{equation}
\hat{ \sigma}_{u,s} = \sigma_{u,s} + O_{P}( m T^{-1/2}) + O_{P}( m^{3/2} n^{-1/2} T^{-1/2}).
\end{equation}
It therefore follows that if
\begin{equation}
m \propto n^{ \delta_{2}} \quad \text{ and } \quad T \propto n^{ \delta_{3}},
\end{equation}
where $\delta_{2} \in (0,1/2]$ and $\delta_{3} > 1/2 + 2 \delta_{2}$ (for a fixed value of $\delta_{2}$), then the error induced from estimating $\sigma_{u}$ does not alter the analysis in Section \ref{section:theory} and \ref{section:bootstrap}. That is, neither Theorem \ref{CLT-BV-Trunc} or \eqref{CLT-Homoskedasticity-Test} are affected, nor is the bootstrap applied to $\check{y}(l,r)_{i}^{n} = | \Delta _{i}^{n}\bar{Y}^{d}|^{l} 1_{\left\{ |\Delta_{i}^{n}\bar{Y}^{d}|<\upsilon _{n}\right\} }| \Delta _{i+k_{n}}^{n}\bar{Y}^{d}|^{r}1_{\left\{ |\Delta _{i+k_{n}}^{n}\bar{Y}^{d}|<\upsilon _{n}\right\} },$
where (with a slight abuse of notation) we redefine
\begin{equation} \label{equation:preavgY-feasible}
\Delta_{i}^{n} \bar{Y}^{d} = \sum_{j=1}^{k_{n}-1}g \bigg( \frac{j}{k_{n}} \bigg) \Delta_{i+j-1}^{n}Y^{d}, \quad i = 1, \ldots, n-k_{n}+2,
\end{equation}
to be based on:
\begin{equation} \label{equation:rescaled-increment-feasible}
\Delta_{i+j-1}^{n} Y^{d} = \frac{ \Delta_{i+j-1}^{n}Y}{\hat{ \sigma}_{u, \frac{i+j-1}{n}}},
\end{equation}
with $\hat{\sigma }_{u, \frac{i+j-1}{n}}$ from \eqref{equation:diurnal-variance-estimator}.

\begin{remark} \label{remark:monte-carlo} If the noise is autocorrelated but not heteroskedastic, $\hat{ \omega}^{2} = \hat{ \text{\upshape{var}}}( \epsilon_{t-1 + (j-1)/m})$ given by \eqref{equation:noise-variance-i.i.d.} is no longer a consistent estimator of $\omega^{2} = \text{\upshape{var}}( \epsilon)$. Indeed, when the noise is a stationary $q$-dependent sequence (for known $q > 0$), the statistic defined in \eqref{equation:noise-variance-i.i.d.} estimates the quantity $2 \big( \rho(0) - \rho(1) \big)$, where $\rho(k) = \text{cov}( \epsilon_{1}, \epsilon_{1+k})$. \citet*[][Lemma 2]{hautsch-podolskij:13a} discuss an estimator of $\rho(k)$, $k = 0, \ldots, q+1$, which is obtained from a simple recursion formula. Building on their result, we can deduce an estimator of $\omega^{2}$ in this alternative setup:
\begin{equation} \label{equation:robust-noise-variance}
\hat{ \omega}^{2} = - \sum_{k=1}^{q+1} k \hat{ \gamma}(k),
\end{equation}
where
\begin{equation} \label{equation:gamma-k}
\hat{ \gamma}(k) = \frac{1}{T} \frac{1}{n/m-k} \sum_{t=1}^{T} \sum_{i=1}^{n/m - k} \Big( \Delta_{i + \left[t-1 + \frac{j-1}{m} \right] n}^{n} Y \Big) \Big( \Delta _{i+k+\left[t-1 + \frac{j-1}{m}\right] n}^{n}Y \Big) ,\text{ for } k = 0, \ldots, q+1.
\end{equation}
Then,
\begin{equation}
\hat{ \omega}^{2} \overset{p}{ \rightarrow} \omega^{2} = \rho(0).
\end{equation}
The Monte Carlo and empirical analysis is based on
\begin{equation}
\hat{ \sigma}_{u,s}^{2} \equiv \frac{1}{T} \sum_{t=1}^{T} ( \sqrt{m} \Delta_{j+(t-1)m}^{m}Y )^{2} - 2m \hat{ \omega}^{2}, \quad \text{for } s \in \lbrack (j-1)/m,j/m),
\end{equation}
where $\hat{ \omega}^{2}$ is \eqref{equation:robust-noise-variance} with $q=3$.
\end{remark}

\section{The bootstrap} \label{section:bootstrap}

In this section, we improve the quality of inference in our test of heteroskedasticity in the noisy jump-diffusion setting by relying on the bootstrap, when computing critical values for the $t$-statistic. This is warranted by the Monte Carlo in Section \ref{section:montecarlo}, which reveals that in small samples, the feasible version of \eqref{CLT-Homoskedasticity-Test} (cf. \eqref{equation:clt}) is poorly approximated by the standard normal. Next, we propose a bootstrap estimator of the conditional covariance matrix of the limiting process $n^{1/4} \big( \check{BV}(l_{1},r_{1})^{n}, \check{BV}(l_{2},r_{2})^{n})^{ \intercal}$, i.e. $\Sigma$. As the bootstrap estimator is positive semi-definite by construction, it renders our test implementable.

We build on a series of papers in the high-frequency volatility area. In particular, \citet*{goncalves-meddahi:09a} propose the wild bootstrap for realized variance, in a framework where the asset price is observed without error \citep*[see also][]{hounyo:19a}. \citet*{goncalves-hounyo-meddahi:14a} and \citet*{hounyo-goncalves-meddahi:17a} extend their work to accommodate noise. The latter studies the pre-averaged realized variance estimator---i.e., $BV(2,0)^{n}$---proposed by \citet*{jacod-li-mykland-podolskij-vetter:09a}, where the pre-averaged returns are both overlapping and heteroskedastic due to stochastic volatility. In this context, a block bootstrap applied to $(\Delta_{i}^{n} \bar{Y}^{d})_{i=1}^{n-k_{n}+2}$ appears natural.

Nevertheless, such a scheme is only consistent if $\sigma_{sv,t}$ is constant. As shown by \citet*{hounyo-goncalves-meddahi:17a}, the problem is that $| \Delta_{i}^{n} \bar{Y}^{d}|^{2}$ are heterogeneously distributed under time-varying volatility.\footnote{This feature is highlighted by the asymptotic distribution of $\Delta_{i}^{n} \bar{Y}^{d}$ in \eqref{equation:distributionYbar} below.} In particular, their mean and variance are unequal. This creates a bias term in the blocks of blocks bootstrap variance estimator. To cope with both dependence and heterogeneity of $| \Delta_{i}^{n} \bar{Y}^{d}|^{2}$, they combine the wild bootstrap with the blocks of blocks bootstrap. The procedure exploits that heteroskedasticity can be handled by the former, while the latter can replicate serial dependence in the data. \citet*{hounyo:17a} generalizes \citet*{hounyo-goncalves-meddahi:17a} to a broad class of covariation estimators in a general setting that accommodates jumps, microstructure noise, irregularly spaced high-frequency data and non-synchronous trading. Also, \citet*{dovonon-goncalves-hounyo-meddahi:19a} develop a new local Gaussian bootstrap for high-frequency jump testing, but market microstructure noise is supposed to be absent. Here, we allow for noise and concentrate on heteroskedasticity.

The bootstrap version of $\check{BV}(l,r)^{n}$ is
\begin{equation}
\check{BV}(l,r)^{n*} = n^{ \frac{l+r}{4} - 1} \frac{1}{\mu_{l} \mu _{r}} \sum_{i=1}^{N_{n}} \check{y}(l,r)_{i}^{n*},
\label{Boot-Estimator}
\end{equation}%
where $( \check{y}(l,r)_{i}^{n*})_{i=1}^{N_{n}}$ is a bootstrap sample from $( \check{y}(l,r)_{i}^{n})_{i=1}^{N_{n}}$.

We apply a bootstrap to $\check{y}(l,r)_{i}^{n}$, which replicates their dependence and heterogeneity. As suggested by \citet*{hounyo-goncalves-meddahi:17a}, we merge the wild bootstrap with block-based resampling. However, our bootstrap is new, and it can be viewed as an overlapping version of their algorithm. We name it ``the overlapping wild blocks of blocks bootstrap.'' We note that the degree of overlap among the blocks to be bootstrapped plays a major role in efficiency: the nonoverlapping block-based approach is less efficient than a partial or full-overlap block \citep*[e.g.,][]{dudek-leskow-paparoditis-politis:14a}.

To describe this approach, let $b_{n}$ be a sequence of integers, which will denote the bootstrap block size, such that for some $\delta_{1} \in (0,1)$:
\begin{equation} \label{Tunpara-Block-Size}
b_{n} = O \big(n^{ \delta _{1}} \big).
\end{equation}
We divide $(\check{y}(l,r)_{i}^{n})_{i=1}^{N_{n}}$ into overlapping blocks of size $b_{n}$. The total number of such blocks is $N_{n}-b_{n}+1$. The bootstrap is based on $N_{n}-2b_{n}+2$ of them. In particular, we look at overlapping blocks within the set $( \check{y}(l,r)_{i}^{n})_{i=1}^{N_{n}-b_{n}}$ (there is $J_{n} = N_{n} - 2b_{n} + 1$ many such blocks) and the last block containing the elements $\check{y}(l,r)_{N_{n}-b_{n}+1}^{n}, \ldots, \check{y}(l,r)_{N_{n}}^{n}$. The bootstrap sample is constructed by properly combining the first $J_{n}$ blocks.

To explain this setup and avoid confusion, note that the main ingredient behind the theoretical validity of the suggested resampling scheme is that we center all bootstrap draws from a block of $b_{n}$ consecutive observations, say the $j$th that holds $\check{y}(l,r)_{j}^{n}, \ldots , \check{y}(l,r)_{j+b_{n}-1}^{n}$, around a local average of data in the $(j + b_{n})$th block (which is thus shifted to the right and consists of $\check{y}(l,r)_{j+b_{n}}^{n}, \ldots, \check{y}(l,r)_{j+2b_{n}-1}^{n}$), as given by $\bar{B}_{j+b_{n}}$ in \eqref{equation:local-average} below. This principle is no longer applicable starting with the block that covers the elements $\check{y}(l,r)_{N_{n}-2b_{n}+2}^{n}, \ldots, \check{y}(l,r)_{N_{n}-b_{n}+1}^{n}$, because the centering here demands a local average to be computed from $\check{y}(l,r)_{N_{n}-b_{n}+2}^{n}, \ldots, \check{y}(l,r)_{N_{n}+1}^{n}$, and the last observation is not available.

Let $u_{1}, \ldots, u_{J_{n}+1}$ be i.i.d. random variables, whose distribution is independent of the original sample. We denote by $\mu_{q}^{*} = E^{*} \big(u_{j}^{q} \big)$ its $q$th order moments.\footnote{As usual in the bootstrap literature, $P^{*}$ ($E^{*} \text{ and var}^{*}$) denotes the probability measure (expected value and variance) induced by the resampling, conditional on a realization of the original time series. In addition, for a sequence of bootstrap statistics $Z_{n}^{*}$, we write (i) $Z_{n}^{*} = o_{p^{*}}(1)$ or $Z_{n}^{*} \overset{p^{*}}{
\rightarrow} 0$, as $n \rightarrow \infty$, if for any $\varepsilon > 0$, $\delta >0$, $\lim_{n \rightarrow \infty} P[ P^{*}( | Z_{n}^{*} | > \delta) > \varepsilon] = 0$, (ii) $Z_{n}^{*} = O_{p^{*}}(1)$ as $n \rightarrow \infty $, if for all $\varepsilon > 0$ there exists an $M_{ \varepsilon } < \infty$ such that $\lim_{n \rightarrow \infty } P[ P^{*} ( |Z_{n}^{*}|
> M_{ \varepsilon }) > \varepsilon] =0$, and (iii) $Z_{n}^{*} \overset{d^{*}}{ \rightarrow}Z$ as $n\rightarrow \infty $, if conditional on the sample $Z_{n}^{*}$ converges weakly to $Z$ under $P^{*}$, for all samples contained in a set with probability $P$ converging to one.} Then,
\begin{equation} \label{equation:local-average}
\bar{B}_{j} = \frac{1}{b_{n}} \sum_{i=1}^{b_{n}} \check{y}(l,r)_{i-1+j}^{n}, \quad j=1, \ldots, N_{n}-b_{n}+1,
\end{equation}
is the average of the data in the $j$th block consisting of $\check{y}(l,r)_{j}^{n}, \ldots, \check{y}(l,r)_{j+b_{n}-1}^{n}$. Next, we generate the overlapping wild blocks of blocks bootstrap observations by:
\begin{equation} \label{Boot-DGP-Jumps}
\check{y}(l,r)_{m}^{n*} - \bar{ \bar{B}}^{N_{n}} = \left\{
\begin{array}{lcl}
\frac{1}{ \sqrt{b_{n}}} \sum_{j=1}^{m} \big( \check{y}(l,r)_{m}^{n} - \bar{B}_{b_{n}+j} \big) u_{j}, & \text{if} & m \in I_{1}^{n}, \\[0.25cm]
\frac{1}{ \sqrt{b_{n}}} \sum_{j=1}^{b_{n}} \big( \check{y}(l,r)_{m}^{n} - \bar{B}_{m+j} \big) u_{m+j-b_{n}}, & \text{if} & m \in I_{2}^{n}, \\[0.25cm]
\frac{1}{ \sqrt{b_{n}}} \sum_{j=1}^{N_{n}-b_{n}+1-m} \big( \check{y}(l,r)_{m}^{n} - \bar{B}_{J_{n}+1-j+b_{n}} \big) u_{J_{n}+1-j}, & \text{if} & m \in I_{3}^{n}, \\[0.25cm]
\frac{1}{ \sqrt{b_{n}}} \big( \check{y}(l,r)_{m}^{n} - \bar{B}_{N_{n}-b_{n}+1} \big) u_{J_{n}+1}, & \text{if} & m \in I_{4}^{n},
\end{array}
\right.
\end{equation}
where
\begin{equation}
\bar{ \bar{B}}^{N_{n}} = \frac{1}{N_{n}} \sum_{i=1}^{N_{n}} \check{y}(l,r)_{i}^{n},
\end{equation}
and
\begin{alignat}{4}
\begin{aligned}
I_{1}^{n} &= \{ 1, \ldots, b_{n}-1 \}, \quad &&I_{2}^{n} = \{ b_{n}, \ldots, J_{n} \}, \\[0.25cm]
I_{3}^{n} &= \{ J_{n}+1, \ldots, N_{n}-b_{n} \}, \quad &&I_{4}^{n} = \{ N_{n}-b_{n}+1, \ldots,N_{n} \} .
\end{aligned}
\end{alignat}
It is interesting to note that if we were to center $\check{y}(l,r)_{m}^{n}$ around the grand mean $\bar{\bar{B}}^{N_{n}}$, instead of the localized block average $\bar{B}_{j+m}$, it would yield a bootstrap observation
\begin{equation}
\check{y}(l,r)_{m}^{n*} - \bar{ \bar{B}}^{N_{n}} = \Big( \check{y}(l,r)_{m}^{n} - \bar{ \bar{B}}^{N_{n}} \Big) \eta_{m},
\end{equation}
for $m \in I_{2}^{n}$ (the main set), where $\eta_{m} = \frac{1}{ \sqrt{b_{n}}} \sum_{j=1}^{b_{n}}u_{m+j-b_{n}}$. Therefore, under the assumption that $E(u_{j}) = 0$ and $\text{var}(u_{j}) = 1$, we find that $E( \eta_{m}) =0$, $\text{var}( \eta_{m}) =1$, and $\text{cov}( \eta_{m}, \eta_{m-k}) = \big( 1-\frac{k}{b_{n}} \big) 1_{ \{ k \leq b_{n} \} }$. Thus, our approach is related to the dependent wild bootstrap of \citet*{shao:10a} (see also, e.g., \citet*{hounyo:14a}), who extends the traditional wild bootstrap of \citet*{wu:86a, liu:88a} to the time series setting, and it is the special case, where the kernel function is assumed to be Bartlett \citep*[see Assumption 2.1 in][]{shao:10a}.

The idea of the new centering $\bar{B}_{j+m}$ is to deal with the mean heterogeneity of $\check{y}(l,r)_{m}^{n}$. As shown by \citet*{hounyo-goncalves-meddahi:17a}, for the case of squared pre-averaged returns $y(2,0)_{m}^{n}$, centering the non-overlapping wild blocks of blocks bootstrap around the corresponding grand mean $N_{n}^{-1} \sum_{i=1}^{N_{n}} y(2,0)_{i}^{n}$ does not work, when $\sigma_{sv,t}$ is time-varying. In this paper, we show that generating the bootstrap observations as in \eqref{Boot-DGP-Jumps} does yield an asymptotically valid bootstrap for $( \check{BV}(l_{1},r_{1})^{n}, \check{BV}(l_{2},r_{2})^{n})^{ \intercal}$, even if $\sigma_{sv,t}$ is not constant.

As in \citet*{shao:10a} and \citet*{hounyo:14a}, the dependence between neighboring observations $\check{y}(l,r)_{m}^{n}$ and $\check{y}(l,r)_{m^{ \prime}}^{n}$ is not only preserved, if $m$ and $m^{ \prime}$ belong to a particular block, as typical in block-based resampling. Indeed, if $|m  -m^{ \prime}| < b_{n}$, $\check{y}(l,r)_{m}^{n*}$ and $\check{y}(l,r)_{m^{ \prime}}^{n*}$ are conditionally dependent (except for the last $b_{n}$ data).

A common feature of the block-based bootstrap, in particular the non-overlapping wild blocks of blocks approach of \citet*{hounyo-goncalves-meddahi:17a}, is that if the sample size $N_{n}$ is not a multiple of $b_{n}$, then one has to either take a shorter bootstrap sample or use a fraction of the last resampled block. This leads to some inaccuracy, when $b_{n}$ is large relative to $N_{n}$. In contrast, for the overlapping version proposed in this paper, the size of the bootstrap sample is always equal to the original sample size.\footnote{We compared the rejection rate of our test based on the overlapping bootstrap suggested here to the non-overlapping version of \citet*{hounyo-goncalves-meddahi:17a}. We found our procedure has better size control, \textit{even} if $b_{n}$ is a multiple of $N_{n}$. Thus, both from a theoretical and practical viewpoint our approach is preferable.}

Write
\begin{equation}
\bar{ \bar{B}}^{N_{n}*} = \frac{1}{N_{n}} \sum_{i=1}^{N_{n}} \check{y}(l,r)_{i}^{n*},
\end{equation}
as the average value of the bootstrap observations. A closer inspection of $\bar{ \bar{B}}^{N_{n}*}$ suggests that we can rewrite the centered bootstrap sample mean $\bar{ \bar{B}}^{N_{n}*} - \bar{ \bar{B}}^{N_{n}}$ as follows
\begin{equation}
N_{n} \Big( \bar{ \bar{B}}^{N_{n}*} - \bar{ \bar{B}}^{N_{n}} \Big) = \frac{1}{ \sqrt{b_{n}}} \sum_{j=1}^{J_{n}} b_{n} \big( \bar{B}_{j} - \bar{B}_{j+b_{n}} \big) u_{j}.
\end{equation}
Thus,
\begin{align} \label{Boot-Estimator-2}
\begin{split}
\check{BV}(l,r)^{n*} &= \check{BV}(l,r)^{n} + n^{ \frac{l+r}{4}-1} \frac{1}{ \mu_{l} \mu_{r}} \frac{1}{ \sqrt{b_{n}}} \sum_{j=1}^{J_{n}}b_{n} \big( \bar{B}_{j} - \bar{B}_{j+b_{n}} \big) u_{j} \\[0.25cm]
&= \check{BV}(l,r)^{n} - \frac{1}{ \sqrt{b_{n}}} \sum_{j=1}^{J_{n}} \check{\Delta B}(l,r)_{j}^{n}u_{j},
\end{split}
\end{align}
where
\begin{equation} \label{Delta-B-Bar-j-trunc}
\check{ \Delta B}(l,r)_{j}^{n} = \check{B}(l,r)_{j+b_{n}}^{n} - \check{B}(l,r)_{j}^{n},
\end{equation}
such that
\begin{equation}
\check{B}(l,r)_{j}^{n} = n^{ \frac{l+r}{4}-1} \frac{1}{ \mu_{l} \mu_{r}} \sum_{i=1}^{b_{n}} \check{y}(l,r)_{i-1+j}^{n}.
\label{B-j-trunc}
\end{equation}
We can now derive the first and second bootstrap moment of $n^{1/4} \bigg( \begin{array}{c} \check{BV}(l_{1},r_{1})^{n*} \\ \check{BV}(l_{2},r_{2})^{n*} \end{array} \bigg)$. The following Lemma states the formulas.

\begin{lemma}
\label{LemmaBootmoments-2} Assume that $\check{y}(l,r)_{m}^{n*}$ are generated as in \eqref{Boot-DGP-Jumps}. Then, it follows that
\begin{equation} \label{equation:unbiased}
E^{*} \big( \check{BV}(l,r)^{n*} \big) = \check{BV}(l,r)^{n} - \frac{1}{ \sqrt{b_{n}}} \sum_{j=1}^{J_{n}} \check{ \Delta B}(l,r)_{j}^{n} E^{*}(u_{j}) ,
\end{equation}
Also, for $1 \leq i,j \leq 2$,
\begin{equation}
\text{\upshape{cov}}^{*} \big( n^{1/4} \check{BV}(l_{i},r_{i})^{n*}, n^{1/4} \check{BV}( l_{j},r_{j})^{n*} \big) = \frac{ \sqrt{n}}{b_{n}} \sum_{k=1}^{J_{n}} \check{ \Delta B} (l_{i},r_{i})_{k}^{n} \check{ \Delta B}(l_{j},r_{j})_{k}^{n} \text{\upshape{var}}^{*}(u_{k}).
\end{equation}
\end{lemma}
Equation \eqref{equation:unbiased} of Lemma \ref{LemmaBootmoments-2} implies that with $E^{*}(u_{j}) = 0$, $\check{BV}(l,r)^{n*}$ is an unbiased estimator of $\check{BV}(l,r)^{n}$, i.e. $E^{*} \big( \check{BV}(l,r)^{n*} \big) = \check{BV}(l,r)^{n}$. The second part shows that the bootstrap covariance of $n^{1/4} \check{BV}( l_{i},r_{i})^{n*}$ and $n^{1/4}\check{BV}(l_{j},r_{j})^{n*}$ depends on the variance of $u$. In particular, if we select $\text{var}^{*}(u) = 1/2$ as in \citet*{hounyo-goncalves-meddahi:17a}:
\begin{equation}
\text{var}^{*} \Bigg( n^{1/4} \Bigg(
\begin{array}{c}
\check{BV}(l_{1},r_{1})^{n*} \\[0.25cm]
\check{BV}(l_{2},r_{2})^{n*}
\end{array}
\Bigg) \Bigg) = \check{ \Sigma}^{n}, \label{Boot-Var-ij-Trunca}
\end{equation}
where $\check{ \Sigma}^{n} = \big( \check{ \Sigma}_{ij}^{l_{1},r_{1},l_{2},r_{2},n} \big) _{1 \leq i,j \leq 2}$ and
\begin{equation}
\check{ \Sigma}_{ij}^{l_{i},r_{i},l_{j},r_{j},n} = \frac{ \sqrt{n}}{2b_{n}} \sum_{k=1}^{J_{n}} \check{ \Delta B}(l_{i},r_{i})_{k}^{n} \check{ \Delta B}(l_{j},r_{j})_{k}^{n}.  \label{Var-Hat-a-ij-Trunca}
\end{equation}
Note that based on \eqref{Var-Hat-a-ij-Trunca}, we can rewrite $\check{\Sigma}^{n}$ as
\begin{equation} \label{Var-Hat-a-Trunc}
\check{ \Sigma}^{n} = \frac{ \sqrt{n}}{2b_{n}} \sum_{j=1}^{J_{n}} \check{ \xi}_{j} \check{ \xi}_{j}^{\intercal},
\end{equation}
where $\check{ \xi}_{j} \equiv \Big( \check{ \Delta B}(l_{1},r_{1})_{j}^{n}, \check{ \Delta B}(l_{2},r_{2})_{j}^{n} \Big)^{ \intercal}$. It follows that if the external random variable $u$ is selected as above, the overlapping wild blocks of blocks bootstrap variance estimator is consistent for the asymptotic variance of $n^{1/4} \big( \check{BV}(l_{1},r_{1})^{n}, \check{BV}(l_{2},r_{2})^{n} \big)^{ \intercal}$ provided $\check{ \Sigma}^{n}$ is a consistent estimator of $\Sigma$, as proved in Theorem \ref{Consistent-PSD-Estimator} below. Note that $\check{ \Sigma}^{n}$ is related to recent work on asymptotic variance estimation by \citet*{mykland-zhang:17a}; see also, e.g., \citet*{christensen-podolskij-thamrongrat-veliyev:17a, jacod-todorov:09a, mancini-gobbi:12a}.

\begin{remark}
\label{Remark1} Note that from \eqref{Var-Hat-a-Trunc}, we can also rewrite $\check{\Sigma}^{n}$ as follows:
\begin{equation} \label{Var-Hat-a-Mean-Trunc}
\check{ \Sigma}^{n} = \frac{1}{b_{n}} \sum_{m=1}^{b_{n}} \check{ \Sigma}_{m}^{n},
\end{equation}
where
\begin{equation} \label{Var-Hat-a-Mean-Summands-Trunc}
\check{ \Sigma}_{m}^{n} = \frac{ \sqrt{n}}{2} \sum_{j=0}^{ \lfloor N_{n} / b_{n} \rfloor - 2} \check{ \xi}_{jb_{n}+m} \check{ \xi}_{jb_{n}+m}^{ \intercal} = \Big( \check{ \Sigma} _{ij,m}^{l_{1},r_{1},l_{2},r_{2},n} \Big) _{1 \leq i,j \leq 2}.
\end{equation}
We deduce that the diagonal elements of $\check{ \Sigma}_{m}^{n}$, i.e. $\check{ \Sigma}_{11,m}^{l_{1},r_{1},l_{2},r_{2},n}$ and $\check{\Sigma}_{22,m}^{l_{1},r_{1},l_{2},r_{2},n}$ are nothing else than the consistent bootstrap variance estimators of the asymptotic variance of $n^{1/4} \check{BV}(l_{1},r_{1})^{n}$ and $n^{1/4} \check{BV}(l_{2},r_{2})^{n}$, as proposed by \citet*{hounyo:17a}.
\end{remark}

The next result shows that under some regularity conditions, the estimator $\check{ \Sigma}^{n}$ converges in probability to $\Sigma$ in a general It\^{o} semimartingale context.

\begin{theorem} \label{Consistent-PSD-Estimator}
Assume that $X$ fulfills Assumption (J) for some $\beta \in [0,2]$. Furthermore, suppose that the conditions of Theorem B.1 in Appendix B hold true, when $X$ is continuous (i.e., $X$ follows (A.7)), and also if $X$ has jumps (i.e., $X$ follows \eqref{equation:one}) with either
\begin{equation}
l_{1}+r_{1}+l_{2}+r_{2} \leq 4(1-\delta_{1}), \quad 0 \leq \beta < 4(1-\delta _{1}), \label{Jumps-Condition-1}
\end{equation}
or
\begin{equation} \label{Jumps-Condition-2}
l_{1}+r_{1}+l_{2}+r_{2} > 4(1- \delta_{1}), \quad 0 \leq \beta < 4(1 - \delta _{1}), \quad \frac{l_{1}+r_{1}+l_{2}+r_{2} - 4( 1-\delta _{1})}{2(l_{1}+r_{1}+l_{2}+r_{2}-\beta)} \leq \varpi < \frac{1}{2}.
\end{equation}
Then, as $n \rightarrow \infty$, it holds that
\begin{equation} \label{Consistent-Var-Hat-b}
\check{ \Sigma}^{n} \overset{p}{ \rightarrow} \Sigma,
\end{equation}
where $\Sigma$ is defined in Appendix A.
\end{theorem}

In our Monte Carlo studies and empirical application, we take $l_{1}=r_{1}=2$ and $l_{2}=r_{2}=1$. Here, \eqref{Jumps-Condition-2} holds provided $\beta < 4( 1 - \delta_{1})$. As $1/2 < \delta_{1} < 2/3$ by assumption (i.e, $4/3 < 4(1 - \delta_{1}) < 2$), it  therefore suffices that $\beta \in [0,4/3)$.

Theorem \ref{Consistent-PSD-Estimator} implies that in finite samples, we get a consistent and nonnegative estimator of $V$:
\begin{equation} \label{Estimator-Var-Homo-test}
\check{V}^{n} = \check{ \Sigma}_{11}^{n} - 2 \bigg( \frac{l_{1}+r_{1}}{l_{2}+r_{2}} \bigg) \big( \check{BV}(l_{2},r_{2})^{n} \big)^{ \frac{l_{1}+r_{1}}{l_{2}+r_{2}}-1} \check{ \Sigma}_{12}^{n} + \bigg( \frac{l_{1}+r_{1}}{l_{2}+r_{2}} \bigg)^{2} \big( \check{BV}(l_{2},r_{2})^{n} \big)^{2 \left( \frac{l_{1}+r_{1}}{l_{2}+r_{2}}-1 \right)} \check{ \Sigma}_{22}^{n}.
\end{equation}

\begin{corollary}
\label{CLT-Feasible-Tests} Assume that the conditions from Theorem \ref{Consistent-PSD-Estimator} hold true. If $X$ is given by \eqref{one-Constant-Vol}, such that Assumption (J) holds for some $\beta \in [0,1)$ and $\Big( \frac{1}{2(2- \beta)} \vee \frac{3}{2(4- \beta)} \Big) \leq \varpi < 1/2$. Then, if $l_{1}+r_{1} > l_{2}+r_{2}$ and as $n \rightarrow \infty$,
\begin{equation} \label{equation:clt}
T^{n} \equiv \frac{{n}^{1/4} \bigg( \check{BV}(l_{1},r_{1})^{n} - \big( \check{BV}(l_{2},r_{2})^{n} \big)^{ \frac{l_{1}+r_{1}}{l_{2}+r_{2}}} \bigg)}{ \sqrt{ \check{V}^{n}}} \overset{d}{ \rightarrow} N(0,1).
\end{equation}
\end{corollary}

Corollary \ref{CLT-Feasible-Tests} delivers the asymptotic normality of the studentized statistic $T^{n}$; the feasible version of \eqref{CLT-Homoskedasticity-Test}. Note that under the alternative presence of heteroskedasticity, $\check{BV}(l_{1},r_{1})^{n} - \big( \check{BV}(l_{2},r_{2})^{n} \big)^{ \frac{l_{1}+r_{1}}{l_{2}+r_{2}}}$ converges to a strictly positive random variable. Moreover, as $\check{V}^{n}$ was shown to be a robust estimator of $V$ even in the presence of stochastic volatility, jumps and noise, we can conclude that the statistic $T^{n} \rightarrow \infty$ if the realization of $X^{d}$ has a heteroskedastic volatility path. Therefore, appealing to the properties of stable convergence, we deduce that
\begin{align}
\lim_{n \rightarrow \infty} P \big( T^{n} > z_{1 - \alpha} \mid \Omega_{ \mathcal{H}_{0}} \big) &= \alpha, \label{Size-hetero-test} \\[0.25cm]
\lim_{n \rightarrow \infty} P \big( T^{n} > z_{1 - \alpha} \mid \Omega_{ \mathcal{H}_{a}} \big) &=      1  \label{Power-hetero-test}
\end{align}
where $z_{\alpha }$ is the $\alpha $-quantile of a standard normal distribution. The implication is that we reject $\mathcal{H}_{0}$, if $T^{n}$ is significantly positive. While the alternative inference procedure based on \eqref{equation:clt} does not require any resampling, it possesses inferior finite sample properties, as shown in Section \ref{section:montecarlo}.

\begin{figure}[t!]
\begin{center}
\caption{Proportion of microstructure noise in $\check{BV}(1,1)^{n}$.
\label{figure:djia-bias.eps}}
\begin{tabular}{cc}
{\footnotesize {Panel A: $\theta = 1/3$.}} & {\footnotesize {Panel B: $\theta = 1$.}} \\
\includegraphics[height=8cm,width=0.48\linewidth]{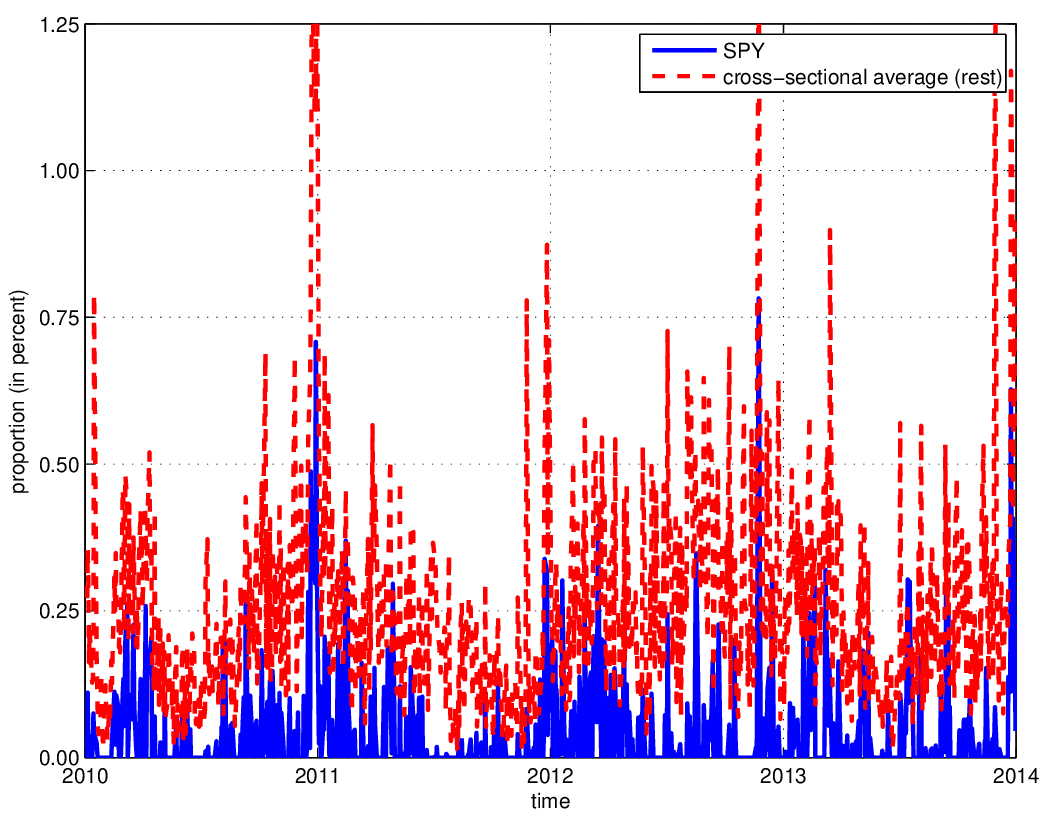} &
\includegraphics[height=8cm,width=0.48\linewidth]{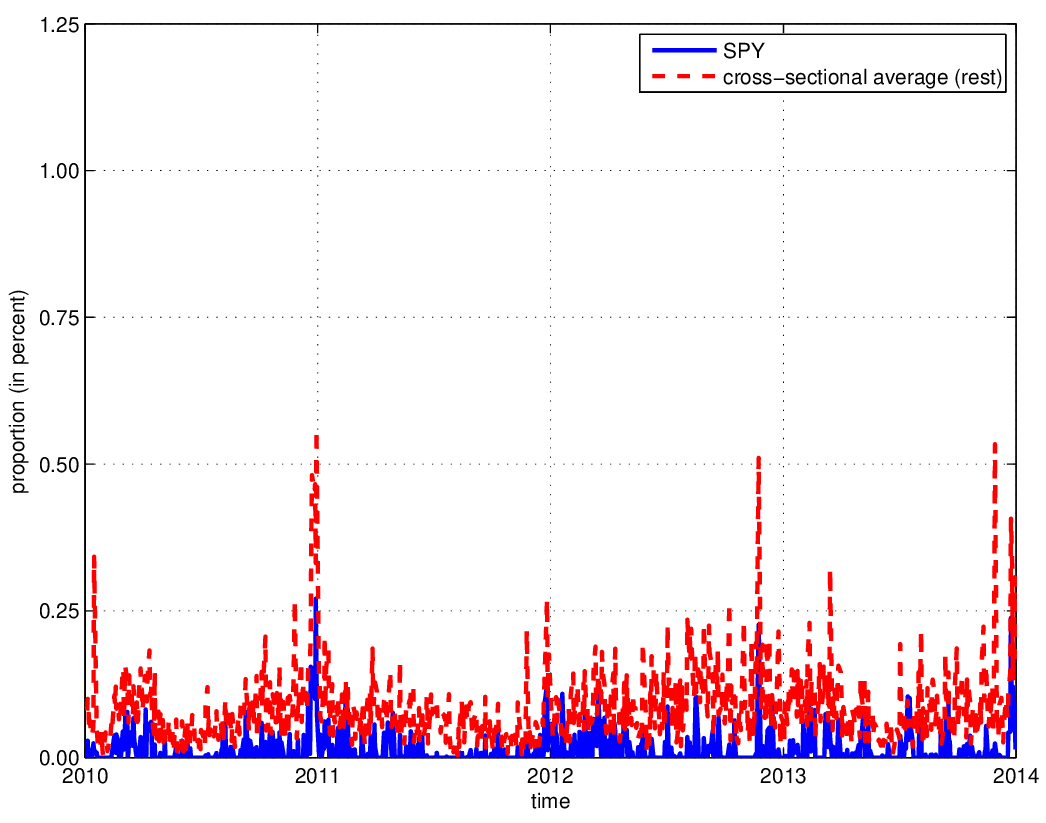}
\end{tabular}
\begin{scriptsize}
\parbox{0.95\textwidth}{\emph{Note.} We plot the proportion of $\check{BV}(1,1)^{n}$ that is due to residual variation (after pre-averaging) in the microstructure noise. $\check{BV}(1,1)^{n}$ is rescaled by $\theta \psi_{2}^{k_{n}}$ to provide an estimate of the integrated variance up to a bias term $\psi_{1}^{k_{n}} \omega^{2} / (\theta^{2} \psi_{2}^{k_{n}})$, see \eqref{Convergence-Proba-BV} and Theorem \ref{CLT-BV-Trunc}. The figure shows the ratio of the bias to the total $\check{BV}(1,1)^{n}$ estimate over time for the ticker symbols included in our empirical analysis. $\omega^{2}$ is replaced by the robust estimator $\hat{ \omega}^{2}$ in \eqref{equation:robust-noise-variance} computed daily with $q = 3$.}
\end{scriptsize}
\end{center}
\end{figure}

\begin{remark}
\label{remark:noise-variance} The results from \citet*{jacod-podolskij-vetter:10a} and \citet*{podolskij-vetter:09a} indicate that some assumptions can be relaxed. In particular, in Corollary \ref{CLT-Feasible-Tests}, if all the powers are even numbers (e.g., $l_{1} = 4, r_{1} = 0, l_{2} = 2$ and $r_{2}=0$), we can prove the results in the general setting of \citet*{jacod-podolskij-vetter:10a} with heteroskedastic noise, which is of a general form $E \big( \epsilon_{t}^{2} \mid X \big) = \omega_{t}^{2}$ and not necessarily restricted to $E \big( \epsilon_{t}^{2} \mid X \big) = \omega^{2} \sigma_{u,t}^{2}$. Here, the null is modified as
\begin{equation}
\mathcal{H}_{0} : \omega \in \Omega_{ \mathcal{H}_{0}} \cap \big\{ \omega : t \longmapsto \text{\upshape{var}} \big( \epsilon_{t}^{d} \mid X \big) \text{ is constant on } [0,1] \big\},
\end{equation}
where $\epsilon_{t}^{d} \equiv \epsilon_{t}/ \sigma_{u,t}$. This suggests that in the presence of heteroskedastic noise of a general form, $\mathcal{H}_{0}$ is a joint statement about the constancy of both diffusive and the rescaled noise variance. Such information should be useful in practice, because it delivers knowledge about the presence of heteroskedasticity irrespective of its origin. Meanwhile, Figure \ref{figure:djia-bias.eps} shows that for our empirical high-frequency data---and the two choices of $\theta$ adopted in the paper---$\check{BV}(1,1)^{n}$ is almost exclusively composed by diffusive volatility. This implies that very little residual noise is left in the data after pre-averaging, which indicates that any rejection of the null is more likely due to genuine time-variation in $\sigma_{sv,t}$. We note that the dampening of the noise is naturally much weaker for $\theta = 1/3$, which is therefore more susceptible to reject $\mathcal{H}_{0}$ on this ground.
\end{remark}

\begin{corollary}
\label{Boot-Var-Estimator-Jumps} Assume that the conditions of Theorem \ref{Consistent-PSD-Estimator} hold true and the external random variable is chosen as $u_{j} \overset{ \text{\upshape{i.i.d.}}}{ \sim} \big( E^{*}(u_{j}), \text{\upshape{var}}^{*} (u_{j}) \big)$, such that $\text{\upshape{var}}^{*} (u_{j}) = 1/2$. Then, as $n \rightarrow \infty$,
\begin{equation} \label{Boot-Var-Cons-b}
\text{\upshape{var}}^{*} \Bigg( n^{1/4} \Bigg(
\begin{array}{c}
\check{BV}(l_{1},r_{1})^{n*} \\[0.25cm]
\check{BV}(l_{2},r_{2})^{n*}
\end{array}
\Bigg) \Bigg) = \check{ \Sigma}^{n} \overset{p}{ \rightarrow} \Sigma,
\end{equation}
both in model (A.7) and \eqref{equation:one}, where $\Sigma$ is defined in Appendix A.
\end{corollary}
Given the consistency of the bootstrap variance estimator, we now prove the associated convergence of the bootstrap distribution of $n^{1/4} \big( \check{BV}(l_{1},r_{1})^{n*}, \check{BV}( l_{2},r_{2})^{n*} \big)^{ \intercal}$.

\begin{theorem}
\label{TheorJointCLT} Assume that all conditions from Corollary \ref{Boot-Var-Estimator-Jumps} hold true and that $E^{*}\big( |{u}_{j}| \big)^{2 + \delta} < \infty$ for some $\delta >0$. Then, as $n \rightarrow \infty$,
\begin{equation} \label{equation:clt-bootstrap}
\big( \check{ \Sigma}^{n} \big)^{-1/2} n^{1/4} \Bigg(
\begin{array}{c}
\check{BV}(l_{1},r_{1})^{n*} - E^{*} \big( \check{BV}(l_{1},r_{1})^{n*} \big) \\[0.25cm]
\check{BV}(l_{2},r_{2})^{n*} - E^{*} \big( \check{BV}(l_{2},r_{2})^{n*} \big)
\end{array}
\Bigg) \overset{d^{*}}{ \rightarrow} N(0,I_{2}),
\end{equation}
in probability-$P$, both in model (A.7) and \eqref{equation:one}. Moreover, let
\begin{equation}
S^{n*} = \frac{n^{1/4} \Bigg[ \check{BV}(l_{1},r_{1})^{n*} - \big( \check{BV}(l_{2},r_{2})^{n*} \big)^{ \frac{l_{1}+r_{1}}{l_{2}+r_{2}}} - \bigg( E^{*} \big( \check{BV}(l_{1},r_{1})^{n*} \big) - \Big( E^{*} \big( \check{BV}(l_{2},r_{2})^{n*} \big) \Big)^{ \frac{l_{1}+r_{1}}{l_{2}+r_{2}}} \bigg) \Bigg] }{\sqrt{V}},
\end{equation}
where $l_{1}+r_{1} > l_{2}+r_{2}$. It holds that
\begin{equation}
V^{n*} \equiv \text{\upshape{var}}^{*} \bigg[ n^{1/4} \Big( \check{BV}(l_{1},r_{1})^{n*} - \big( \check{BV}(l_{2},r_{2})^{n*} \big)^{ \frac{l_{1}+r_{1}}{l_{2}+r_{2}}} \Big) \bigg]
\overset{p}{ \rightarrow} V,
\end{equation}
and
\begin{equation}
S^{n*} \overset{d^{*}}{ \rightarrow} N(0,1),
\end{equation}
in probability-$P$, both in model \eqref{one-Constant-Vol} and \eqref{equation:one}.
\end{theorem}

Theorem \ref{TheorJointCLT} shows that the normalized statistic $S^{n*}$ is asymptotically normal both in model \eqref{one-Constant-Vol} and \eqref{equation:one}. This implies, independently of whether $\mathcal{H}_{0}$ or $\mathcal{H}_{a}$ is true, $S^{n*} \overset{d^{*}}{ \rightarrow} N(0,1)$, in probability-$P$. This ensures that the following bootstrap test both controls the size and is consistent under the alternative. Let
\begin{equation}
\mathcal{Z}^{n*} \equiv n^{1/4} \Big[ \check{BV}(l_{1},r_{1})^{n*} - \big(BV(l_{2},r_{2})^{n*} \big)^{ \frac{l_{1}+r_{1}}{l_{2}+r_{2}}} - \Big( E^{*} \big(BV(l_{1},r_{1})^{n*} \big) - \big( E^{*}(BV(l_{2},r_{2})^{n*}) \big)^{ \frac{l_{1}+r_{1}}{l_{2}+r_{2}}} \Big) \Big]
\end{equation}
and
\begin{equation}
\mathcal{Z}^{n} \equiv n^{1/4} \bigg( \check{BV}(l_{1},r_{1})^{n} - \big( \check{BV}(l_{2},r_{2})^{n} \big)^{ \frac{l_{1}+r_{1}}{l_{2}+r_{2}}} \bigg).
\end{equation}

\begin{remark} \label{Remark3}
We reject $\mathcal{H}_{0}$ at level $\alpha$, if $\mathcal{Z}^{n} > p_{1- \alpha}^{*}$, where $p_{1- \alpha }^{*}$ is the $(1- \alpha)$-percentile of the bootstrap distribution of $\mathcal{Z}^{n*}$. Under the conditions of Theorem \ref{TheorJointCLT}, the statistic $\mathcal{Z}^{n*} \overset{d^{*}}{ \rightarrow} N(0,V)$, in probability-$P$. Note that as $\mathcal{Z}^{n} \overset{d_{st}}{ \rightarrow} N(0,V)$ on $\Omega_{ \mathcal{H}_{0}}$, the fact that $\mathcal{Z}^{n*} \overset{d^{*}}{ \rightarrow} N(0,V)$, in probability-$P$, ensures that the test has correct size, as $n \rightarrow \infty$. On the other hand, under the alternative (i.e. on $\Omega_{ \mathcal{H}_{a}}$), as $\mathcal{Z}^{n}$ diverges at rate $n^{1/4}$, but we still have that $\mathcal{Z}^{n*} \overset{d^{*}}{ \rightarrow} N(0,V) = O_{p^{*}}(1)$, the test has unit power asymptotically.
\end{remark}

The above bootstrap test is convenient, as it does not require estimation of the asymptotic variance-covariance matrix $\Sigma$, but it may not lead to asymptotic refinements. In order to attain such improvement (or at least be able to prove it), we should base the bootstrap on an asymptotically pivotal $t$-statistic. To this end, we propose a consistent bootstrap estimator of $\check{ \Sigma}^{n} = \text{var}^{*} \Big( n^{1/4} \big( \check{BV}(l_{1},r_{1})^{n*}, \check{BV}( l_{2},r_{2})^{n*} \big)^{ \intercal} \Big)$. We look at the following adjusted bootstrap version of $\check{ \Sigma}^{n}$ given by $\check{ \Sigma}^{n*} = \Big( \check{ \Sigma}_{ij}^{l_{1},r_{1},l_{2},r_{2},n*} \Big) _{1 \leq i,j \leq 2}$, where the individual entries of $\check{ \Sigma}^{n*}$ are
\begin{equation} \label{Var-Hat-a-ij-Trunca-star}
\check{ \Sigma}_{ij}^{l_{i},r_{i},l_{j},r_{j},n*} = \frac{ \sqrt{n}}{b_{n}} \frac{\text{var}^{*}(u)}{E^{*}(u^{2})} \sum_{k=1}^{J_{n}} \check{ \Delta B}(l_{i},r_{i})_{k}^{n*} \check{ \Delta B}^{*}(l_{j},r_{j})_{k}^{n},
\end{equation}
with
\begin{equation}
\check{ \Delta B}(l,r)_{j}^{n*} = \check{ \Delta B}(l,r)_{j}^{n}u_{j},
\end{equation}
where $\check{ \Delta B}(l,r)_{j}^{n}$ is from \eqref{Delta-B-Bar-j-trunc} and $(u_{j})_{j = 1}^{J_{n}}$ are the external random variables used to generate the bootstrap observations in \eqref{Boot-DGP-Jumps}. We can also write
\begin{equation}
\check{ \Sigma}^{n*} = \frac{ \sqrt{n}}{b_{n}} \frac{\text{var}^{*}(u)}{E^{*}(u^{2})} \sum_{j=1}^{J_{n}} \check{ \xi}_{j}^{*} \check{ \xi}_{j}^{*^{\intercal}},
\end{equation}
where $\check{ \xi}_{j}^{*} \equiv u_{j} \Big( \check{ \Delta B}(l_{1},r_{1})_{j}^{n}, \check{ \Delta B}(l_{2},r_{2})_{j}^{n} \Big)^{ \intercal}$.  We can show that $\check{ \Sigma}^{n*}$ consistently estimates $\check{ \Sigma}^{n}$ for any choice of external random variable $u$ with $E^{*} \big( |{u}_{j}|^{4} \big) < \infty$. Next, based on $\check{ \Sigma}^{n*}$ we construct a bootstrap studentized variant of \eqref{equation:clt}:
\begin{equation} \label{Feasible-Boot-Jumps-Homo-Tests}
T^{n*} \equiv \frac{ \mathcal{Z}^{n*}}{ \sqrt{ \check{V}^{n*}}},
\end{equation}
where
\begin{equation}
\check{V}^{n*} = \check{ \Sigma}_{11}^{n*} - 2 \bigg( \frac{l_{1}+r_{1}}{l_{2}+r_{2}} \bigg) \big( \check{BV}(l_{2},r_{2})^{n} \big)^{ \frac{l_{1}+r_{1}}{l_{2}+r_{2}}-1} \check{ \Sigma}_{12}^{n*} + \bigg( \frac{l_{1}+r_{1}}{l_{2}+r_{2}} \bigg)^{2} \big( \check{BV}(l_{2},r_{2})^{n} \big)^{2 \left( \frac{l_{1}+r_{1}}{l_{2}+r_{2}}-1 \right)} \check{ \Sigma}_{22}^{n*}.
\end{equation}

\begin{theorem} \label{TheorBoot-Student-test}
Assume that the conditions of Corollary \ref{Boot-Var-Estimator-Jumps} are true and the external random variable is chosen as $u_{j} \overset{\text{\upshape{i.i.d.}}}{ \sim} \big( E^{*}(u_{j}), \text{\upshape{var}}^{*}(u_{j}) \big)$, such that $E^{*} \big( | {u}_{j} |^{4 + \delta} \big) < \infty$ for some $\delta > 0$. Then, as $n \rightarrow \infty$,
\begin{equation}
\big( \check{ \Sigma}^{n*} \big)^{-1/2} n^{1/4} \Bigg(
\begin{array}{c}
\check{BV}(l_{1},r_{1})^{n*} - E^{*} \big( \check{BV}(l_{1},r_{1})^{n*} \big) \\[0.25cm] \check{BV}(l_{2},r_{2})^{n*} - E^{*} \big( \check{BV}(l_{2},r_{2})^{n*} \big)
\end{array}
\Bigg) \overset{d^{*}}{ \rightarrow } N(0,I_{2}),
\end{equation}
in probability-$P$, both in model (A.7) and \eqref{equation:one}. Also,
\begin{equation}
T^{n*} \overset{d^{*}}{ \rightarrow} N(0,1),
\end{equation}
in probability-$P$, both in model \eqref{one-Constant-Vol} and \eqref{equation:one}.
\end{theorem}

Theorem \ref{TheorBoot-Student-test} shows the asymptotic normality of the studentized statistic $T^{n*}$. An implication of results in Theorem \ref{TheorBoot-Student-test} is that we reject $\mathcal{H}_{0}$ at significance level $\alpha$, if $T^{n} > q_{1 - \alpha}^{*}$, where $q_{1 - \alpha}^{*}$ is the $(1-\alpha)$-percentile of the bootstrap distribution of $T^{n*}$.

\section{Monte Carlo analysis} \label{section:montecarlo}

We here assess the properties of the nonparametric noise- and jump-robust test of deterministic versus stochastic variation in the intraday volatility coefficient that was proposed in Section \ref{section:theory}. We also highlight the refinements that can potentially be offered by the bootstrap, as outlined in Section \ref{section:bootstrap}, in sample sizes that resemble those, we tend to encounter in practice. We do so via detailed and realistic Monte Carlo simulations, and we start by describing the design of the study.

To simulate the efficient log-price $X_{t}$, we adopt the model:
\begin{equation} \label{sim:X}
\text{d}X_{t} = a_{t} \text{d}t + \sigma_{t} \text{d}W_{t} + \text{d}J_{t},
\end{equation}
where $X_{0} = 0$, $a_{t} = 0.03$ (per annum) and the other components are defined below.

As above, $\sigma_{t} = \sigma_{sv,t} \sigma_{u,t}$. To describe $\sigma_{u,t}$, we follow earlier work of \citet*{hasbrouck:99a} and \citet*{andersen-dobrev-schaumburg:12a} by using the specification:
\begin{equation} \label{equation:true-diurnal-variance}
\sigma _{u,t} = C + A e^{-a_{1}t} + B e^{-a_{2} (1-t)}.
\end{equation}
We set $A = 0.75$, $B = 0.25$, $C = 0.88929198$ and $a_{1} = a_{2} = 10$.\footnote{The calibration of $C$ ensures that $\int_{0}^{1} \sigma_{u,t}^{2} \text{d}t = 1$.} This produces a pronounced, asymmetric reverse J-shape in $\sigma_{u,t}$ with a value of about 1.8 (1.1) times higher at the start (end) of each simulation compared to the observations in the middle. This is also a good description of the actual intraday volatility pattern observed in our empirical high-frequency data (cf. Panel B in Figure \ref{figure:djia-diurnal-variance.eps}).

We assume that $\sigma_{sv,t}$ follows a stochastic volatility two-factor structure (SV2F):\footnote{The s-exp function is used to denote the exponential function that has been spliced with a polynomial of linear growth at high values of its argument, i.e. $s$-$\exp (x)= e^{x}$ if $x\leq x_{0}$ and $s$-$\exp (x)= \frac{ e^{x_{0}}}{ \sqrt{x_{0}-x_{0}^{2}+ x^{2}}}$, if $x > x_{0}$. As advocated by \citet*{chernov-gallant-ghysels-tauchen:03a}, we set $x_{0} = \ln(1.5)$.}
\begin{equation}
\sigma _{sv,t} = s\text{-} exp \left( \beta _{0}+ \beta _{1} \tau _{1,t} + \beta_{2} \tau _{2,t} \right),
\end{equation}
where
\begin{equation}
\text{d} \tau_{1,t} = \alpha_{1} \tau_{1,t} \text{d}t + \text{d}B_{1,t}, \qquad \text{d} \tau_{2,t} = \alpha_{2} \tau_{2,t} \text{d}t + (1 + \phi \tau _{2,t}) \text{d}B_{2,t}.
\end{equation}
Here, $B_{1,t}$ and $B_{1,t}$ are two independent standard Brownian motions with $E \left( \text{d}W_{t} \text{d}B_{1,t} \right) = \rho_{1} \text{d}t$ and $E \left( \text{d}W_{t} \text{d}B_{2,t}\right) = \rho_{2} \text{d}t$.

We follow \citet*{huang-tauchen:05a} and use the parameters $\beta_{0}=-1.2$, $\beta _{1}=0.04$, $\beta _{2}=1.5$, $\alpha _{1}=-0.00137$, $\alpha_{2}=-1.386$, $\phi = 0.25$ and $\rho_{1}=\rho_{2}=-0.3$.\footnote{Note that these parameters are annualized. We assume there are 250 trading days in a year.} This means that the first factor becomes a slowly-moving component, which generates persistence in volatility, while the second is a fast mean-reverting process that allows for a sufficient amount of volatility-of-volatility. At the start of each simulation, we initialize $\tau_{1}$ at random from its stationary distribution, i.e. $\tau_{1,0} \sim N \big( 0, -[2\alpha_{1}]^{-1} \big)$. Meanwhile, $\tau_{2}$ is started at $\tau _{2,0} = 0$ \citep*[e.g.,][]{barndorff-nielsen-hansen-lunde-shephard:08a}.

In absence of stochastic volatility, i.e. under the null hypothesis of deterministic diurnal variation, we freeze $\sigma_{sv,t}$ at a value equal to the unconditional expectation implied by the above SV2F model, i.e. $\sigma^{2} = E( \sigma_{sv,t}^{2})$.

$J_{t}$ is a symmetric tempered stable process with L\'{e}vy measure:
\begin{equation}
\nu(\text{d}x) = c\frac{e^{- \lambda x}}{x^{1+\beta}} \text{d}x,
\end{equation}
where $c > 0$, $\lambda > 0$, and $\beta \in [0,2)$ measures the degree of jump activity. We assume $\lambda = 3$ and $\beta = 0.5$. The choice of $\beta$ produces an infinite-activity, finite-variation process and is consistent with Theorem \ref{Consistent-PSD-Estimator}. The idea is to subdue $X_{t}$ to a stream of small jumps that, in contrast to the large ones, are typically difficult to filter via truncation, and which can be confused by the $t$-statistic with stochastic volatility. We therefore anticipate that this setup induces some size distortions in the test. $c$ is calibrated so that $J_{t}$ accounts for 20\% of the quadratic variation. This parameterization aligns with other papers \citep*[e.g.,][]{ait-sahalia-jacod-li:12a, ait-sahalia-xiu:16a, huang-tauchen:05a}.

We approximate the continuous time representation of $\sigma_{sv,t}$ using an Euler scheme, while $J_{t}$ is generated as the difference between two spectrally positive tempered stable processes, which are simulated using the acceptance-rejection algorithm of \citet*{baeumer-meerschaert:10a}, as described in \citet*{todorov-tauchen-grynkiv:14a}.\footnote{We thank Viktor Todorov for sharing Matlab code to simulate a tempered stable process.} Note that the latter is exact, if $\beta < 1$, as is the case here.

We simulate data for $t \in [0,1]$ (this is thought of as corresponding to a trading session on a US stock exchange, which spans 6.5 hours), where the discretization step is $\Delta t = 1/23,400$ (i.e., time runs on a one second grid).

\begin{figure}[t!]
\caption{Illustration of a simulation. \label{figure:illustration}}
\begin{center}
\begin{tabular}{cc}
{\footnotesize {Panel A: volatility.}} & {\footnotesize {Panel B: jump process.}} \\
\includegraphics[height=8cm,width=0.48\linewidth]{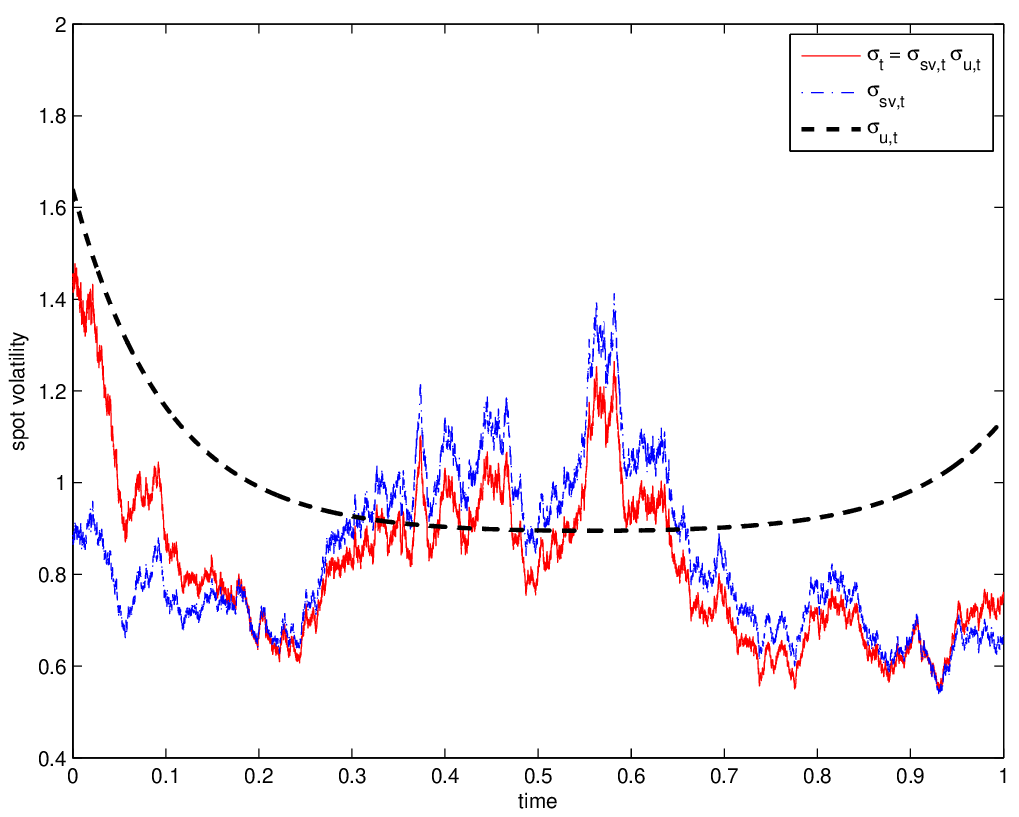} &
\includegraphics[height=8cm,width=0.48\linewidth]{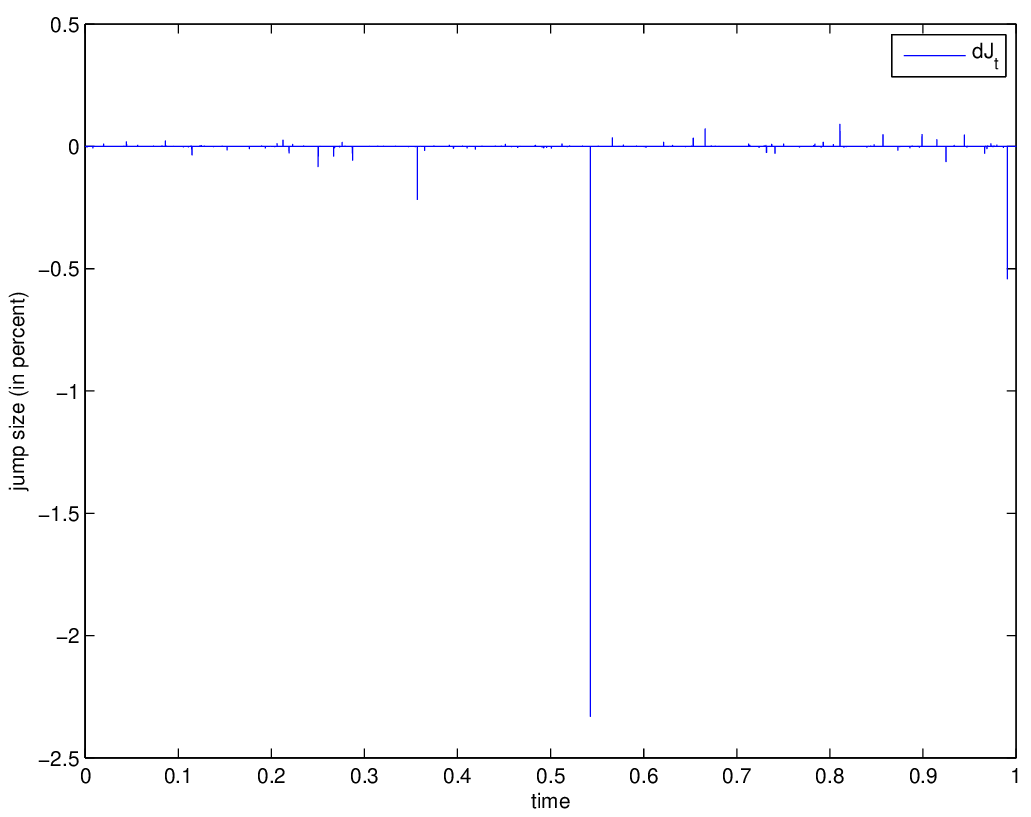}
\end{tabular}
\begin{scriptsize}
\parbox{0.925\textwidth}{\emph{Note.} The figure shows a sample path of the two main ingredients in the jump-diffusion model (from the first simulation of 1,000 replica in total). In Panel A, the volatility is measured relative to its unconditional average. In Panel B, as barely noticeable, the tempered stable jump process has many small increments that are close, but not equal, to zero.}
\end{scriptsize}
\end{center}
\end{figure}

In Figure \ref{figure:illustration}, we provide an illustration of a realization of the volatility and jump process from the full model.

A total of $T = 1,000$ Monte Carlo replica is generated. In each simulation, we pollute the efficient price with an additive noise term by setting $Y_{i/n} = X_{i/n} + \epsilon_{i/n}$. To capture the well-known negative serial correlation in log-returns induced by bid-ask bounce in transaction prices and apparent second-order effects present in our real data (cf. Panel A in Figure \ref{figure:djia-diurnal-variance.eps}), we follow \citet*{kalnina:11a} and model $\epsilon_{i/n}$ (for a given observation frequency $n$) as an MA(1):
\begin{equation} \label{equation:ar-noise}
\epsilon_{i/n} = \epsilon_{i/n}^{\prime }+ \varphi \epsilon_{(i-1)/n}^{ \prime}, \quad \text{where} \quad \epsilon_{i/n}^{\prime }\mid ( \sigma_{t} )_{t \in [0,1]} \overset{ \text{i.i.d.}}{ \sim} N \bigg(0, \frac{ \omega^{2}}{1 + \varphi^{2}} \bigg),
\end{equation}
so that $\text{var}( \epsilon) = \omega^{2}$.

To gauge how the strength of autocorrelation in $\epsilon$ affects our results, we consider $\varphi = 0$, $-0.3$, $-0.5$, and $-0.9$. Of course, the first value corresponds to the i.i.d. noise case. To model the magnitude of $\epsilon$, we set $\omega^{2} = \xi^{2} \sqrt{ \int_{0}^{1} \sigma_{t}^{4} \text{d}t}$, such that the variance of the market microstructure component scales with volatility \citep*[e.g.,][]{bandi-russell:06a,kalnina-linton:08a}. As in \citet*{barndorff-nielsen-hansen-lunde-shephard:08a}, we fix $\xi^{2} = 0.0001$, $0.001$ and $0.01$, as motivated by the empirical work of \citet*{hansen-lunde:06b}, who find these to be typical sizes of noise contamination for the 30 stocks in the Dow Jones Industrial Average index \citep*[see also, e.g.,][]{ait-sahalia-yu:09a}.

We construct $\Delta_{i}^{n}Y^{d} \equiv Y_{i/n}^{d} - Y_{(i-1)/n}^{d}$ at sampling frequency $n = 390$, $780$, $1560$, $4680$, $7800$, $11700$ and $23400$, thereby varying the sample size across a broad range of selections. With the above interpretation of time, the smallest (largest) value of $n$ amounts to observing a new price every minute (second). Such a number of trade arrivals is not unrealistic compared to real high-frequency data, as reported in Section \ref{section:empirical}.

\begin{figure}[t!]
\caption{Estimation of the diurnal component. \label{figure:sim-diurnal-variance}}
\begin{center}
\begin{tabular}{cc}
{\footnotesize {Panel A: $\mathcal{H}_{0}: \sigma_{t} = \sigma \sigma_{u,t}$.}} & {\footnotesize {Panel B: $\mathcal{H}_{a}: \sigma_{t} = \sigma_{sv,t} \sigma_{u,t}$.}} \\
\includegraphics[height=8cm,width=0.48\textwidth]{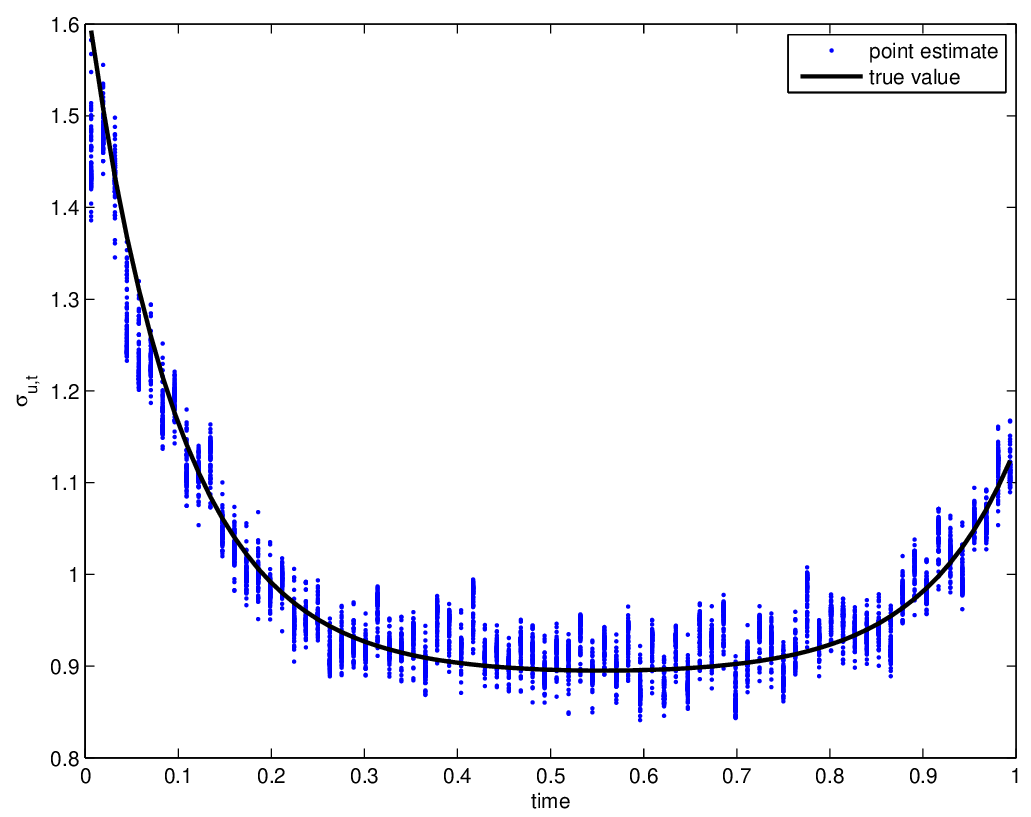} & \includegraphics[height=8cm,width=0.48\textwidth]{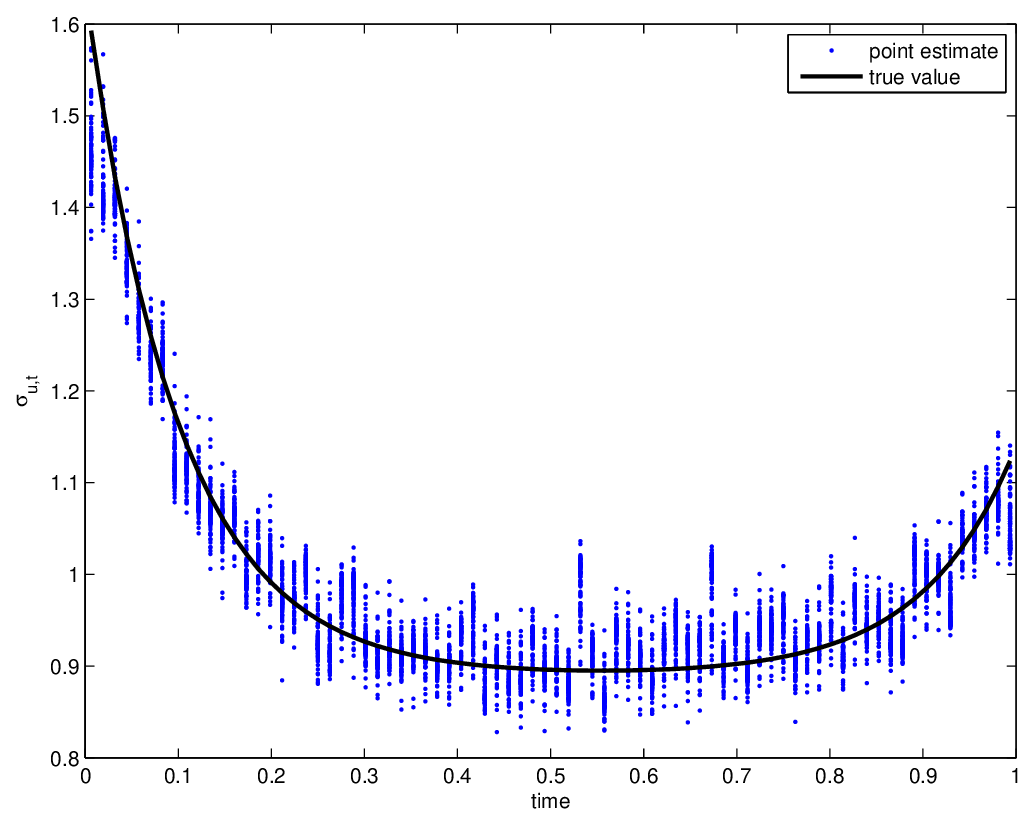}
\end{tabular}
\begin{scriptsize}
\parbox{0.925\textwidth}{\emph{Note.} We plot $\sigma_{u,t}$ from \eqref{equation:true-diurnal-variance}. An estimate is recovered from $\hat{ \sigma}_{u,t}$ proposed in \eqref{equation:diurnal-variance-estimator} based on $m = 78$. The bias-correction is done via $\hat{ \omega}^{2}$ in \eqref{equation:robust-noise-variance} with $q = 3$. Panel A depicts estimates under $\mathcal{H}_{0} : \sigma_{t} = \sigma \sigma_{u,t}$, while Panel B is for $\mathcal{H}_{a} : \sigma_{t} = \sigma_{sv,t} \sigma_{u,t}$. Each point on a vertical line corresponds to an estimate in that time interval for some combination of $n$, $\xi^{2}$ and $\varphi$.}
\end{scriptsize}
\end{center}
\end{figure}

The estimator $\hat{ \sigma}_{u,t}$ in \eqref{equation:diurnal-variance-estimator} is applied to deflate $\Delta_{i}^{n} Y$, as in \eqref{equation:rescaled-increment-feasible}.\footnote{In unreported results (available in the web appendix), we found that replacing $\sigma_{u,t}$ with $\hat{ \sigma}_{u,t}$ does not modify the properties of the $t$-statistic much. In particular, the rejection rates are only marginally higher with the latter. An increase was to be expected due to the inherent sampling error in the estimator. This means $\hat{ \sigma}_{u,t}$ does not completely control for the true seasonal pattern---to the extend it deviates from $\sigma_{u,t}$---leaving $\Delta_{i}^{n} Y^{d}$ modestly heteroskedastic even under $\mathcal{H}_{0}$. The effect is rather benign, which is remarkable given our naive configuration of $\hat{ \sigma}_{u,t}$ with a relatively small choice of $m$ even for larger sample sizes $n$.} We compute $\hat{ \sigma}_{u,t}$ based on the $T$ simulations with the associated value of $n$ and $m = 78$ (i.e., 5-minute data).\footnote{The noise plays a limited role at this sampling frequency, but we prefer a low value of $m$ to accommodate all the choices of $n$ in the simulation design. In our empirical work, we exploit 30-second data (i.e., $m = 780$), as the instruments analyzed there are very liquid and, as a result, sample sizes are generally large.} We bias-correct using $\hat{ \omega}^{2}$ from \eqref{equation:robust-noise-variance} with $q = 3$ (even if the noise is at most 1-dependent here). This matches the empirical implementation, where this value of $q$ suffices to capture the autocorrelation found in our real high-frequency data, as evident from Panel A in Figure \ref{figure:djia-diurnal-variance.eps}. In Figure \ref{figure:sim-diurnal-variance}, we plot $\hat{ \sigma}_{u,t}$ alongside $\sigma_{u,t}$ as estimated both under $\mathcal{H}_{0}$ and $\mathcal{H}_{a}$. As seen, $\hat{ \sigma}_{u,t}$ is roughly unbiased, but it exhibits higher variation around $\sigma_{u,t}$ in the presence of stochastic volatility, which adds measurement error to the calculations.\footnote{To illustrate the robustness of our approach, we do not filter $\hat{ \sigma}_{u,t}$, although it appears natural to exploit the smoothness condition in Assumption (D3) to reduce sampling errors further.}

We pre-average using \eqref{equation:preavgY} and \eqref{equation:preavgY-feasible}, which we do locally on a window of size $k_{n} = [ \theta \sqrt{n}]$. We work with $\theta = 1/3$ and $\theta = 1$ \citep*[as also done in, e.g.,][]{christensen-kinnebrock-podolskij:10a}.\footnote{As this introduces a small rounding effect in the relation between $\theta$ and $k_{n}$, we therefore reset $\theta = k_{n} / \sqrt{n}$ following the determination of $k_{n}$. We apply this ``effective'' $\theta$ in all the subsequent computations, as also advocated in \citet*{jacod-li-mykland-podolskij-vetter:09a}.} As standard, the weight function is $g(x) = \min(x,1-x)$.

The $t$-statistic is based on comparing $\check{BV}(l,r)^{n}$ with $l_{1} = r_{1} = 2$ and $l_{2} = r_{2} = 1$. To truncate, we set $v_{n}=cu_{n}^{\varpi }$ with $u_{n}=k_{n}/n$ in \eqref{equation:truncation}, which is adapted to an estimate of the local spot volatility. As in, e.g., \citet*{li-todorov-tauchen:13a,li-todorov-tauchen:16a}, we fix the ``rate'' parameter $\varpi =0.49$, while we determine the ``scale'' dynamically as $c = \Phi (0.999)\sqrt{BV(1,1)^{n}}$, where $\Phi (0.999)$ is the 99.9\%-quantile from the standard normal distribution and $BV(1,1)^{n} $ is the non-truncated estimator in \eqref{equation:preavgBV}. The intuition behind this construction is as follows. Assume there are no jumps in the interval $\left[ i/n,(i+k_{n})/n\right]$. Then, under mild regularity conditions:
\begin{equation} \label{equation:distributionYbar}
n^{1/4} \Delta_{i}^{n} \bar{Y}^{d} \overset{a}{ \sim} N \bigg(0, \theta \psi_{2} \sigma_{sv,i/n}^{2} + \frac{1}{ \theta} \psi_{1} \omega^{2} \bigg).
\end{equation}
It follows from \eqref{Convergence-Proba-BV} that $BV(1,1)^{n}\overset{p}{ \rightarrow} \int_{0}^{1}\big(\theta \psi_{2} \sigma_{sv,s}^{2} + \frac{1}{ \theta } \psi_{1} \omega^{2} \big) \text{d}s$, so that $\sqrt{BV(1,1)^{n}}$ is a (jump-robust) measure of the average dispersion (i.e., standard deviation) of the sequence $\Delta_{i}^{n} \bar{Y}^{d}$, while $\Phi (\cdot )$ controls how far out in the tails of the distribution truncation is enforced.\footnote{While the pre-averaged (1,1)-bipower variation is robust to the presence of jumps in the p-lim, as $n \rightarrow \infty$, in practice it tends to be slightly upward biased for a finite value of $n$, because the jumps are not completely eliminated, see, for example, \citet*{christensen-oomen-podolskij:14a}.} On the other hand, while $\Delta_{i}^{n} \bar{Y}^{d}$ is of order $O_{p} \big(n^{-1/4} \big)$, $\varpi \in (0,1/2)$ implies that $u_{n}^{ \varpi}$ shrinks at a slower pace than $\Delta_{i}^{n} \bar{Y}^{d}$. Therefore, purely ``continuous'' returns fall within the boundary of the threshold asymptotically. In contrast, if there are jumps in $\left[ i/n,(i+k_{n})/n\right] $, $\Delta_{i}^{n} \bar{Y}^{d}$ usually has order $O_{p}(1)$, and such ``discontinuous'' returns are, eventually, discarded.

The bootstrap inference is done as follows. We resample the pre-averaged high-frequency data $B = 999$ times for each Monte Carlo replication. Application of our bootstrap also requires the selection of the external random variable $u$. This is an important choice in practice, and consistent with previous work \citep*[e.g.,][]{hounyo:17a, hounyo-goncalves-meddahi:17a}
we examine the robustness of our approach by adopting two candidate distributions:\footnote{We also experimented with a third external random variable using an alternative formulation of the two-point distribution, where $u_{j} = \pm 1$ with probability $p = 1-p = 1/2$. The outcome was more or less identical to the results we report based on (2.), so we decided to exclude these results to save space.}

\begin{description}
\item[\hspace*{0.5cm} (1.)] $u_{j} \sim N(0,1/2)$.

\item[\hspace*{0.5cm} (2.)]
\begin{equation}
u_{j}=%
\begin{cases}
\displaystyle \frac{1}{\sqrt{2}}\Bigg(\frac{1-\sqrt{5}}{2}\Bigg), & \text{with probability}\quad \displaystyle p=\frac{\sqrt{5}+1}{2\sqrt{5}} \\[0.50cm]
\displaystyle \frac{1}{\sqrt{2}}\Bigg(\frac{1+\sqrt{5}}{2}\Bigg), & \text{with probability}\quad \displaystyle1-p=\frac{\sqrt{5}-1}{2\sqrt{5}}.
\end{cases}
\end{equation}
\end{description}

In both cases, $E^{ \ast}( u_{j}) = 0$ and $\text{var}^{\ast }(u_{j}) = 1/2$, so these are asymptotically valid choices of $u_{j}$ for the purpose of constructing a bootstrap test based on studentized and unstudentized statistics. The two-point distribution in (2.) was originally proposed by \citet*{mammen:93a}, and here we just scale it such that its variance is a half.

Estimation of the asymptotic variance-covariance matrix $\Sigma$ depends on the block size $b_{n}=O(n^{\delta })$ with $1/2<\delta <2/3$. Of course, this means nothing other than eventually $b_{n}=cn^{\delta }$, for some constant $c$. There is no available theory, which can help us find optimal choices of $c$ and $\delta $ (e.g., via a MSE criterion). Moreover, in finite samples any fixed block size $b_{n}$ can be achieved from many combinations of $c$ and $\delta $. Set against this upshot, we propose the following heuristic. We fix $\delta =2/3$ at the upper bound (again, the inequality constraint is only binding in the limit). We set $b_{n}^{\text{min}}=[2n^{\delta }]$ and $b_{n}^{\text{max}}=[\min (3n^{\delta },N_{n}/2)]$. The first choice is motivated, since we need at least $b_{n} \geq 2k_{n}$ for the estimator to capture the dependence in $(\check{y}(l,r)_{i}^{n})_{i=1}^{N_{n}}$, while the latter amounts to saying $b_{n}$ should also not be too large compared to $N_{n}$. We partition $[b_{n}^{\text{min}},b_{n}^{\text{max}}]$ into 30 equidistant subintervals and loop $b_{n}$ over the integers that are closest to the endpoints. We then select a value of $b_{n}$ by using the minimum variance criterion of \citet*{politis-romano-wolf:99a} with a two-sided averaging window of length $d=2$.

\input{tables/sim-theta=0.333-beta=-0.5-f.tex}

\input{tables/sim-theta=1.000-beta=-0.5-f.tex}

In Table \ref{table:sim-theta=0.333-beta=-0.5-f.tex} -- \ref{table:sim-theta=1.000-beta=-0.5-f.tex}, we report the rejection rates---averaged across simulations---of the above jump- and noise-robust test of $\mathcal{H}_{0}$ at the 5\% level of significance. The critical value in each test is found either via the 95\% quantile of the standard normal distribution function (labeled CLT), as motivated by the asymptotic theory in Corollary \ref{CLT-Feasible-Tests}, or with the help of the bootstrap-based percentile and percentile-$t$ approach---with the headings $z_{ \text{wb} \cdot}$ and $t_{ \text{wb} \cdot}$---for the two external random variates $u$ introduced above.

Throughout, we highlight the setting with $\varphi = -0.5$, while noting the simulated size and power for other values of $\varphi$ are generally within $\pm 1$\%-point of the numbers reported here (the latter are omitted, but available at request). This is also true for the noise variance parameter, $\xi$, which changes the results in a limited way, if at all, as
gauged by inspection of Panel A -- C in each table. As such, neither of the parameters associated to noise has a material effect on the outcome of the $t$-statistic, illustrating its robustness to market frictions. The only exception to this rule is for $\theta = 1/3$ and $\xi^{2} = 0.01$, where a visible drop in the rejection rate under $\mathcal{H}_{a}$ is noticed, suggesting that narrow pre-averaging is inadequate to counter the microstructure contamination in the presence of an (abnormally) large noise. As expected, the block size $b_{n}$ increases monotonically with $n$.

On the other hand, by comparing the two tables we note the pre-averaging window itself, via $\theta$, has a more significant impact on the test, though mostly in small samples. We comment further on that below.

Turning to the analysis of the rejection rates under $\mathcal{H}_{0}$ of deterministic volatility (size), the tables show the test is oversized. In particular, the CLT-based approach has a pronounced distortion in finite samples, starting at about 25.3\% (20.5\%) for $\theta = 1/3$ ($\theta = 1$). This is more than five (four) times larger than the nominal level. These rates improve and decline towards 5\% as $n$ increases, but remain notably elevated even in fairly large samples.

In contrast, the bootstrap-based approaches are much less biased relative to inference with the asymptotic critical value. The refinement brought about by bootstrapping is often substantial, when the sample size is limited, and the rejection rates are closer to the significance level across the board, albeit they are also mildly inflated initially. The percentile approach appears to possess better size properties compared to the percentile-$t$, and it settles around 5 -- 6\% fast. As noted above, the former procedure has the added advantage that it does not require the user to input a---potentially imprecise---estimate of $\Sigma$. This also helps to make it slightly less computationally intensive, so as a practical choice we advocate the percentile approach. It is interesting to see that the difference between the two external random variables, in terms of controlling size, is negligible, perhaps with a very weak preference for the one based on the discrete two-point distribution. In the empirical application below, we base our investigation on $z_{ \text{wb2}}$.

Next, we look at the simulation results under $\mathcal{H}_{a}$ with stochastic volatility (power). The power exhibited by the various tests is not overwhelming for small $n$, but it improves steadily towards 100\% as $n$ grows large. Still, it stays somewhat less than unity even for $n = 23,400$. It appears the CLT-based test has good power, but this is largely due to the sheer amount of Type I errors committed with this statistic. We observe a notable drop in the rejection rates going from Table \ref{table:sim-theta=0.333-beta=-0.5-f.tex} (with $\theta = 1/3$) vis-\`{a}-vis to Table \ref{table:sim-theta=1.000-beta=-0.5-f.tex} (with $\theta = 1$), caused by the heavy increase in pre-averaging. While this generally renders our testing procedures more resilient to the detrimental effects of microstructure noise, it also smooths out the underlying volatility path and thereby diminishes our ability to uncover genuine heteroskedasticity in the data. It thus highlights a crucial trade-off in practice in terms of selecting $\theta$.

The above suggests our test is not always powerful enough to pick up variation in $\sigma_{sv,t}$. There are several possible explanations of this finding. Firstly, the choice of the tuning parameter $\theta$ has a significant impact, as we highlighted above and inspect further below. Secondly, the problem is not trivial. It may just be hard to detect fluctuations in $\sigma_{sv,t}$ from noisy high-frequency data, leaving the jump distortion aside. Thirdly, and although $\sigma_{sv,t}$ is time-varying under the alternative, it may be so persistent that its sample path---which of course differs between simulations---moves about so little (in relative terms) that it appears essentially homoskedastic on an intraday time frame. While this feature partially justifies regarding $\sigma_{sv,t}$ as ``almost constant,'' it also makes it hard for the test to discriminate $\mathcal{H}_{a}$ from natural sampling variation under $\mathcal{H}_{0}$ (which it rightfully should do here), at least for the simulated sample sizes.

\begin{figure}[t!]
\begin{center}
\caption{Simulation properties of $t$-statistic and H-index. \label{figure:H-index}}
\begin{tabular}{cc}
\footnotesize{Panel A: assessment of power.} & \footnotesize{Panel B: relative bias of $\hat{\text{H}}$-index.} \\
\includegraphics[height=8cm,width=0.48\textwidth]{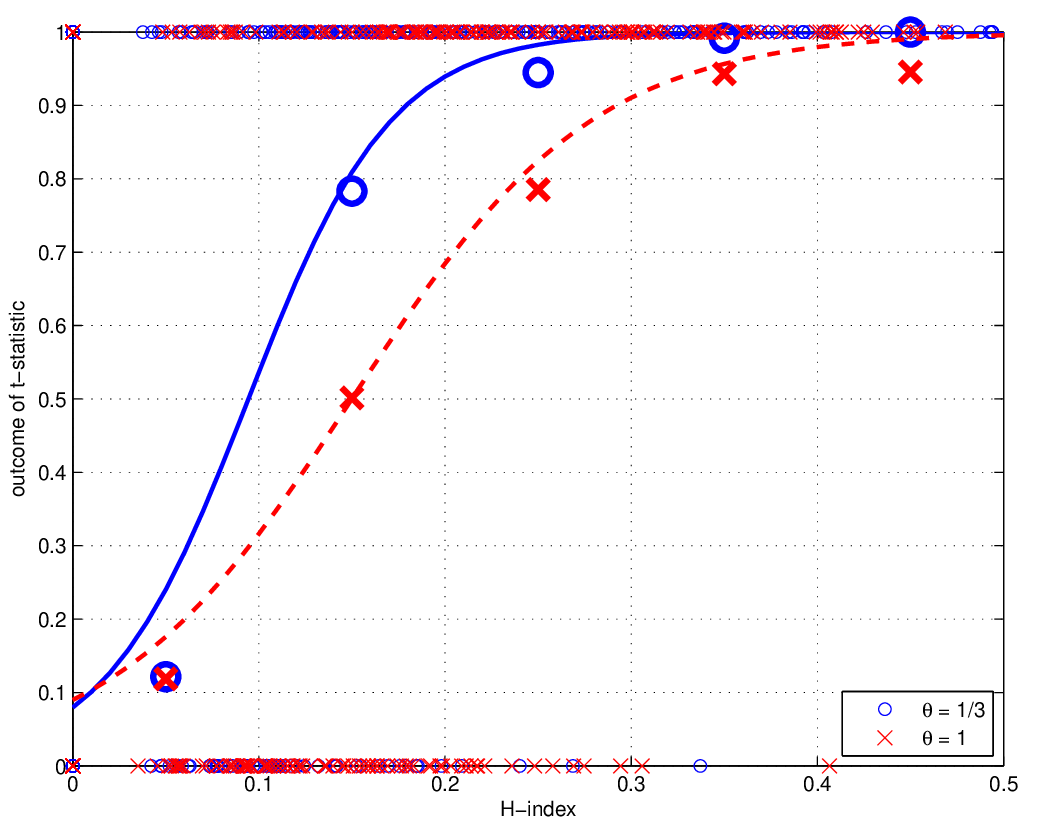} & \includegraphics[height=8cm,width=0.48\textwidth]{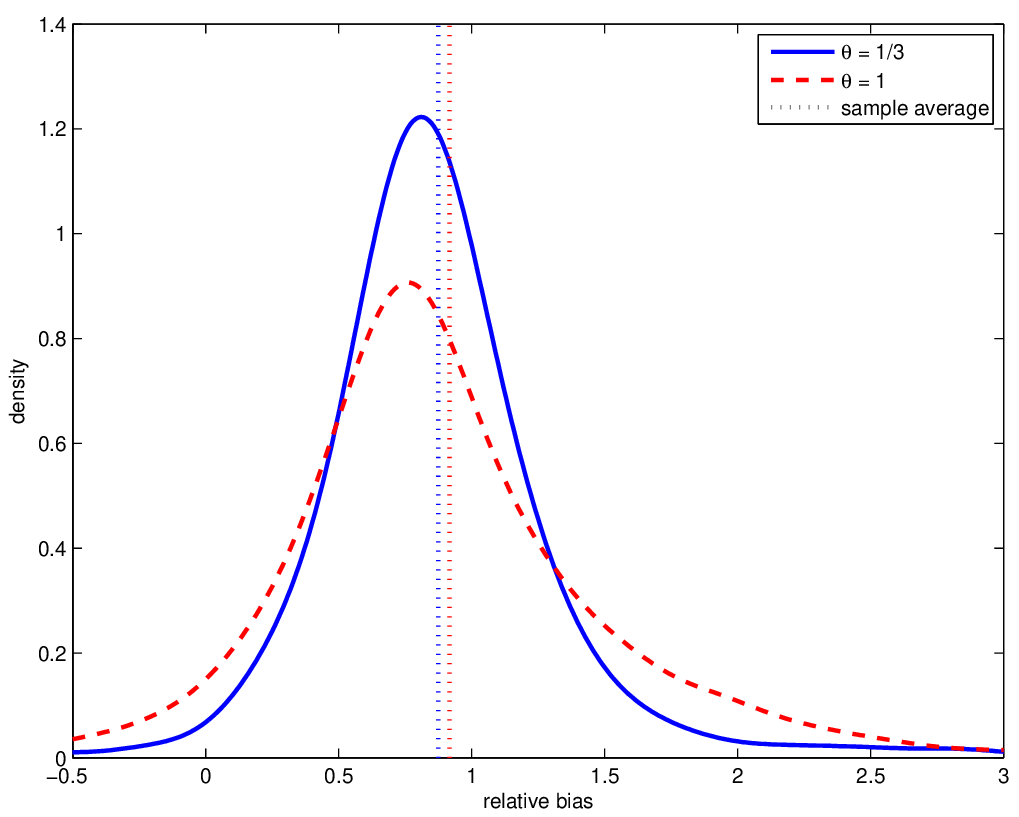} \\
\end{tabular}
\begin{scriptsize}
\parbox{0.925\textwidth}{\emph{Note.} In Panel A, we create an indicator variable I, which equals one if the $t$-statistic (based on $z_{ \text{wb2}}$) is significant at the 5\% nominal level, zero otherwise. We plot it against the true H-index from \eqref{equation:H-index} (small symbol). The curve is from a logistic regression between the two. Also shown are local averages of I (large symbol). In Panel B, we plot the distribution of $\text{$\widehat{\text{H}}$-index}$---defined in \eqref{equation:H-index-hat}---scaled by the H-index. Throughout, the setting is $n = 23,400$, $\xi^{2} = 0.001$ and $\varphi = -0.5$.}
\end{scriptsize}
\end{center}
\end{figure}

To shed light on these aspects, we compute:
\begin{equation}
\label{equation:H-index}
\text{H-index} = 1 - \displaystyle \frac{ \big( \int_{0}^{1} \sigma _{sv,s}^{2} \text{d}s \big)^{2}}{ \int_{0}^{1} \sigma_{sv,s}^{4} \text{d}s}.
\end{equation}
The H-index compares the square of integrated variance to the integrated quarticity of $X^{d}$. It has the intuitive interpretation that it describes how much $\sigma_{sv,t}$ deviates from $\mathcal{H}_{0}$ in percent, see, e.g., \citet*{podolskij-wasmuth:13a}.\footnote{The statistic has been used in earlier work to test for the parametric form of volatility \citep*[e.g.,][]{dette-podolskij-vetter:06a,vetter-dette:12a}. In contrast to our paper, the former operate with continuous $X$. Moreover, the ratio appears---sometimes in a different format---in other strands of the literature, for example asymptotic variance reduction \citep*[e.g.,][]{clinet-potiron:17a}, estimation of integrated variation \citep*[e.g.,][]{andersen-dobrev-schaumburg:14a,barndorff-nielsen-hansen-lunde-shephard:08a,xiu:10a}, or in the context of jump-testing \citep*[e.g.,][]{barndorff-nielsen-shephard:06a,kolokolov-reno:16a}.} We note that H-index $\in [0,1]$ by construction and that it equals zero if and only if $\sigma_{sv,t} = \sigma$ is constant. Strictly positive values imply $\sigma_{sv,t}$ is to some extent time-varying (not necessarily random, though). The H-index is therefore a natural measure of heteroskedasticity in our framework.\footnote{While it is possible to base the $t$-statistic on the H-index by transforming the CLT in \eqref{equation:clt} via the delta rule, we refrain from doing so due to the severe amount of time it takes to run the code.}

The two-factor stochastic volatility process used in this paper has an average H-index value of about 0.20 (based on a large number of paths drawn from the model). It falls below 0.10 20\% of the times, while it is rarely smaller than 0.05.

In Panel A of Figure \ref{figure:H-index}, we report the outcome of the $t$-statistic both for the set of experiments with deterministic and stochastic volatility. We define an indicator variable I, which takes the value one if $\mathcal{H}_{0}$ was rejected (on the basis of $z_{ \text{wb2}}$), and zero otherwise. The figure is a scatter plot of I versus the H-index. The fitted line originates from a logit regression of I on the H-index, which can be interpreted as the power of the test, conditional on H-index. As expected, the tendency to discard $\mathcal{H}_{0}$ is an increasing function of the H-index. When the deviation from the null is 0.15, the $t$-statistic is significant about 80\% of the time for $\theta = 1/3$, which falls down to 50\% for $\theta = 1$. Meanwhile, an H-index above 0.35 ($\theta = 1/3$) -- 0.45 ($\theta = 1$) implies it more or less always lies in the rejection region. It thus requires rather convincing evidence against the null to firmly reject it, more so for $\theta = 1$.

In practice, we estimate the H-index based on
\begin{equation}
\label{equation:H-index-hat}
\text{$\hat{\text{H}}$-index}=1-\displaystyle\frac{\widehat{\text{IV}}^{2}}{\widehat{\text{IQ}}},
\end{equation}
where
\begin{equation} \label{equation:IV-hat}
\widehat{\text{IV}} = \frac{1}{\theta \psi_{2}^{n}} \check{BV}(1,1)^{n} - \frac{ \psi_{1}^{n}}{ \theta^{2} \psi_{2}^{n}} \hat{ \rho}^{2}, \qquad \widehat{\text{IQ}} = \frac{1}{( \theta \psi_{2}^{n})^{2}} \check{BV}(2,2)^{n} - 2 \frac{ \psi_{1}^{n}}{ \theta^{2} \psi_{2}^{n}} \hat{ \rho}^{2} \widehat{\text{IV}} - \bigg( \frac{ \psi_{1}^{n}}{ \theta^{2} \psi_{2}^{n}} \bigg)^{2} \hat{ \rho}^{4},
\end{equation}
and
\begin{equation} \label{equation:rho-hat}
\hat{ \rho}^{2} = \hat{ \rho}(0) + 2 \sum_{k = 1}^{q} \hat{ \rho}(k)
\end{equation}
estimates the asymptotic bias in $\check{BV}(l,r)^{n}$ in the presence of $q$-dependent measurement error \citep*[e.g.,][Lemma 2]{hautsch-podolskij:13a},
\begin{equation}
\hat{ \rho}(k) = - \sum_{j = 1}^{q-k+1}j \hat{ \gamma}(k+j), \qquad \text{and} \qquad \hat{ \gamma}(k) = \frac{1}{n-k} \sum_{i=1}^{n-k} \Delta_{i}^{n}Y^{d} \Delta_{i+k}^{n}Y^{d},
\end{equation}
for $k = 0, \ldots, q+1$.

In Panel B of Figure \ref{figure:H-index}, we plot the distribution of the relative bias $\text{$\widehat{\text{H}}$-index} / \text{$\text{H}$-index}$ as a function of $\theta$ across simulations under the alternative. Note that an unbiased estimator has the distribution centered at one. As apparent, $\text{$\widehat{\text{H}}$-index}$ is slightly downward biased both for $\theta = 1/3$ and $\theta = 1$, which is mainly caused by a modest underestimation of $\text{IQ}$.\footnote{There is also an attenuation effect induced by the non-linear transformation of $\widehat{\text{IV}}$.} This leads to conservative statements about the true level of heteroskedasticity in the data, thereby reducing the rejection rate. The distribution is more dispersed and has a higher probability of being close to zero or even outright negative, when $\theta = 1$. This, in part, can help to explain why the simulated power is smaller for $\theta = 1$.

Overall, our noise- and jump-robust test of heteroskedasticity in diurnally-corrected diffusive volatility implemented via the bootstrap percentile-approach has good properties. In contrast to the CLT-based version of the test, it is almost unbiased, also for very small values of $n$, while it has decent---albeit not perfect---power under the presence of stochastic volatility.

\section{Empirical application} \label{section:empirical}

We apply our framework to a large cross-sectional panel of US equity high-frequency data. It includes the 30 stocks of the Dow Jones Industrial Average---following the update of its constituent list on March 18, 2015---and the SPDR S\&P 500 trust. The latter is an ETF with a price of about 1/10 the cash market value of the S\&P 500 index. The sample period is January 4, 2010 through December 31, 2013 for a total of $T = 1,006$ official exchange trading days. Table \ref{table:djia-descriptive} presents a list of ticker symbols along with a few summary statistics.\footnote{Notice that the variance of the noise, as captured by $\hat{ \xi}^{2}$, is generally smaller than what we assumed in the simulations. This is consistent with the notion that the noise has decreased over time.}

The data were extracted from the TAQ database and comprise a complete transaction record for each stock. They were cleaned with the algorithm developed by \citet*{christensen-oomen-podolskij:14a}, who build on earlier work of \citet*{brownless-gallo:06a} and \citet*{barndorff-nielsen-hansen-lunde-shephard:09a}. It is a standard way of preparing data for analysis in the high-frequency volatility literature.

\input{tables/djia-descriptive.tex}

\begin{figure}[t!]
\begin{center}
\caption{Properties of equity high-frequency data. \label{figure:djia-diurnal-variance.eps}}
\begin{tabular}{cc}
{\footnotesize {Panel A: ACF of $\Delta_{i}^{n} Y$.}} & {\footnotesize {Panel B: $\hat{\sigma}_{u,t}$.}} \\
\includegraphics[height=8cm,width=0.48\textwidth]{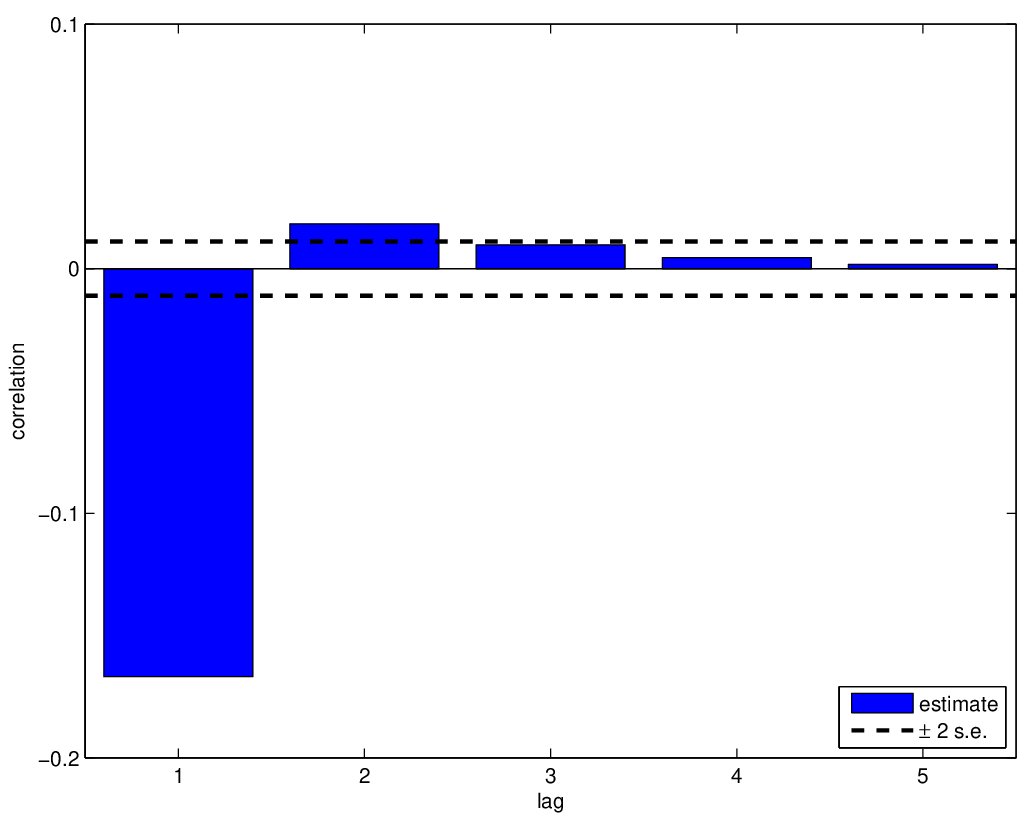} &
\includegraphics[height=8cm,width=0.48\textwidth]{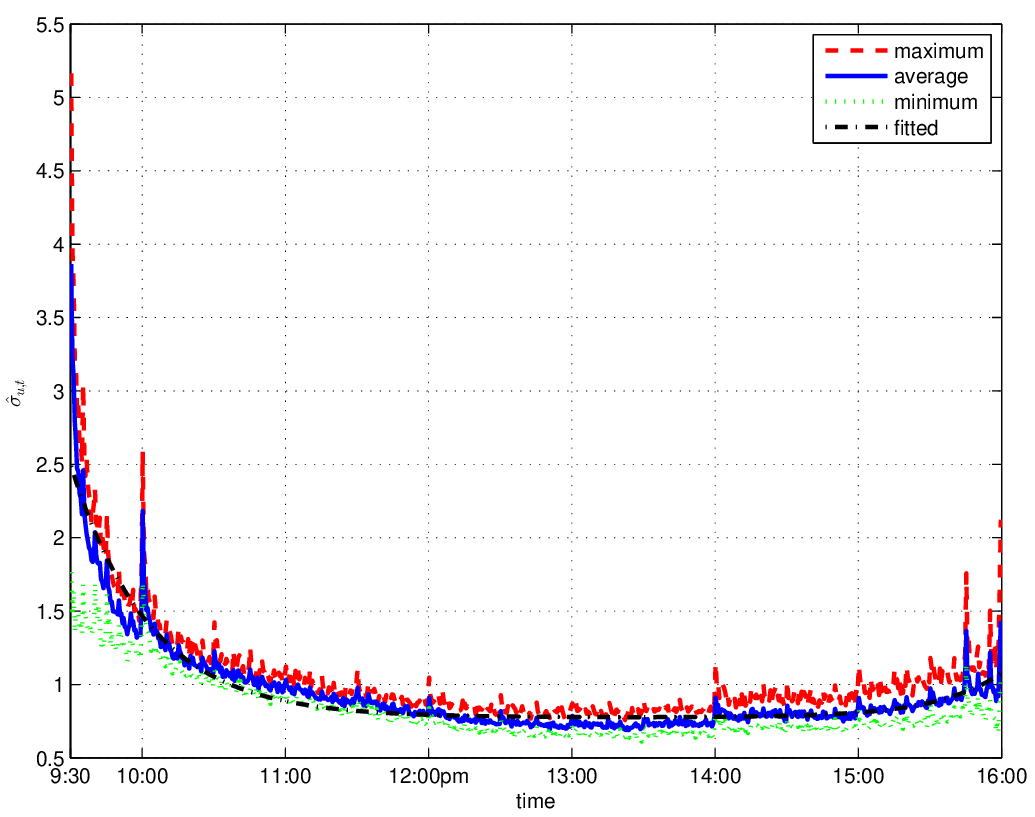}
\end{tabular}
\begin{scriptsize}
\parbox{0.95\textwidth}{\emph{Note.} In Panel A, we plot the ACF of our TAQ high-frequency equity data averaged across assets and over time. In Panel B, we present our estimator $\hat{ \sigma}_{u,t}$ of within-day volatility. The cross-sectional average is reported, along with the highest and smallest point estimate. As a comparison, we also fit the parametric form of diurnal variation in \eqref{equation:true-diurnal-variance} via non-linear least squares.}
\end{scriptsize}
\end{center}
\end{figure}

To compute $\hat{ \sigma}_{u,t}$, we create an equidistant log-price series for each asset pre-ticked to a 5-second resolution, i.e. $n = 4,680$. We then set $m = 780$---or $n/m = 6$---to retrieve a local estimate $\hat{ \sigma}_{u,t}$ that covers a 30-second interval.\footnote{To estimate $\sigma_{u,t}$, we further delete a few outliers from the sample. Firstly, the Flash Crash of May 6, 2010. Secondly, for each 30-second interval we remove the top 1\% of data for each stock---as measured by $|\Delta_{i}^{m} Y|$---to filter out observations typically associated with idiosyncratic news announcements. These events exert an unduly influence on the estimates due to our relatively small value of $T$. In a larger sample this should not be necessary.} It ensures that we recover a detailed view of the diurnality in volatility, while still being able to purge the associated noise with decent accuracy. On each block, we bias-correct with the robust estimator in \eqref{equation:robust-noise-variance} using $q = 3$.\footnote{$\omega^{2}$ is sometimes estimated to be negative, due to the sampling distribution of $\hat{ \omega}^{2}$. This cannot be true, of course, and we therefore truncate $\hat{ \omega}^{2}$ at zero throughout the paper. This happens more often if the noise is really small relative to the underlying volatility of the asset, as demonstrated by the index-tracker SPY in Figure \ref{figure:djia-bias.eps}.} This is motivated by Panel A in Figure \ref{figure:djia-diurnal-variance.eps}, which reports the autocorrelation function (ACF) of $\Delta_{i}^{n} Y$. As shown, while the first few autocorrelations are significant, the ACF dies out fast and is generally insignificant after lag three and negligible beyond lag five (not shown in figure), so this choice of $q$ suffices to capture the observed serial dependence in the noise.

The cross-sectional average of $\hat{ \sigma}_{u,t}$ is reported along with the minimum and maximum value in Panel B of Figure \ref{figure:djia-diurnal-variance.eps}. We note $\hat{ \sigma}_{u,t}$ features the reverted J-shape as reported in prior work, but also that it is very rough with sharp increases around pre-scheduled macroeconomic announcements (e.g., at 10:00am or 2:00pm). The latter is consistent with empirical findings in, e.g., \citet*{todorov-tauchen:11a}, who note that jumps in volatility are strongly correlated with large moves in market prices.\footnote{It also suggests that $\sigma_{u,t}$ may not be as smooth as stipulated by Assumption (D3). Note, however, that although volatility peaks at the announcement, it does not necessarily jump. In our data it actually starts to increase around 1-minute to 30-seconds \textit{prior} to the time, where the numbers are officially slated for release. This is consistent with the findings of \citet*{jiang-lo-verdelhan:11a} from the U. S. Treasury market.} There are also some notable spikes in within-day volatility prior to the close of the exchange. Overall, diurnal variation is remarkably constant across assets. To assess the parametric model used in the simulations, we estimate the parameters of \eqref{equation:true-diurnal-variance} via non-linear least squares based on the cross-sectional average. The fitted equation $\sigma_{u,t} = 0.78 + 1.71 e^{11.85t} + 0.30 e^{14.17(1-t)}$ is a good approximation to our nonparametric estimates, although it does not track the sharp initial decay in early trading.

\begin{figure}[t!]
\begin{center}
\caption{The distribution of the $\hat{\text{H}}$-index. \label{figure:djia-hi-distribution.eps}}
\begin{tabular}{cc}
{\footnotesize {Panel A: before correction.}} & {\footnotesize {Panel B: after correction.}} \\
\includegraphics[height=8cm,width=0.48\textwidth]{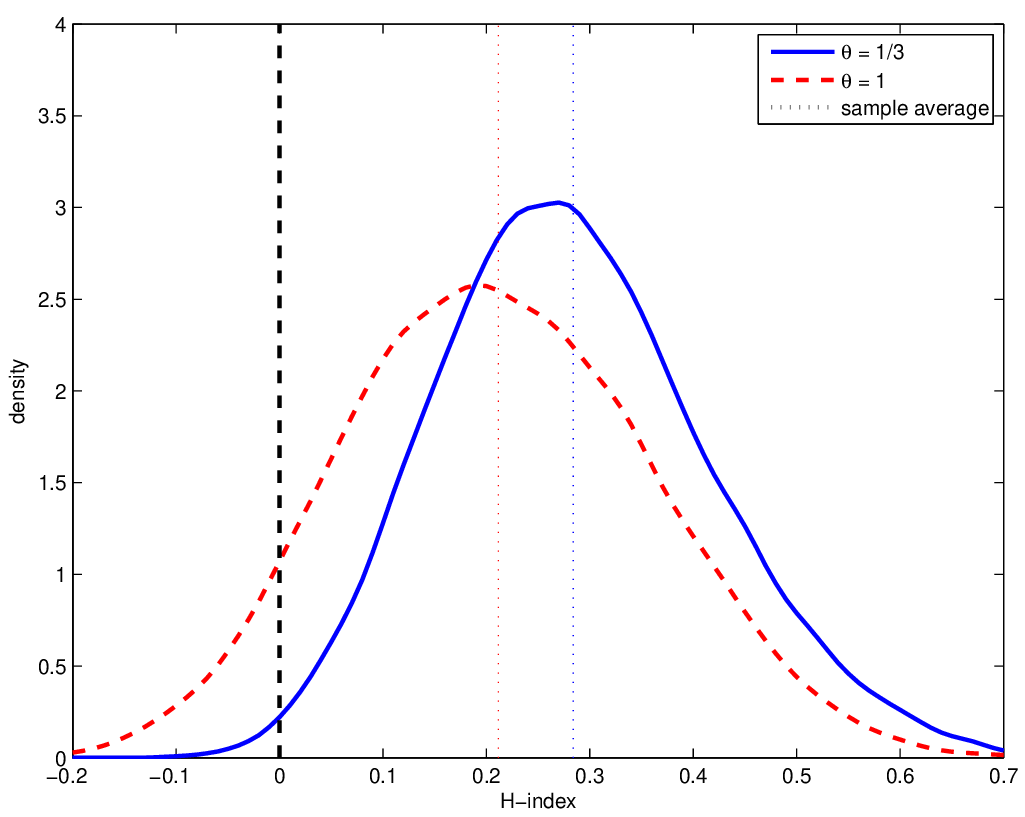} &
\includegraphics[height=8cm,width=0.48\textwidth]{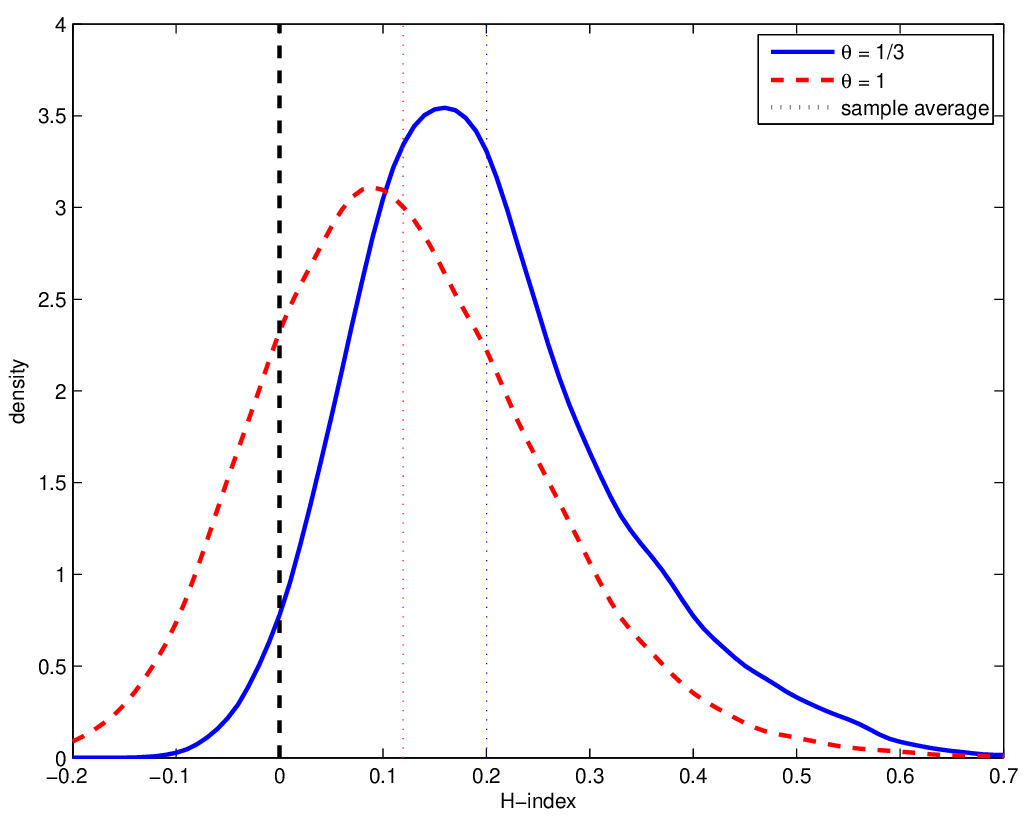}
\end{tabular}
\begin{scriptsize}
\parbox{0.95\textwidth}{\emph{Note.} We plot the cross-sectional distribution of the $\hat{\text{H}}$-index before and after diurnal correction with our nonparametric estimator from \eqref{equation:diurnal-variance-estimator}, also shown in Panel B of Figure \ref{figure:djia-diurnal-variance.eps}. The dashed vertical line at zero indicates the theoretical lower bound.}
\end{scriptsize}
\end{center}
\end{figure}

On this basis, we construct the deflated log-returns and compute the test for each stock and day in the sample. As above, we base the investigation on $\check{BV}(2,2)^{n}$ and $\check{BV}(1,1)^{n}$ with $\theta = 1/3$ and $\theta = 1$. The bootstrap percentile approach is applied to evaluate the significance of our $t$-statistic, i.e. $z_{ \text{wb2}}$. We apply a standard Bonferroni-type correction to account for multiple testing and control the family-wise error rate. That is, we run each individual test at significance level $\alpha / T$ with $\alpha = 0.05$.

In the right-hand side of Table \ref{table:djia-descriptive}, we report the average rejection rate of $\mathcal{H}_{0}$ and associated H-index measurement. As a comparison, we also retrieve the corresponding results from the raw data prior to diurnal correction. Looking at the table, we observe that $\mathcal{H}_{0}$ is discarded about half of the times for the typical stock if $\theta = 1/3$ and there is no seasonal adjustment. The levels are lower for $\theta = 1$, where about one-third of the tests are rejected. This is consistent with the decrease in power if $\theta$ is higher, as uncovered in the simulation section. On the other hand, there is also evidence of the caveat raised in Remark \ref{remark:noise-variance}, i.e. the pre-averaging estimator is affected harder by residual microstructure noise if $\theta$ is small, so that here the test is prone to discredit $\mathcal{H}_{0}$ in the presence of general forms of heteroscedasticity in the variance of the noise. Indeed, there is a tendency for stocks with very negative first-order return autocorrelation and large levels of noise---as measured by $\hat{ \xi}^{2}$--- to reject more frequently.

If we control for diurnal variation, the rejection rate drops to about 30\% (for $\theta = 1/3$) to almost 5\% (for $\theta = 1$). This implies the diurnal pattern is a first-order effect, which captures a large fraction of variation in intraday volatility. As readily seen, however, important and potent sources of heteroskedasticity remain present in the data.

\begin{figure}[t!]
\begin{center}
\caption{Empirical properties of $t$-statistic and $\hat{\text{H}}$-index. \label{figure:djia-logit.eps}}
\begin{tabular}{cc}
{\footnotesize {Panel A: before correction.}} & {\footnotesize {Panel B: after correction.}} \\
\includegraphics[height=8cm,width=0.48\textwidth]{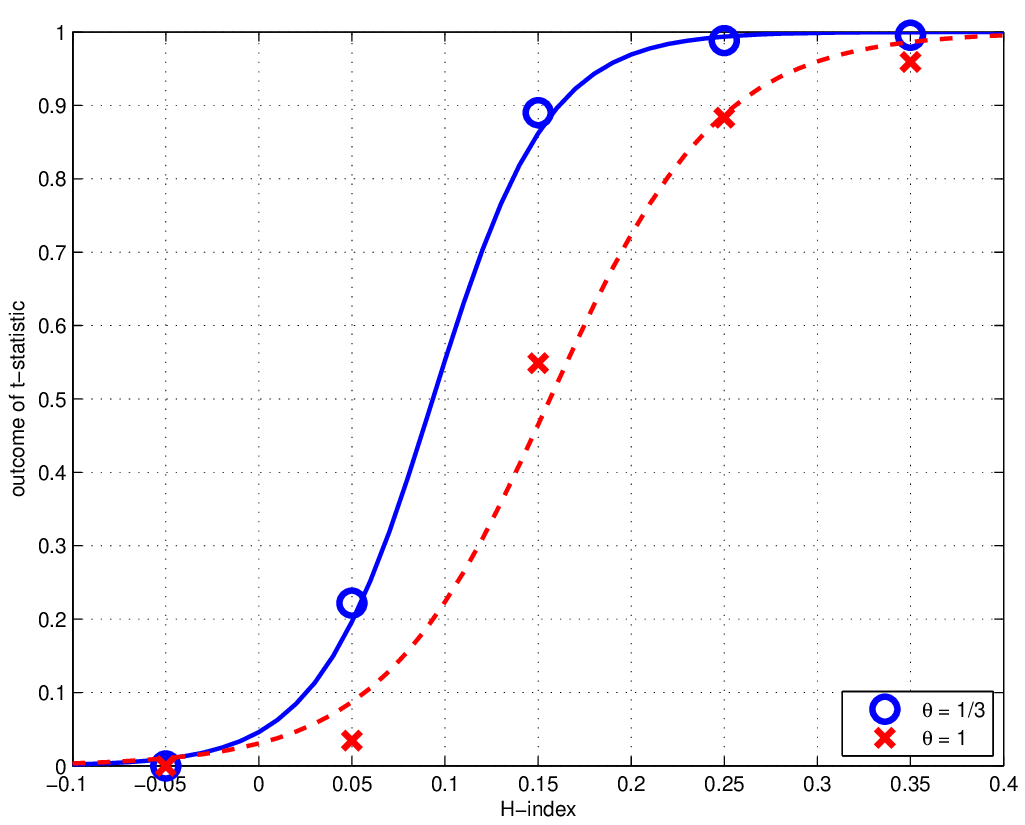} &
\includegraphics[height=8cm,width=0.48\textwidth]{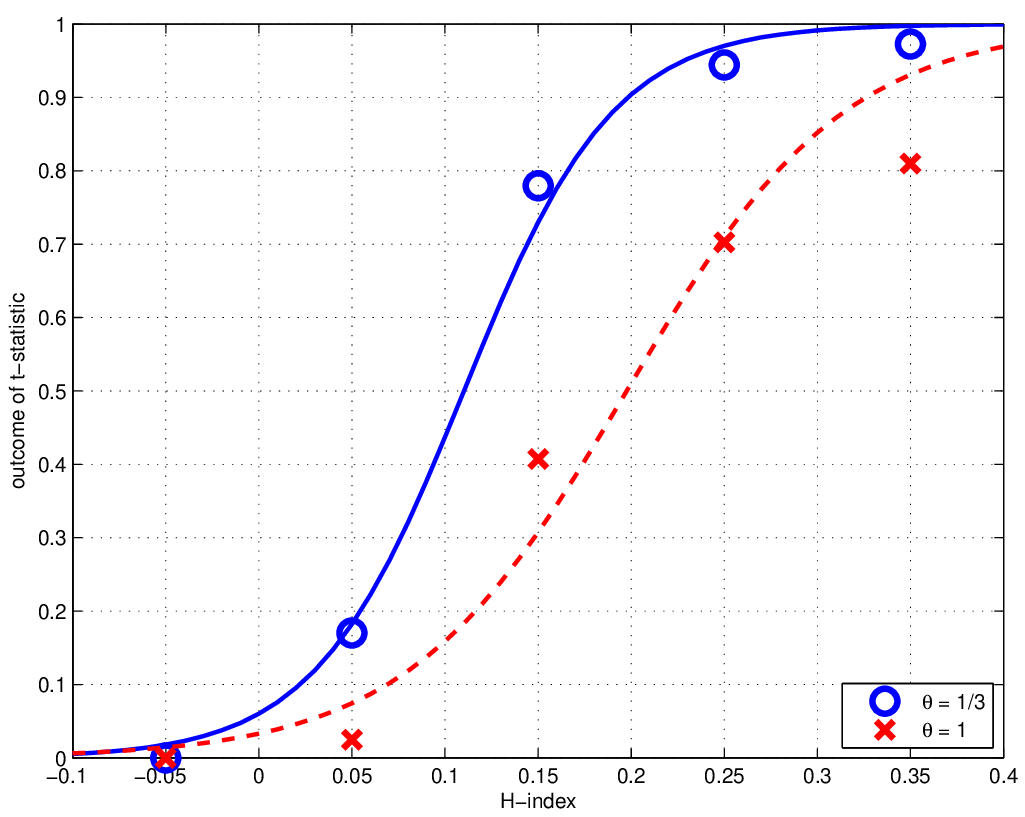}
\end{tabular}
\begin{scriptsize}
\parbox{0.925\textwidth}{\emph{Note.} We create an indicator variable I, which equals one if the $t$-statistic is significant at the 5\% nominal level, zero otherwise. We plot I against the $\hat{\text{H}}$-index before and after diurnal correction (i.e., based on $(\hat{\text{H}}\text{-index}, z_{ \text{wb2}})$ or $(\hat{\text{H}}\text{-index}^{d}, z_{ \text{wb2}}^{d})$) and as a function of $\theta$. The curve is from a logistic regression between the two. We also show local averages of I.}
\end{scriptsize}
\end{center}
\end{figure}

This is corroborated by the $\hat{\text{H}}$-index, for which we also plot the cross-sectional distribution before and after diurnal correction in Figure \ref{figure:djia-hi-distribution.eps}. The distribution shifts to the left and displays less sampling variation after diurnal correction, while we again notice a slight increase in the dispersion by moving $\theta$ up, although the effect is weak. As measured by the cross-sectional average shown in Panel B of the figure, the strength of residual heteroskedasticity present in $\Delta_{i}^{n} Y^{d}$ is broadly comparable to that of the two-factor stochastic volatility model from Section \ref{section:montecarlo} for $\theta = 1/3$, while it is somewhat less for $\theta = 1$.

At last, in Figure \ref{figure:djia-logit.eps} we model the empirical rejection rate of the $t$-statistic (with no Bonferroni correction here to be comparable with Figure \ref{figure:H-index} and improve the fit of the logistic regression) as a function of the $\hat{\text{H}}$-index. The logit fit is consistent with Section \ref{section:montecarlo} in that it takes a relatively high reading of $\hat{\text{H}}$-index to confidently reject $\mathcal{H}_{0}$. Note that the rejection rates are close to each other if $\hat{\text{H}}$-index $\simeq$ $\hat{\text{H}}$-index$^{d}$, as can be gauged by comparing Panel A and B. The intuitive explanation is that the power of the test depends only on the level of heteroskedasticity in the data, which is captured by the $\text{H}$-index, and not as such on whether one has diurnally-corrected it or not.

\section{Conclusion} \label{section:conclusion}

In this paper, we study a new approach to determine if changes in intraday spot volatility of a discretely sampled noisy jump-diffusion model can be attributed solely to a deterministic cyclical component (i.e., the so-called U- or reverse J-shape) against an alternative of further variation induced by a stochastic process.

We propose to construct a test of this hypothesis from an asset return series, which has been deflated by the diurnal component and, as such, is homoskedastic under the null. The $t$-statistic diverges to infinity, if the deflated return series is heteroskedastic, and it has a standard normal distribution otherwise. To get a feasible test, we develop a---surprisingly robust---nonparametric estimator of unobserved diurnal volatility, which (in contrast to the test itself) is computed directly from noisy high-frequency data without pre-averaging or jump-truncation. It requires only a trivial bias-correction to eliminate the noise variance. Our estimator is consistent and has a sampling error, which is of small enough order that replacing the true diurnal factor with it does not alter the asymptotic theory.

We inspect the properties of the test in a Monte Carlo simulation. We note the theory-based version has gross size distortions in the presence of infinite-activity price jumps, thus motivating a bootstrap. We validate the bootstrap and confirm it helps to improve inference by making the test almost correctly sized. The test also has acceptable power, but can fail to reject the null even in large samples, if a wide pre-averaging window is applied. The estimation of the diurnal factor has a limited impact, but it raises the rejection rate slightly.

We implement our nonparametric estimator of diurnal variance and test of heteroskedasticity on real high-frequency data. The diurnal pattern explains a sizable portion of within-day variation in the volatility in practice, as inferred by the notable drop in the rejection rate of the test and the reduction in the H-index---a descriptive statistic that measures the strength of time-varying volatility---once we control for intraday seasonality. This suggests that the rescaled log-returns are often close to homoskedastic. Still, important sources of variation remain present in the data. So the answer to the title of the paper appears to be ``no.'' The diurnal pattern does not explain \textit{all} intraday variation in volatility, but it does capture a rather significant portion of it.

\clearpage

\renewcommand{\baselinestretch}{1.0}
\small
\bibliographystyle{rfs}
\bibliography{userref}

\end{document}

%% file: tables/sim-theta=0.333-beta=-0.5-f.tex
\begin{sidewaystable}[p!]
\setlength{ \tabcolsep}{0.09cm}
\begin{center}
\caption{Rejection rate at 5\% level of significance with $\theta = 0.333$ and $\varphi = -0.5$.
\label{table:sim-theta=0.333-beta=-0.5-f.tex}}
\vspace*{-0.25cm}
\begin{small}
\begin{tabular}{lrrrrrrrrrrrrrrrrrrrrrrrrrrrrrrrrrrrrrrrrrrrrrrrrrrrrrrrrrrrr}
\hline \hline
& & & \multicolumn{15}{c}{$\mathcal{H}_{0}$ : deterministic volatility} & & \multicolumn{15}{c}{$\mathcal{H}_{a}$ : stochastic volatility}\\ \cline{4-18} \cline{20-34}
& & & \multicolumn{7}{c}{size} & & \multicolumn{7}{c}{avg. block length} & & \multicolumn{7}{c}{power} & & \multicolumn{7}{c}{avg. block length}\\ \cline{4-10} \cline{12-18} \cline{20-26} \cline{28-34}
& & & CLT & &  $z_{ \text{wb1}}$ & $z_{ \text{wb2}}$ && $t_{ \text{wb1}}$ & $t_{ \text{wb2}}$
  & & CLT & &  $z_{ \text{wb1}}$ & $z_{ \text{wb2}}$ && $t_{ \text{wb1}}$ & $t_{ \text{wb2}}$
  & & CLT & &  $z_{ \text{wb1}}$ & $z_{ \text{wb2}}$ && $t_{ \text{wb1}}$ & $t_{ \text{wb2}}$
  & & CLT & &  $z_{ \text{wb1}}$ & $z_{ \text{wb2}}$ && $t_{ \text{wb1}}$ & $t_{ \text{wb2}}$\\
\multicolumn{10}{l}{Panel A: $\xi^{2} = 0.0001$}\\
$n = $& 390 && 25.3 && 6.9 & 7.3 && 10.1 & 10.1&& 136&& 130&  131&& 130&  131 && 62.0 && 34.9 & 35.7 && 42.1 & 41.1&& 136&& 131&  131&& 132&  131\\
& 780 && 18.5 && 5.6 & 5.5 && 8.0 & 8.3&& 215&& 210&  211&& 212&  211 && 64.2 && 43.2 & 41.9 && 46.2 & 47.0&& 215&& 210&  211&& 210&  208\\
& 1560 && 13.5 && 5.1 & 4.9 && 6.4 & 6.3&& 342&& 335&  335&& 334&  334 && 66.7 && 50.5 & 50.1 && 52.5 & 52.4&& 341&& 335&  334&& 335&  335\\
& 4680 && 12.0 && 5.7 & 5.9 && 6.5 & 7.2&& 712&& 698&  701&& 700&  700 && 71.5 && 61.6 & 61.9 && 62.3 & 62.1&& 707&& 700&  698&& 699&  700\\
& 7800 && 11.7 && 7.5 & 7.4 && 9.1 & 9.2&& 999&& 978&  984&& 984&  981 && 76.4 && 68.3 & 68.2 && 68.4 & 68.6&& 995&& 979&  978&& 978&  982\\
& 11700 && 11.2 && 6.4 & 6.7 && 8.4 & 8.2&& 1,297&& 1,284&  1,283&& 1,283&  1,295 && 79.5 && 73.3 & 72.8 && 73.5 & 73.2&& 1,308&& 1,291&  1,293&& 1,288&  1,284\\
& 23400 && 10.7 && 7.1 & 6.9 && 8.0 & 7.8&& 2,061&& 2,052&  2,049&& 2,047&  2,051 && 85.7 && 81.2 & 82.1 && 82.1 & 82.3&& 2,062&& 2,043&  2,042&& 2,042&  2,049\\
\\
\multicolumn{10}{l}{Panel B: $\xi^{2} = 0.0010$}\\
$n = $& 390 && 24.6 && 7.0 & 6.6 && 10.7 & 11.1&& 136&& 131&  131&& 132&  131 && 61.6 && 33.3 & 33.1 && 39.2 & 40.2&& 135&& 131&  131&& 131&  131\\
& 780 && 19.0 && 5.4 & 5.7 && 8.0 & 8.9&& 215&& 211&  211&& 211&  210 && 63.7 && 41.3 & 39.9 && 45.6 & 44.9&& 216&& 211&  212&& 211&  210\\
& 1560 && 14.6 && 4.8 & 4.5 && 6.3 & 6.2&& 343&& 336&  336&& 335&  335 && 64.6 && 48.5 & 48.1 && 51.7 & 51.6&& 341&& 336&  337&& 334&  336\\
& 4680 && 12.3 && 5.8 & 5.9 && 7.1 & 7.4&& 705&& 699&  700&& 702&  696 && 70.6 && 60.9 & 60.2 && 60.5 & 60.3&& 708&& 702&  702&& 699&  702\\
& 7800 && 11.8 && 6.6 & 6.8 && 8.3 & 8.1&& 998&& 982&  985&& 983&  981 && 75.0 && 66.6 & 66.6 && 66.9 & 66.2&& 998&& 984&  983&& 978&  973\\
& 11700 && 11.1 && 6.4 & 6.8 && 8.3 & 7.8&& 1,311&& 1,285&  1,292&& 1,287&  1,289 && 79.5 && 72.8 & 72.8 && 71.8 & 72.5&& 1,306&& 1,283&  1,293&& 1,285&  1,295\\
& 23400 && 10.1 && 6.3 & 6.5 && 7.8 & 8.4&& 2,057&& 2,039&  2,041&& 2,049&  2,038 && 84.5 && 81.0 & 80.6 && 80.5 & 81.2&& 2,056&& 2,040&  2,054&& 2,037&  2,044\\
\\
\multicolumn{10}{l}{Panel C: $\xi^{2} = 0.0100$}\\
$n = $& 390 && 24.7 && 6.6 & 6.7 && 10.1 & 10.6&& 136&& 131&  130&& 131&  131 && 49.0 && 20.6 & 21.1 && 26.3 & 26.7&& 135&& 131&  131&& 131&  131\\
& 780 && 18.0 && 3.6 & 4.0 && 5.8 & 6.0&& 216&& 211&  210&& 211&  211 && 52.3 && 26.3 & 25.1 && 31.3 & 31.6&& 216&& 210&  211&& 209&  212\\
& 1560 && 15.0 && 5.1 & 5.1 && 7.7 & 7.8&& 343&& 335&  334&& 334&  335 && 50.2 && 31.3 & 30.7 && 35.2 & 35.0&& 341&& 339&  337&& 337&  336\\
& 4680 && 12.3 && 6.2 & 6.4 && 7.8 & 8.0&& 706&& 702&  699&& 699&  697 && 61.9 && 48.4 & 48.1 && 49.7 & 49.6&& 707&& 702&  699&& 703&  700\\
& 7800 && 10.6 && 5.4 & 4.8 && 6.5 & 6.4&& 999&& 978&  989&& 979&  985 && 65.9 && 57.7 & 57.0 && 58.0 & 58.1&& 997&& 982&  981&& 984&  980\\
& 11700 && 9.4 && 6.1 & 6.3 && 7.0 & 6.9&& 1,305&& 1,287&  1,291&& 1,285&  1,293 && 69.8 && 62.4 & 61.4 && 60.7 & 61.0&& 1,301&& 1,293&  1,291&& 1,279&  1,286\\
& 23400 && 9.9 && 6.4 & 6.3 && 8.4 & 8.1&& 2,050&& 2,033&  2,051&& 2,026&  2,041 && 75.5 && 69.4 & 69.1 && 69.9 & 70.3&& 2,058&& 2,032&  2,036&& 2,038&  2,041\\
\hline \hline
\end{tabular}
\end{small}\smallskip
\begin{scriptsize}
\parbox{0.98\textwidth}{\emph{Note.} 
We simulate from a model with drift, volatility, infinite-activity jumps and microstructure noise. We test the hypothesis that $\sigma_{t} = \sigma \sigma_{u,t}$ is a deterministic function of time (induced by diurnal variation) and report rejection rates both under $\mathcal{H}_{0}$ (size) and $\mathcal{H}_{a}$ (power). In the latter, $\sigma_{t}  = \sigma_{sv,t} \sigma_{u,t}$ is also time-varying due to a two-factor SV structure. $\theta$ is a tuning parameter that is used to compute the pre-averaging window $k_{n} = [\theta \sqrt{n}]$, $\varphi$ is the MA(1) coefficient in the noise process, $n$ is the sample size, and $\xi^{2}$ controls the magnitude of noise relative to volatility. CLT is for the asymptotic theory from \eqref{equation:clt}, while $z_{\text{wb}\cdot}$ and $t_{\text{wb}\cdot}$ are rejection rates based on the percentile and percentile-$t$ bootstrap test for two choices of the external random variable $u$. We made 1,000 Monte Carlo trials with 999 bootstrap replica in each simulation. Further details can be found in Section \ref{section:montecarlo}.} 
\end{scriptsize}
\end{center}
\end{sidewaystable}

%% file: tables/sim-theta=1.000-beta=-0.5-f.tex
\begin{sidewaystable}[p!]
\setlength{ \tabcolsep}{0.09cm}
\begin{center}
\caption{Rejection rate at 5\% level of significance with $\theta = 1.000$ and $\varphi = -0.5$.
\label{table:sim-theta=1.000-beta=-0.5-f.tex}}
\vspace*{-0.25cm}
\begin{small}
\begin{tabular}{lrrrrrrrrrrrrrrrrrrrrrrrrrrrrrrrrrrrrrrrrrrrrrrrrrrrrrrrrrrrr}
\hline \hline
& & & \multicolumn{15}{c}{$\mathcal{H}_{0}$ : deterministic volatility} & & \multicolumn{15}{c}{$\mathcal{H}_{a}$ : stochastic volatility}\\ \cline{4-18} \cline{20-34}
& & & \multicolumn{7}{c}{size} & & \multicolumn{7}{c}{avg. block length} & & \multicolumn{7}{c}{power} & & \multicolumn{7}{c}{avg. block length}\\ \cline{4-10} \cline{12-18} \cline{20-26} \cline{28-34}
& & & CLT & &  $z_{ \text{wb1}}$ & $z_{ \text{wb2}}$ && $t_{ \text{wb1}}$ & $t_{ \text{wb2}}$
  & & CLT & &  $z_{ \text{wb1}}$ & $z_{ \text{wb2}}$ && $t_{ \text{wb1}}$ & $t_{ \text{wb2}}$
  & & CLT & &  $z_{ \text{wb1}}$ & $z_{ \text{wb2}}$ && $t_{ \text{wb1}}$ & $t_{ \text{wb2}}$
  & & CLT & &  $z_{ \text{wb1}}$ & $z_{ \text{wb2}}$ && $t_{ \text{wb1}}$ & $t_{ \text{wb2}}$\\
\multicolumn{10}{l}{Panel A: $\xi^{2} = 0.0001$}\\
$n = $& 390 && 20.5 && 8.9 & 8.6 && 11.6 & 11.9&& 135&& 131&  130&& 131&  130 && 49.1 && 24.0 & 23.7 && 28.7 & 28.4&& 136&& 131&  131&& 132&  130\\
& 780 && 14.4 && 5.2 & 5.0 && 7.4 & 7.5&& 215&& 211&  210&& 210&  209 && 42.8 && 24.9 & 25.1 && 27.6 & 27.3&& 216&& 210&  209&& 209&  211\\
& 1560 && 14.2 && 6.4 & 6.5 && 7.9 & 8.1&& 341&& 335&  337&& 335&  334 && 48.4 && 35.3 & 34.8 && 37.0 & 36.9&& 342&& 334&  335&& 336&  334\\
& 4680 && 11.7 && 6.9 & 6.1 && 8.3 & 8.1&& 710&& 703&  698&& 700&  699 && 54.1 && 44.3 & 44.2 && 44.5 & 44.4&& 710&& 698&  700&& 694&  695\\
& 7800 && 10.9 && 6.5 & 5.8 && 7.0 & 7.4&& 996&& 983&  985&& 982&  975 && 60.3 && 51.4 & 50.6 && 51.1 & 51.8&& 990&& 981&  979&& 981&  976\\
& 11700 && 10.4 && 6.2 & 7.0 && 7.7 & 7.9&& 1,292&& 1,295&  1,296&& 1,294&  1,291 && 64.2 && 55.6 & 56.1 && 56.2 & 55.7&& 1,302&& 1,284&  1,281&& 1,285&  1,292\\
& 23400 && 10.5 && 7.8 & 8.0 && 8.0 & 7.8&& 2,066&& 2,053&  2,045&& 2,064&  2,038 && 69.0 && 63.6 & 63.9 && 63.0 & 63.0&& 2,066&& 2,046&  2,047&& 2,049&  2,036\\
\\
\multicolumn{10}{l}{Panel B: $\xi^{2} = 0.0010$}\\
$n = $& 390 && 21.0 && 9.6 & 8.5 && 11.3 & 11.3&& 135&& 130&  130&& 131&  130 && 49.9 && 24.5 & 24.3 && 28.3 & 29.1&& 135&& 131&  130&& 131&  130\\
& 780 && 14.4 && 5.1 & 5.3 && 7.4 & 8.1&& 215&& 209&  210&& 209&  211 && 44.9 && 25.6 & 25.3 && 28.8 & 29.2&& 215&& 211&  211&& 211&  210\\
& 1560 && 14.1 && 6.4 & 6.3 && 8.3 & 8.7&& 339&& 335&  337&& 336&  334 && 48.1 && 34.6 & 34.6 && 37.3 & 36.9&& 341&& 335&  333&& 335&  334\\
& 4680 && 11.5 && 7.2 & 6.1 && 8.1 & 8.0&& 708&& 700&  700&& 696&  699 && 53.2 && 44.2 & 44.2 && 44.4 & 44.7&& 706&& 700&  699&& 700&  702\\
& 7800 && 10.2 && 5.6 & 5.7 && 6.9 & 7.3&& 999&& 984&  987&& 983&  985 && 60.2 && 50.5 & 49.9 && 49.6 & 50.0&& 996&& 979&  991&& 975&  983\\
& 11700 && 10.7 && 5.8 & 6.3 && 7.2 & 7.6&& 1,295&& 1,288&  1,291&& 1,287&  1,285 && 63.2 && 55.8 & 54.9 && 55.2 & 54.6&& 1,302&& 1,293&  1,291&& 1,285&  1,291\\
& 23400 && 9.4 && 7.1 & 7.6 && 7.8 & 7.8&& 2,056&& 2,044&  2,057&& 2,048&  2,047 && 68.6 && 63.6 & 63.4 && 62.6 & 62.6&& 2,074&& 2,056&  2,052&& 2,056&  2,047\\
\\
\multicolumn{10}{l}{Panel C: $\xi^{2} = 0.0100$}\\
$n = $& 390 && 20.6 && 8.5 & 8.2 && 12.2 & 12.2&& 135&& 130&  130&& 131&  130 && 49.2 && 25.4 & 25.3 && 29.3 & 29.3&& 136&& 131&  130&& 131&  130\\
& 780 && 14.4 && 6.0 & 5.8 && 7.9 & 8.4&& 215&& 210&  211&& 210&  211 && 41.7 && 24.2 & 24.6 && 27.8 & 27.3&& 216&& 210&  210&& 210&  211\\
& 1560 && 14.3 && 6.8 & 7.0 && 8.8 & 8.5&& 342&& 335&  337&& 334&  334 && 47.4 && 33.2 & 33.2 && 34.4 & 34.5&& 340&& 336&  334&& 335&  333\\
& 4680 && 10.0 && 6.3 & 6.0 && 7.4 & 7.7&& 708&& 698&  700&& 699&  695 && 53.1 && 42.6 & 41.7 && 43.8 & 43.8&& 708&& 698&  702&& 693&  701\\
& 7800 && 9.9 && 6.2 & 6.0 && 7.2 & 7.2&& 996&& 983&  984&& 986&  983 && 57.1 && 47.3 & 45.9 && 47.0 & 46.9&& 994&& 976&  980&& 977&  982\\
& 11700 && 10.1 && 5.8 & 6.3 && 7.2 & 7.2&& 1,296&& 1,287&  1,286&& 1,284&  1,288 && 61.1 && 52.7 & 52.5 && 52.8 & 52.8&& 1,297&& 1,284&  1,287&& 1,287&  1,279\\
& 23400 && 9.9 && 7.0 & 7.3 && 8.3 & 8.3&& 2,061&& 2,040&  2,045&& 2,047&  2,039 && 67.1 && 60.2 & 61.5 && 60.5 & 60.6&& 2,069&& 2,039&  2,042&& 2,050&  2,058\\
\hline \hline
\end{tabular}
\end{small}\smallskip
\begin{scriptsize}
\parbox{0.98\textwidth}{\emph{Note.} 
We simulate from a model with drift, volatility, infinite-activity jumps and microstructure noise. We test the hypothesis that $\sigma_{t} = \sigma \sigma_{u,t}$ is a deterministic function of time (induced by diurnal variation) and report rejection rates both under $\mathcal{H}_{0}$ (size) and $\mathcal{H}_{a}$ (power). In the latter, $\sigma_{t}  = \sigma_{sv,t} \sigma_{u,t}$ is also time-varying due to a two-factor SV structure. $\theta$ is a tuning parameter that is used to compute the pre-averaging window $k_{n} = [\theta \sqrt{n}]$, $\varphi$ is the MA(1) coefficient in the noise process, $n$ is the sample size, and $\xi^{2}$ controls the magnitude of noise relative to volatility. CLT is for the asymptotic theory from \eqref{equation:clt}, while $z_{\text{wb}\cdot}$ and $t_{\text{wb}\cdot}$ are rejection rates based on the percentile and percentile-$t$ bootstrap test for two choices of the external random variable $u$. We made 1,000 Monte Carlo trials with 999 bootstrap replica in each simulation. Further details can be found in Section \ref{section:montecarlo}.} 
\end{scriptsize}
\end{center}
\end{sidewaystable}

%% file: tables/djia-descriptive.tex
\begin{sidewaystable}
\setlength{ \tabcolsep}{0.18cm}
\begin{center}
\caption{Descriptive statistics of TAQ high-frequency data.
\label{table:djia-descriptive}}
\vspace*{-0.25cm}
\begin{small}
\begin{tabular}{lcrcccccrcccccccrcccccc}
\hline \hline
& & & & & & & & \multicolumn{7}{c}{$\theta = 1/3$} & &  \multicolumn{7}{c}{$\theta = 1$}\\
\cline{9-15} \cline{17-23}
& & & & & & & & & & \multicolumn{2}{c}{$\#\mathcal{Z}^{n} > p_{1-\alpha/T}^{*}$} & & & & & & & \multicolumn{2}{c}{$\#\mathcal{Z}^{n} > p_{1-\alpha/T}^{*}$} \\
\cline{11-12} \cline{19-20}
ticker & & \multicolumn{1}{c}{$n$} & & $\hat{ \sigma}$ & \multicolumn{1}{c}{$\hat{ \rho}_{1}$} & $\hat{ \xi}^{2} \times 10^{4}$ & & \multicolumn{1}{c}{$k_{n}$} & & $z_{ \text{wb2}}$ & $z_{\text{wb2}}^{d}$ & & $\hat{\text{H}}$-index & $\hat{\text{H}}$-index$^{d}$ & & \multicolumn{1}{c}{$k_{n}$} & & $z_{ \text{wb2}}$ & $z_{\text{wb2}}^{d}$ & & $\hat{\text{H}}$-index & $\hat{\text{H}}$-index$^{d}$\\
\hline
AAPL & & 14,576 & & 18.9 & -0.12 & 0.12 & & 40 & & 0.50 & 0.11 & & 0.23 & 0.12 & & 120 & & 0.17 & 0.01 & & 0.18 & 0.06\\
AXP & & 6,843 & & 19.5 & -0.07 & 0.29 & & 27 & & 0.39 & 0.12 & & 0.23 & 0.15 & & 82 & & 0.12 & 0.02 & & 0.17 & 0.09\\
BA & & 6,486 & & 18.7 & -0.10 & 0.39 & & 27 & & 0.44 & 0.15 & & 0.27 & 0.17 & & 80 & & 0.19 & 0.03 & & 0.23 & 0.12\\
CAT & & 8,510 & & 21.7 & -0.09 & 0.18 & & 30 & & 0.34 & 0.07 & & 0.22 & 0.11 & & 91 & & 0.15 & 0.00 & & 0.18 & 0.05\\
CSCO & & 11,884 & & 16.5 & -0.38 & 1.74 & & 36 & & 0.88 & 0.76 & & 0.41 & 0.35 & & 108 & & 0.42 & 0.22 & & 0.29 & 0.21\\
CVX & & 8,568 & & 16.1 & -0.05 & 0.29 & & 31 & & 0.31 & 0.07 & & 0.20 & 0.12 & & 92 & & 0.10 & 0.01 & & 0.15 & 0.07\\
DD & & 6,521 & & 18.4 & -0.11 & 0.44 & & 27 & & 0.41 & 0.14 & & 0.24 & 0.15 & & 80 & & 0.14 & 0.02 & & 0.18 & 0.09\\
DIS & & 7,422 & & 17.1 & -0.14 & 0.34 & & 29 & & 0.52 & 0.24 & & 0.26 & 0.18 & & 86 & & 0.14 & 0.02 & & 0.18 & 0.09\\
GE & & 12,990 & & 16.5 & -0.41 & 2.02 & & 38 & & 0.89 & 0.82 & & 0.42 & 0.36 & & 114 & & 0.44 & 0.25 & & 0.30 & 0.22\\
GS & & 7,940 & & 22.2 & -0.10 & 0.22 & & 29 & & 0.42 & 0.09 & & 0.24 & 0.14 & & 88 & & 0.16 & 0.01 & & 0.19 & 0.08\\
HD & & 7,798 & & 17.0 & -0.16 & 0.27 & & 29 & & 0.54 & 0.27 & & 0.28 & 0.19 & & 88 & & 0.17 & 0.03 & & 0.21 & 0.11\\
IBM & & 7,193 & & 13.9 & -0.14 & 0.38 & & 28 & & 0.37 & 0.11 & & 0.23 & 0.15 & & 84 & & 0.13 & 0.01 & & 0.18 & 0.10\\
INTC & & 12,515 & & 17.3 & -0.37 & 1.24 & & 37 & & 0.87 & 0.74 & & 0.39 & 0.32 & & 112 & & 0.37 & 0.14 & & 0.29 & 0.18\\
JNJ & & 8,766 & & 11.3 & -0.18 & 0.32 & & 31 & & 0.59 & 0.37 & & 0.28 & 0.21 & & 93 & & 0.17 & 0.03 & & 0.20 & 0.12\\
JPM & & 12,141 & & 21.9 & -0.14 & 0.11 & & 37 & & 0.58 & 0.22 & & 0.26 & 0.16 & & 110 & & 0.18 & 0.02 & & 0.19 & 0.08\\
KO & & 7,991 & & 12.2 & -0.21 & 0.46 & & 30 & & 0.61 & 0.36 & & 0.30 & 0.22 & & 89 & & 0.21 & 0.04 & & 0.21 & 0.12\\
MCD & & 7,192 & & 12.2 & -0.13 & 0.28 & & 28 & & 0.43 & 0.16 & & 0.26 & 0.16 & & 84 & & 0.17 & 0.03 & & 0.20 & 0.11\\
MMM & & 5,466 & & 14.9 & -0.08 & 0.64 & & 24 & & 0.37 & 0.10 & & 0.24 & 0.15 & & 73 & & 0.15 & 0.02 & & 0.19 & 0.10\\
MRK & & 8,455 & & 14.8 & -0.25 & 0.49 & & 31 & & 0.69 & 0.48 & & 0.33 & 0.26 & & 92 & & 0.24 & 0.07 & & 0.23 & 0.14\\
MSFT & & 12,814 & & 16.2 & -0.34 & 0.72 & & 38 & & 0.87 & 0.74 & & 0.37 & 0.31 & & 113 & & 0.38 & 0.13 & & 0.27 & 0.18\\
NKE & & 4,847 & & 18.0 & -0.07 & 0.35 & & 23 & & 0.41 & 0.10 & & 0.27 & 0.16 & & 69 & & 0.18 & 0.01 & & 0.22 & 0.11\\
PFE & & 11,253 & & 14.6 & -0.38 & 2.12 & & 35 & & 0.90 & 0.81 & & 0.43 & 0.38 & & 106 & & 0.48 & 0.25 & & 0.31 & 0.23\\
PG & & 7,967 & & 11.9 & -0.16 & 0.28 & & 30 & & 0.53 & 0.30 & & 0.27 & 0.20 & & 89 & & 0.15 & 0.03 & & 0.19 & 0.12\\
TRV & & 4,779 & & 14.9 & -0.11 & 0.88 & & 23 & & 0.42 & 0.16 & & 0.26 & 0.17 & & 68 & & 0.17 & 0.03 & & 0.19 & 0.12\\
UNH & & 6,730 & & 20.6 & -0.09 & 0.28 & & 27 & & 0.59 & 0.26 & & 0.31 & 0.19 & & 81 & & 0.22 & 0.04 & & 0.25 & 0.12\\
UTX & & 5,938 & & 16.1 & -0.09 & 0.47 & & 25 & & 0.42 & 0.13 & & 0.26 & 0.16 & & 76 & & 0.15 & 0.02 & & 0.21 & 0.11\\
V & & 6,213 & & 19.3 & -0.12 & 0.53 & & 26 & & 0.48 & 0.18 & & 0.28 & 0.19 & & 78 & & 0.19 & 0.03 & & 0.24 & 0.14\\
VZ & & 9,191 & & 13.4 & -0.29 & 0.68 & & 32 & & 0.78 & 0.56 & & 0.36 & 0.28 & & 95 & & 0.28 & 0.06 & & 0.25 & 0.16\\
WMT & & 8,085 & & 12.4 & -0.17 & 0.25 & & 30 & & 0.60 & 0.33 & & 0.29 & 0.21 & & 89 & & 0.17 & 0.03 & & 0.21 & 0.11\\
XOM & & 10,693 & & 14.7 & -0.07 & 0.17 & & 34 & & 0.38 & 0.15 & & 0.19 & 0.14 & & 103 & & 0.11 & 0.01 & & 0.14 & 0.07\\
SPY & & 18,154 & & 10.4 & -0.05 & 0.10 & & 45 & & 0.56 & 0.32 & & 0.21 & 0.16 & & 135 & & 0.13 & 0.02 & & 0.14 & 0.08\\
\hline \hline
\end{tabular}
\end{small}\smallskip
\begin{scriptsize}
\parbox{\textwidth}{\emph{Note.} This table reports descriptive statistics for our TAQ high-frequency data computed daily and averaged over time. The sample covers January 4, 2010 through December 31, 2013 for a total of $T = 1,006$ days. $n$ is the number of transaction data available after filtering, $\widehat{ \sigma} = \sqrt{ 256 \times \widehat{ \text{IV}}}$ is an annualized jump-robust measure of volatility based on \eqref{equation:IV-hat} (with $\theta = 1/3$), $\hat{ \rho}_{1}$ is the first-order autocorrelation of $\Delta_{i}^{n} Y$, and $\hat{ \xi}^{2}$ is the noise level $\times 10^{4}$. $k_{n}$ is the pre-averaging window, while $\#\mathcal{Z}^{n} > p_{1-\alpha/T}^{*}$ is the fraction of $t$-statistics for testing $\mathcal{H}_{0}$ larger than the ($1-\alpha/T$)-quantile of the bootstrap distribution of $\mathcal{Z}^{n*}$ (based on $z_{ \text{wb2}}$) with $\alpha = 0.05$. $\hat{\text{H}}$-index is the heteroscedasticity measure defined in \eqref{equation:H-index-hat}. The latter three are computed for both $\theta = 1/3$ and $\theta = 1$. A superscript $d$ refers to the average value of a statistic based on $\Delta_{i}^{n} Y^{d}$, where the rescaling is implemented via $\hat{ \sigma}_{u,t}$ from \eqref{equation:diurnal-variance-estimator}.}
\end{scriptsize}
\end{center}
\end{sidewaystable}